\documentclass{udthesis}

\usepackage{amsmath}
\usepackage{amssymb}
\usepackage{soul}
\usepackage{natbib}
\usepackage{graphicx}
\usepackage{caption}
\usepackage{subcaption}
\usepackage{enumerate}

\begin{document}
\newcommand{\aj}{AJ}			% Astronomical Journal
\newcommand{\araa}{ARA\&A}		% Annual Review of Astron and Astrophys
\newcommand{\apj}{ApJ}			% Astrophysical Journal
\newcommand{\apjl}{ApJLett}		% Astrophysical Journal, Letters
\newcommand{\apjlett}{ApJLett}		% Astrophysical Journal, Letters
\newcommand{\apjs}{ApJS}		% Astrophysical Journal, Supplement
\newcommand{\apjsupp}{ApJS}		% Astrophysical Journal, Supplement
\newcommand{\ao}{Appl.~Opt.}		% Applied Optics
\newcommand{\apss}{Ap\&SS}		% Astrophysics and Space Science
\newcommand{\aap}{A\&A}			% Astronomy and Astrophysics
\newcommand{\astap}{A\&A}		% Astronomy and Astrophysics
\newcommand{\aapr}{A\&A~Rev.}		% Astronomy and Astrophysics Reviews
\newcommand{\aaps}{A\&AS}		% Astronomy and Astrophysics, Supplement
\newcommand{\azh}{AZh}			% Astronomicheskii Zhurnal
\newcommand{\baas}{BAAS}		% Bulletin of the AAS
\newcommand{\jrasc}{JRASC}		% Journal of the RAS of Canada
\newcommand{\memras}{MmRAS}		% Memoirs of the RAS
\newcommand{\mnras}{MNRAS}		% Monthly Notices of the RAS
\newcommand{\na}{NewA}                   %  New Astronomy
\newcommand{\nar}{NewAR}                % New Astronomy Reviews
\newcommand{\pra}{Phys.~Rev.~A}		% Physical Review A: General Physics
\newcommand{\prb}{Phys.~Rev.~B}		% Physical Review B: Solid State
\newcommand{\prc}{Phys.~Rev.~C}		% Physical Review C
\newcommand{\prd}{Phys.~Rev.~D}		% Physical Review D
\newcommand{\pre}{Phys.~Rev.~E}		% Physical Review E
\newcommand{\prl}{Phys.~Rev.~Lett.}	% Physical Review Letters
\newcommand{\pasp}{PASP}		% Publications of the ASP
\newcommand{\pasj}{PASJ}		% Publications of the ASJ
\newcommand{\qjras}{QJRAS}		% Quarterly Journal of the RAS
\newcommand{\revmodphys}{Rev.\ Mod.\ Phys.} %Rev Mod Phys
\newcommand{\skytel}{S\&T}		% Sky and Telescope
\newcommand{\solphys}{Sol.~Phys.}	% Solar Physics
\newcommand{\sovast}{Soviet~Ast.}	% Soviet Astronomy
\newcommand{\ssr}{Space~Sci.~Rev.}	% Space Science Reviews
\newcommand{\zap}{ZAp}			% Zeitschrift fuer Astrophysik
\newcommand{\nat}{Nature}		% Nature
\newcommand{\iaucirc}{IAU~Circ.}       	% IAU Circulars
\newcommand{\aplett}{Astrophys.~Lett.} 	% Astrophysics Letters
\newcommand{\apspr}{Astrophys.~Space~Phys.~Res.}% Astrophysics Space Physics Research
\newcommand{\fcp}{Fund.~Cosmic~Phys.}  % Fundamental Cosmic Physics
\newcommand{\gca}{Geochim.~Cosmochim.~Acta}   % Geochimica Cosmochimica Acta
\newcommand{\grl}{Geophys.~Res.~Lett.} % Geophysics Research Letters
\newcommand{\jcp}{J.~Chem.~Phys.}	% Journal of Chemical Physics
\newcommand{\jgr}{J.~Geophys.~Res.}	% Journal of Geophysics Research
\newcommand{\nphysa}{Nucl.~Phys.~A}   % Nuclear Physics A
\newcommand{\physrep}{Phys.~Rep.}   % Physics Reports
\newcommand{\physscr}{Phys.~Scr}   % Physica Scripta
\newcommand{\planss}{Planet.~Space~Sci.}   % Planetary Space Science
\newcommand{\procspie}{Proc.~SPIE}   % Proceedings of the SPIE

\newcommand{\gtsim}{\mbox{{\raisebox{-0.4ex}{$\stackrel{>}{{\scriptstyle\sim}}$}}}}
\newcommand{\ltsim}{\mbox{{\raisebox{-0.4ex}{$\stackrel{<}{{\scriptstyle\sim}}$}}}}
\newcommand{\bs}{\boldsymbol}
\newcommand{\beq}{\begin{equation}}
\newcommand{\eeq}{\end{equation}}

\definecolor{jamie}{rgb}{1,0,0}
\newcommand{\jamie}[1]{\textcolor{jamie}{\bf #1}}

\definecolor{asif}{rgb}{0,0,1}
\newcommand{\asif}[1]{\textcolor{asif}{\bf #1}}

\definecolor{dylan}{rgb}{0,0.5,0.8}
\newcommand{\dylan}[1]{\textcolor{dylan}{\bf #1}}

\renewcommand{\bibname}{BIBLIOGRAPHY}

\title[Massive Star Disks]{Radiative Ablation \\of Disks Around Massive Stars}
\author{Nathaniel Dylan Kee}
\type{dissertation}
\degree{Doctor of Philosophy}
\majorfieldtrue\majorfield{Physics and Astronomy}
\degreedate{Fall, 2015}
% Optional PDF properties
%\keywords{Keyword,Keyword,Keyword}
%\subject{Subject}
\maketitlepage % Generates Title Page

\begin{approvalpage}
\chair{Edmund Nowak, Ph.D.}{Chair of the Department of Physics and Astronomy}
\dean{George Watson, Ph.D.}{Dean of the College of Arts and Sciences}
\end{approvalpage}
\begin{signedpage} % Up to 4 signatures
\profmember{Stanley Owocki, Ph.D.}
\member{Michael Shay, Ph.D.}
\member{Asif ud-Doula, Ph.D.}
\member{Dermott Mullan, Ph.D.}
\end{signedpage}

% For additional signatures beyond 4, uncomment and use
 \begin{signedpagecont}
 \member{Sarah Dodson-Robinson, Ph.D.}
 \member{Jamie Holder, Ph.D.}
 \end{signedpagecont}
 
 \begin{front} % Starts front material (Roman style page numbers)
\prefacesection{Acknowledgements}
In memory of Paul J. White, without whom none of this would have been possible.

Extensive thanks also go to my advisor, Dr. Stan Owocki, who tirelessly coached me through this process, and to my parents, Tara and David Kee, who always had words of encouragement.

Additional thanks go to my entire committee for their comments and critiques. % This file (acknowl.tex) contains the text
% for the acknowledgments or type text here.
% Table of Contents is always created, but you
% may set \tablespagefalse and \figurespagefalse
% if you don't want these generated automatically
% (i.e. List of Tables and List of Figures).
% These are set to true by default (i.e. \tablespagetrue,
% \figurespagetrue).
% Uncomment if you do not want a List of Figures.
%\figurespagefalse
% Uncomment if you do not want a List of Tables.
%\tablespagefalse
\maketocloflot
\prefacesectiontoc{Abstract}
Hot, massive stars (spectral types O and B) have extreme luminosities ($10^4-10^6L_\odot$) that drive strong stellar winds through UV line-scattering. Some massive stars also have disks, formed by either decretion from the star (as in the rapidly rotating ``Classical Be stars''), or accretion during the star's formation. Extending the wind-developed Sobolev methods for line radiative transfer, this dissertation examines the role of stellar radiation in driving (ablating) material away from these circumstellar disks.

A key result is that the observed month to year decay of optically thin Be disks can be explained by line-driven ablation without, as was done in previous work, appealing to anomalously strong viscous diffusion. Moreover, the higher luminosity of O stars leads to ablation of optically thin disks on dynamical timescales of order a day, providing a natural explanation for the lack of observed Oe stars.
In addition to the destruction of Be disks, this dissertation also introduces a model for their formation via ``Pulsationally Driven Orbital Mass Ejection''. This ``PDOME'' model couples observationally inferred non-radial pulsation modes and rapid stellar rotation to launch material into orbiting Keplerian disks of Be-like densities.

In contrast to such Be decretion disks, star-forming \emph{accretion} disks are much denser and so are generally optically thick to continuum processes like electron scattering. To circumvent the computational challenges associated with long-characteristic radiation hydrodynamics through optically thick media, we develop an approximate method for treating optically thick continuum absorption in the limit of geometrically thin disks. The comparison of ablation with and without continuum absorption shows that accounting for disk optical thickness leads to less than a 50\% reduction in ablation rate, implying that ablation rate is largely independent of disk mass, and depends mainly on stellar properties like luminosity.

Finally, as a side problem, we discuss the role of ``thin-shell mixing'' in reducing X-rays from colliding wind binaries. Laminar, adiabatic shocks produce well understood X-ray emission, but the emission from radiatively cooled shocks is more complex due to thin-shell instabilities. The parameter study conducted here systematically varies colliding wind binary shock densities to determine scaling relations for this emission. A key result is that, in the limit of strongly radiatively cooled shocks, emission is reduced by a fixed factor $\sim50$ from analytic scalings that ignore thin-shell structure.
 % This file (abstract.tex) contains the text
% for an abstract or type text here.
\end{front}

\chapter{Introduction}\label{chap:intro}

Luminous, massive stars play a crucial role in the chemistry and energetics of the galaxy. 
Through their high bolometric luminosity and peak flux of ultraviolet photons, they ionize their surroundings, and their violent demise in supernovae both energizes the interstellar medium while also enriching it with heavy elements synthesized in their interiors.
The high luminosity of massive stars also drives powerful stellar winds, with mass loss rates ranging up to $\sim10^{-5}\, M_\odot$/yr, more than billion times the much weaker, gas-pressure-driven solar wind
\citep{PulVin08}.
%The associated reduction of the star's own mass, often by a factor of two or more, can significantly alter its evolution.

These radiation forces likely even play a fundamental role in governing the upper mass limit of stars. 
As summarized in chapter \ref{chap:line}, if one takes a standard mass-luminosity scaling $L \propto M^3$, 
then the breaching of the classical Eddington limit \citep{Edd30}, wherein the basal radiation acceleration from just free electron scattering exceeds the stellar gravity, corresponds roughly with
the observationally inferred stellar upper mass limit\footnote{There is some contention as to whether the current record holder, the $300M\odot$ star R139a, is indeed a single star or an unresolved binary. 
Nevertheless, it seems clears that the limit is of order a few hundred solar masses and, by observational constraints, is certainly neither tens or thousands of times the mass of the sun.}
of $\sim 200-300 M_\odot$ \citep{CroSch10}.
While establishing a strict theoretical upper mass limit requires further detailed stellar structure and opacity modeling, the intense luminosity of massive stars remains a fundamental obstacle in stars reaching high masses. 
What remains unclear, however, is whether this upper mass limit is predominantly a stellar structure constraint, or perhaps instead a consequence of the formation mechanics of massive stars.

Initial star-formation models assuming spherically symmetric accretion concluded that radiation forces acting on dust opacity would limit the gravitationally induced accretion for cloud cores as low as 20 $M_\odot$ \citep{Kah74,WolCas87}.
More recent studies \citep{KruKle07,ComHen11,KuiYor15} account for the natural tendency of contracting clouds to collapse into an extended accretion {\em disk};
the associated optically thick shielding of the central equatorial layers of the disk from radiative forces then allows the gravitational accretion in the outer disk to proceed for cloud cores approaching the inferred stellar mass limit of a few hundred $M_\odot$.
However, in focussing on the accretion of dust-forming disks at scales of 10 to 100 au or more, there has been relatively little study of how such accretion proceeds down to scales of a few radii of the protostar, where the strong UV radiation is likely to ionize much of the disk, and also drive a stellar wind {\em outflow} through line-scattering of the stellar radiation by heavy ions.

Motivated by this, a central goal of this dissertation is to develop methods for treating the effects of such strong line-scattering forces in driving (or ablating) material from an orbiting disk in these near-star regions.

\section{Background Formalism on Winds and Disks}

Chapter \ref{chap:line} reviews the basic formalism of line-driven stellar winds. 
Drawing on the expertise of the massive star group at the University of Delaware, we use the approach of \cite{Cra96}, who generalized the standard, spherically symmetric line-driven wind model of \cite{CasAbb75} to non-spherical winds from rapidly rotating stars, including the effects of stellar oblateness and gravity darkening \citep{CraOwo95,OwoCra96}. 
These models retain the assumption that the background wind is optically thin in the continuum, but use the Sobolev approximation \citep{Sob60} to derive the radiative driving from scattering by a mixture of optically thin and thick lines.
Within this approximation, it is possible to calculate all three vector components of the line-acceleration in a fully three dimensional medium, assuming strong line-of-sight velocity gradients.
While the traditional implementation assumes that these velocity gradients arise in the radial direction from an expanding wind, they can also arise along non-radial rays from Keplerian shear in a disk (see, e.g. figure \ref{fig:vel_grad}).
Equation \ref{eq:gcak_3d_full} gives the line-acceleration prescription used in the remainder of the thesis to model line-driven ablation of circumstellar disks.

Chapter \ref{chap:disks} reviews the structure of the disks used in this dissertation.
For simplicity, the disks are assumed to be isothermal, which makes their vertical stratification Gaussian, with a scale height $H$ ($=R a/v_{orb}$) set by the ratio of sound speed $a$ to local orbital velocity $v_{orb}=\sqrt{GM/R}$ (where $M$ is the stellar mass and $R$ is the local cylindrical radius).
The radial variation of density is not set a priori, so we assume the equatorial density to be a power-law in cylindrical radius, $\rho_{eq} = \rho_\mathrm{o} (R/R_\ast)^{-n}$, where $\rho_\mathrm{o}$ is the equatorial density at the stellar surface\footnote{For accretion disks, one often takes $n\sim1.5$ $R_\ast$. As discussed below, we use $n=3.5$ as is appropriate for \emph{decretion} disks.}.
For constant opacity (a.k.a. mass absorption coefficient) $\kappa$, the radial optical depth through the disk midplane is then given by $\tau=\kappa\rho_\mathrm{o}R_\ast/(n-1)$.
For protostellar accretion disks, $\tau$ is extremely large ($>10^6$), which presents a fundamental problem for applying the standard approach of line-driving from an optically thin continuum.

\section{Optically Thin Disk Ablation}

In applying these line-driving methods in chapter \ref{chap:ab_thin}, the initial focus is thus not on the accretion disks of protostars, but on the lower density, marginally optically thin \emph{decretion} disks of Classical Be stars.
These rapidly rotating, main-sequence objects are capable of dynamically launching and destroying low density disks on time-scales of only a few months or years.
As such, Be disks provide their own unique set of challenging research problems.
For example, their observational signatures indicate the disks can decay on timescales of months to a year, which, within the standard model of mass transport by viscous diffusion, requires an anomalously high viscous coefficient\footnote{In the formalism introduced by \cite{ShaSun76} where the coefficient of kinematic viscosity is given by $\nu=\alpha a H$, \cite{CarBjo12} required $\alpha\sim1$.} \citep{CarBjo12}.

The broad parameter study of chapter \ref{chap:ab_spec_type} shows that the timescale for ablation for such a Be star can readily explain the month to year disk decay without invoking such an anomalously high viscosity.
Moreover, the decay of disks around hotter, more luminous O stars occurs on a dynamical timescale of order days.
This provides a natural explanation for the relative rarity of observed Oe stars.

\section{Disk Production Modeling}

To complement the modeling of disk destruction by ablation, chapter \ref{chap:pdome} introduces a ``Pulsationally Driven Orbital Mass Ejection'' (``PDOME'') model for disk production.
This couples rapid stellar rotation and non-radial pulsation modes to launch material into circumstellar orbit, and thereby form a Keplerian decretion disk with Be-like densities. This is done in the absence of the competing radiative-driving, leaving to future work a study of the direct competition between such production by mass ejection and destruction by radiative ablation.

\section{Optically Thick Disk Ablation}

While Classical Be stars provide a useful laboratory for star-disk interactions, recall that a central motivation of this thesis is to investigate the interaction of radiation with optically thick accretion disks within the last few protostellar radii.
To circumvent the computational challenges associated with long-characteristic radiation hydrodynamics through optically thick media, chapter \ref{chap:ab_thick} presents an approximate method for treating optically thick continuum absorption, in the limit of geometrically thin disks.
By applying this method to disks of accretion densities, and comparing to models that omit continuum optical depth effects, we find that optical depth effects lead to a modest, less than 50\% reduction on the rate of disk ablation. 
Furthermore, comparison of these results with those for the optically thin Classical Be stars demonstrates that disk ablation rate is largely independent of disk mass, and instead, like wind mass loss rate, depends mainly on stellar properties such as luminosity.

This chapter also can be seen as a method for modeling B[e] stars \citep{KraMir06}. This rare class of B supergiants is observed to host optically thick circumstellar disks, diagnosed through the presence of forbidden line emission. While the origin of these disks is uncertain, such disk-star systems can be seen, in many ways, as naked analogues to the high density star formation disks.

\section{X-ray Reduction by Thin-Shell Mixing of Radiative Shocks in Colliding Wind Binaries}

Chapter \ref{chap:thin_shell} presents a side discussion\footnote{While thematically somewhat peripheral, this work was instrumental in developing the skill sets necessary for the rest of this dissertation.} of X-ray diagnostics of massive star winds.
When stellar wind material collides supersonically, it can be shock heated to tens of millions of Kelvin and emit hard X-rays of energies up to tens of keV.
By studying this X-ray emission, we can learn about the gas that generated it.
These X-ray emitting shocks arise in a variety of astrophysical scenarios, but our discussion focusses on ``colliding wind binaries'', in which the winds of two massive stars collide head-on.
In situations where the binary separation is large, or the winds are relatively low density, these colliding wind binary shocks tend to be thick, laminar, and adiabatic, generating X-ray emission which has been readily interpreted \citep{SteBlo92,RauMos16}.
When the shocked layers are much denser, as occurs in close binary systems, the shock rapidly cools by radiation over a geometrically thin layer.
In multi-dimensional models, thin-shell instabilities break these layers into small, dense knots.
Our results show that this reduces the X-ray emission from radiative shocks by about a factor 50 relative to an analytic, laminar model without mixing.

\section{Conclusions and Future Work}

Chapter \ref{chap:conc} concludes the body of the dissertation with a summary and outline for future work. This includes immediate extensions and modifications applicable to optically thin Be disks, and longer term developments appropriate to the central goal of modeling much denser accretion disks of massive protostars. The latter will likely be a major focus of postdoctoral work by this doctoral candidate.

\chapter{Physics of Line-Driven Stellar Winds}\label{chap:line}

The regions around high-mass, luminous stars are characterized by outflows, referred to as stellar winds, with speeds of $\sim1000$ km/s and carrying away as much as $10^{-5}\, M_\odot$/yr. These winds are driven by the scattering of photons in spectral lines. To illuminate the issues surrounding the interaction of radiation and gas that form the core of this dissertation, discussing the launching of these winds provides an excellent jumping off point.

\section{Radiative Acceleration \label{sec:rad_acc}}

 To begin, let us consider the most general form of a radiative acceleration. For intensity $I$ at frequency $\nu$ and position $\mathbf{r}$ in direction $\mathbf{\hat{n}}$, absorption or scattering by material with mass absorption coefficient (a.k.a. opacity) $\kappa$ gives a vector acceleration \citep[see e.g.][]{Mih78}

\beq\label{eq:g_rad}
\mathbf{g}_{rad}=\frac{1}{c}\oint\int_0^\infty \kappa(\nu) \,I(\nu,\mathbf{r},\mathbf{\hat{n}}) \,d\nu \,\mathbf{\hat{n}} \,d\Omega\, ,
\eeq
where $c$ is the speed of light and $\Omega$ is solid angle. 

\subsection{Acceleration from electron scattering}

In general, $\kappa$ can arise either from continuum or line opacity. The atmospheres of main sequence massive stars have temperatures greater than $10^4\, K$, hot enough that they are effectively fully ionized, implying that continuum opacity is dominated by electron scattering. Since electron scattering is frequency independent, or gray, and, when averaged over many photons, isotropic, $\kappa$ can be pulled out of both integrals in equation \ref{eq:g_rad}, leaving only the scaling with stellar bolometric flux, $\mathbf{F}_\ast$:

\beq\label{eq:g_e}
\mathbf{g}_e = \frac{\kappa_e \mathbf{F}_\ast}{c} = \frac{\kappa_e L_\ast}{4\pi r^2 c} \mathbf{\hat{r}}\, .
\eeq
The second equality applies in the case of a spherically symmetric star with luminosity, $L_\ast$ viewed from a radius $r$.

Since the acceleration due to electron scattering has the same $1/r^2$ scaling as gravity, it is useful to define their ratio \citep{Edd30},

\beq\label{eq:Edd_lim}
\Gamma_e \equiv \frac{g_e}{g_{grav}}= \frac{\kappa_e L_*}{4\pi G M_* c}\, ,
\eeq
where $M_\ast$ is the stellar mass and $G$ is Newton's gravitation constant.
As $\Gamma_e\rightarrow 1$ (referred to as the classical Eddington limit), the star formally becomes gravitationally unbound and is subject to highly unstable mass loss \citep{OwoGay97,Owo15}. For the Sun, $\Gamma_{e,\odot}\approx 2.5\times10^{-5}$; if one takes the standard scaling that $L_\ast \propto M_\ast^3$ on the main sequence, one infers that $\Gamma_e\approx1$ for $M_\ast\approx200\,M_\odot$, in good agreement with the observationally inferred upper mass limit.\footnote{In practice the mass-luminosity scaling becomes nearly linear for high mass star as radiation reduces the gravitational force by a factor of $(1-\Gamma_e)$. Nevertheless, this simple calculation gives a good first-order estimate of the observed stellar mass limit.}

For the stars considered in this thesis, $\Gamma_e$ ranges from $10^{-3}$ to $0.5$. To drive material away from a star in a stellar wind -- or by ablation of a circumstellar disk -- requires the radiative acceleration to exceed gravity in these circumstellar regions, with associated $\Gamma>1$. As discussed in the next section, the inclusion of line opacity can readily achieve this.

%In the more general case of $\Gamma_{rad}$, where the subscript ``rad'' refers now to any radiation force and denotes the replacement of $\kappa_e$ with $\kappa_{rad}$ (the opacity associated with the chosen radiation force), $\Gamma_{rad}$ increasing from the stellar interior to pass unity at some position above the star is the condition necessary to drive a stellar wind by radiation. Scattering off of the spectral lines which appear due to recombination in the atmospheres of high mass stars produces an opacity which behaves in just such a way.

\subsection{Acceleration from a single, isolated spectral line}\label{sec:single_line_force}

In contrast to the continuum scattering of free electrons, line-scattering from bound electrons is confined to a very narrow range around a resonance frequency, $\nu_\mathrm{o}$. However, the resonant nature means that the frequency-integrated opacity can be much stronger than the opacity of free electrons. When the line resonance is removed from its own shadow by the Doppler shift of a flow acceleration, the associated radiative force can also be much stronger than for free electrons.

To quantify this, let us define a direction- and frequency-independent line center opacity, $\kappa_L$, and a frequency-dependent shape of the line-profile, given by the normalized profile function $\phi(\nu,\mathbf{n},\mathbf{r},\mathbf{v}(\mathbf{r}))$. The radiative acceleration due to a single, isolated line can then be cast as
%Considering the scattering of UV photons off of spectral lines as our chosen opacity source, $\kappa$ is no longer frequency independent but can be assumed to be able to be split into 

\beq\label{eq:g_line_gen_1}
\mathbf{g}_{line}=\frac{\kappa_L}{c} \oint \int_0^\infty \phi(\nu,\mathbf{n},\mathbf{r},\mathbf{v}(\mathbf{r})) \,I(\nu,\mathbf{r},\mathbf{n}) \,d\nu \,\mathbf{\hat{n}} \,d\Omega\, .
\eeq
The line center opacity can be calculated by \citep[e.g.][]{PulOwo93},

\beq\label{eq:chiL}
\kappa_L(\mathbf{r})=\frac{\sigma_L(\mathbf{r}) n_l(\mathbf{r})}{\rho(\mathbf{r})}\, ,
\eeq
where $\rho$ is the mass density and $n_l$ is the number population of atoms that can be photo-excited in the transition. The line cross-section $\sigma_L$ depends on the classical oscillator, $\pi e^2/m_e c=\pi r_e c$ (given in terms of the mass, radius, and charge of an electron, $m_e$, $r_e$, and $e$ respectively), and the associated quantum mechanical oscillator strength of the transition, $f_{lu}$,

\beq\label{eq:sigL}
\sigma_L(\mathbf{r})=\frac{\pi e^2}{m_e c} \frac{1}{\Delta \nu_D} f_{lu}\, .
\eeq
The inverse proportionality to the Doppler width of the line, $\Delta \nu_D \equiv \nu_\mathrm{o} v_{th}/c$ -- which describes the characteristic scale of thermal broadening of the spectral line about its natural frequency, $\nu_0$, in terms of the ion thermal speed, $v_{th}$ -- converts the frequency times cross-section units of the classical oscillator to the cross-section units of $\sigma_L$.

To illustrate the relative strength of line scattering to scattering off of free electrons, let us define the ratio q to be

\beq
q\equiv \frac{\kappa_L}{\kappa_e}\frac{v_{th}}{c}\, ,
\eeq
which, recalling the inverse dependence of $\kappa_{L}$ on $\Delta \nu_D$, is \emph{independent} of $v_{th}$. In parallel with equations \ref{eq:chiL} and \ref{eq:sigL}, electron scattering opacity scales as 
\beq
\kappa_e =\frac{\sigma_{Th} n_e}{\rho}\, ,
\eeq
where

\beq
\sigma_{Th}=\frac{8}{3} \pi r_e^2\, ,
\eeq
is the Thompson cross-section. For an allowed transition with $f_{lu}\sim1$, $q$ can then be cast in terms of the mean mass per electron, $\mu_e=n_e/\rho$, and the mean mass per ion that can be excited, $\mu_l=n_l/\rho$, such that

\beq
q=\frac{3}{8}\frac{\lambda_\mathrm{o}}{r_e}\frac{\mu_e}{\mu_l}\, .
\eeq
For a line in the optical, $\lambda_\mathrm{o}/r_e\sim10^8$ but, for the metal ion resonance lines that provide the dominant contribution to line opacity for massive stars, $\mu_e/\mu_l\sim10^{-4}-10^{-5}$. Nonetheless, this still gives a characteristic value of several thousand for $q$, confirming the inherent strength of resonance scattering lines noted above.

For an optically thin line, the intensity is not significantly attenuated by the line at any frequency and can be pulled out of the frequency integration. Since the profile-function is normalized, we can now write

\beq\label{eq:g_thin}
g_{thin}=\frac{\kappa_L \Delta \nu_D L_{*,\nu}}{4\pi r^2 c}=\frac{\kappa_e q w_\nu L_\ast}{4 \pi r^2 c}=q w_\nu g_e\, ,
\eeq
where the second equality introduces the weighting factor $w_\nu \equiv\nu_\mathrm{o}L_{\ast,\nu}/L_\ast$ which characterizes the placement of the line in the stellar spectrum. For lines near the peak of the flux spectrum $w_\nu \lesssim 1$, we now see that the acceleration from a \emph{single} spectral line can be boosted by up to a factor $q\sim10^3$ over the electron scattering force integrated over the whole continuum.

In practice, such strong lines will be optically thick, and one must account for the attenuation of the intensity by self absorption within the spectral line. In general, this can involve complicated non-local integrations. However, in the presence of strong velocity gradients \cite{Sob60} showed that this line transfer can be reduced to a purely local calculation. For example, for the simplified case of radially streaming photons and a spherically symmetric outflow undergoing strong radial acceleration $dv_r/dr$, a photon emitted at a frequency $\nu\gtrsim\nu_\mathrm{o}$ can only interact with a line over a short resonance zone of width,

\beq
l_{Sob}\equiv \frac{v_{th}}{dv_r/dr}\, ,
\eeq
and centered on a radius $r_{res}$, where the photon has been brought into resonance by the Doppler shift of the outflow, i.e. with $v_r(r_{res})=c(\nu/\nu_\mathrm{o}-1)$.
For supersonic outflow with $v_r\gg v_{th}$, this ``Sobolev length'' is quite small, allowing one to obtain a local expression for the integrated line optical depth
\beq
\tau_r\equiv \frac{\kappa_L \rho v_{th}}{dv_r/dr}\,.
\eeq
In the general case of a vector velocity $\mathbf{v}$, the analogous optical depth along a direction $\mathbf{\hat{n}}$ is
\beq
\tau_n\equiv \frac{\kappa_L \rho v_{th}}{dv_n/dn}\,,
\eeq
where $dv_n/dn\equiv\hat{\mathbf{n}}\cdot\nabla(\hat{\mathbf{n}}\cdot\mathbf{v})$. While the general expression for this gradient is algebraically cumbersome to express, its components in various geometries are available, for instance, in \cite{Bat67}.

This ``Sobolev optical depth'' allows one to account for the line attenuation of a given continuum intensity\footnote{Intensity is usually assumed to be the unattenuated stellar intensity. However, this approximation breaks down in the presence of very dense circumstellar disks, as is addressed in chapter \ref{chap:ab_thick}.} $I_{\nu,\ast}$ that is interacting with the spectral line. Defining a frequency displacement from line center in units of the thermal Doppler width
\beq
x\equiv \frac{\nu-\nu_\mathrm{o}}{\Delta\nu_D}\, ,
\eeq 
and velocity in units of the thermal velocity $\mathbf{u}\equiv\mathbf{v}/v_{th}$, the local intensity in a direction $\mathbf{\hat{n}}$ at vector location $\mathbf{r}$ is the continuum intensity exponentially attenuated by an optical depth term, which in the Sobolev approximation is given by
\beq\label{eq:exp_attenuated_int}
I(\nu,\mathbf{r},\mathbf{\hat{n}})=I_{\nu,\ast}e^{-\tau_n \Phi(x-\mathbf{\hat{n}}\cdot\mathbf{u})}\, .
\eeq
Here we have introduced the integrated profile function,
\beq
\Phi(x)\equiv\int_{x}^\infty \phi(x')dx'\,,
\eeq
which, for a Gaussian profile is given by the complementary error function.

Applying \ref{eq:exp_attenuated_int} in equation \ref{eq:g_line_gen_1} gives, for the vector line force

\beq\label{eq:g_line_gen_3}
\mathbf{g}_{line}=\frac{\kappa_L \Delta \nu_D}{c}\oint  I_{*,\nu} \int_{-\infty}^\infty \phi(x-\mathbf{\hat{n}}\cdot\mathbf{u}) e^{-\tau_n \Phi(x-\mathbf{\hat{n}}\cdot\mathbf{u})} \,dx \,\mathbf{\hat{n}} \,d\Omega\, .
\eeq
Noting that $\phi dx=-d\Phi$, the frequency integration reduces to
\beq
\int_{0}^1 e^{-\tau_{n}\Phi} \,d\Phi = \frac{1-e^{-\tau_{n}}}{\tau_{n}}\, ,
\eeq
giving, then
\beq
\mathbf{g}_{line}=\frac{\kappa_L \Delta \nu_D}{c}\oint_{4 \pi} I_{*,\nu}\frac{1-e^{-\tau_{n}}}{\tau_{n}} \,\mathbf{\hat{n}} \,d\Omega\, .\label{eq:g_line}
\eeq

%For future chapters, this full three-dimensional force from a single line will be used. However, as it can not in general be analytically evaluated (due to the inclusion of $dv_n/dn$), for the remainder of this and the following section, only the radial component of the line-force will be used. This will allow for derivation of analytic scaling relations. However, we will return to this full 3D expression in the final section of this chapter.

For radial streaming radiation, with $\mathbf{\hat{n}}=\mathbf{\hat{r}}$ as from a point star, the line acceleration is only in the radial direction $\mathbf{g}_{line}=g_{line} \mathbf{\hat{r}}$, where 
%then $\tau_n$ is replaced by $\tau_r$ and the integral over solid angle reduces to $F_{*,\nu}$ such that

\beq\label{eq:g_line_simp}
g_{line}=g_{thin}\frac{1-e^{-\tau_{r}}}{\tau_{r}}\, .
\eeq
For weak lines with low optical depth, $\tau_r\ll1$, this recovers the optically thin expression, $g_{thin}$, given by \ref{eq:g_thin}. On the other hand, for strong lines with high optical depth, we obtain 

\beq
g_{thick} \approx \frac{g_{thin}}{\tau_{r}} = \frac{w_{\nu} L_{*}}{4 \pi r^2 c^2 \rho} \frac{dv_r}{dr}\, \:\: ; \:\: \tau_r\gg1 .
\eeq
Note that in the second expression, since both $g_{thin}$ and $\tau_{r}$ depend linearly on $\kappa_L$, $g_{thick}$ is \emph{independent} of $\kappa_L$. This can be understood by recognizing that once a line is optically thick it can scatter all photons that encounter it, regardless of its intrinsic strength.

\subsection{Acceleration from an ensemble of lines}\label{sec:line_ens}

The analyses above have only considered the acceleration from a single line. However, spectra of massive stars show many spectral lines that can contribute to radiation driving. \cite{CasAbb75} (hereafter CAK) developed a general formalism to account for a large number of non-overlapping lines. Following CAK\footnote{The form here replaces the fiducial thermal speed $v_{th}$, as used by CAK, with the speed of light $c$. This removes the artificial dependance of both $t$ and the line acceleration on $v_{th}$.}, let us introduce the optical depth for a line with $q=1$,
\beq
t=\frac{\kappa_e \rho c}{dv_r/dr}\, .
\eeq
 This has the advantage of encapsulating the physics of Doppler shift into $t$ and line strength into $q$, such that $\tau_{r}=qt$. Making this substitution, the total line force is formally given by

\beq\label{eq:gtot_sum}
g_{tot} = \sum g_{thin}\frac{1-e^{-\tau_{r}}}{\tau_{r}}= g_e \sum w_\nu  q \frac{1-e^{-qt}}{qt}\, .
\eeq
As there are typically thousands of lines contributing to this sum, the sum can be well approximated by an integral over a flux-weighted number distribution\footnote{This is defined such that each line contributes its associated $w_\nu$ to the integration over lines.} $dN/dq$ in line strength $q$,

\beq\label{eq:gtot_int}
g_{tot} = \frac{g_e}{t} \int_{0}^{\infty} \left(1-e^{-qt}\right) \frac{dN}{dq}dq\, .
\eeq
Following CAK (see their appendix), the number distribution in opacity can be well approximated by a power-law.
%by plotting the associated finite differenced quantity for a detailed line list\footnote{A variety of distribution functions have been used in the literature \citep[e.g.][]{Gay95,OwoCas88,PulSpr00}, but all have had a power-law functional form. Appendix \ref{app:kappa_to_q} discusses the translation between some of the more prevalent of the notations used in such distribution functions.}. For purposes of this derivation, a slightly modified version of the distribution given by 
Using a notation analogous to that introduced by \cite{Gay95}, we express this as

\beq\label{eq:gayley_powlaw}
\frac{dN}{dq} = \frac{1}{\Gamma(\alpha)Q}\left(\frac{q}{Q}\right)^{\alpha-2} \,,
\eeq
where $Q$ can be interpreted as a flux and population weighted $q$, and $\alpha$ is a temperature dependant fit parameter between about 0.5 and 0.65 \citep{PulSpr00}. 

Evaluating equations \ref{eq:gtot_int} and \ref{eq:gayley_powlaw} yields the CAK line-acceleration,

\beq\label{eq:g_CAK}
\mathbf{g}_{CAK} = \frac{g_e Q^{1-\alpha}}{(1-\alpha) t^\alpha}=\left[\frac{(\kappa_e Q)^{1-\alpha} L_\ast}{4 \pi(1-\alpha) c^{1+\alpha}}\right]\frac{1}{r^2}\left(\frac{1}{\rho}\frac{dv_r}{dr}\right)^\alpha\mathbf{\hat{r}}\, ,
\eeq
where the square bracket term is constant, showing then explicitly the spatial dependance of the acceleration, specifically its dependance on the velocity gradient to the $\alpha$ power. CAK cast this through a ``force multiplier'', defined such that $g_{CAK}/g_e\equiv M(t)=k t^{-\alpha}$, where now $k=Q^{1-\alpha}/(1-\alpha)$.

\section{Line-Driven Stellar Winds}\label{sec:ldsw}

%\section{CAK wind solutions in the zero sound speed limit}
In addition to describing the behavior of line acceleration, the CAK formalism provides a basis for deriving stellar wind solutions.
In the limit of zero sound speed, which neglects the relatively small contribution of gas pressure in accelerating the wind outflow, the steady-state, spherically symmetric equation of motion takes the form

\beq\label{eq:sph_sym_eq_mom}
v_r\frac{d v_r}{d r} = -\frac{G M_*(1-\Gamma_e)}{r^2} +\frac{g_e Q^{1-\alpha}}{(1-\alpha) t^\alpha}\, .
\eeq
The left hand term represents the contribution of inertia, while the two terms on the right respectively represent the inward pull of gravity (reduced by continuum electron scattering) and the outward push of lines. 

\begin{figure}
\centering
\includegraphics[width=0.85\textwidth]{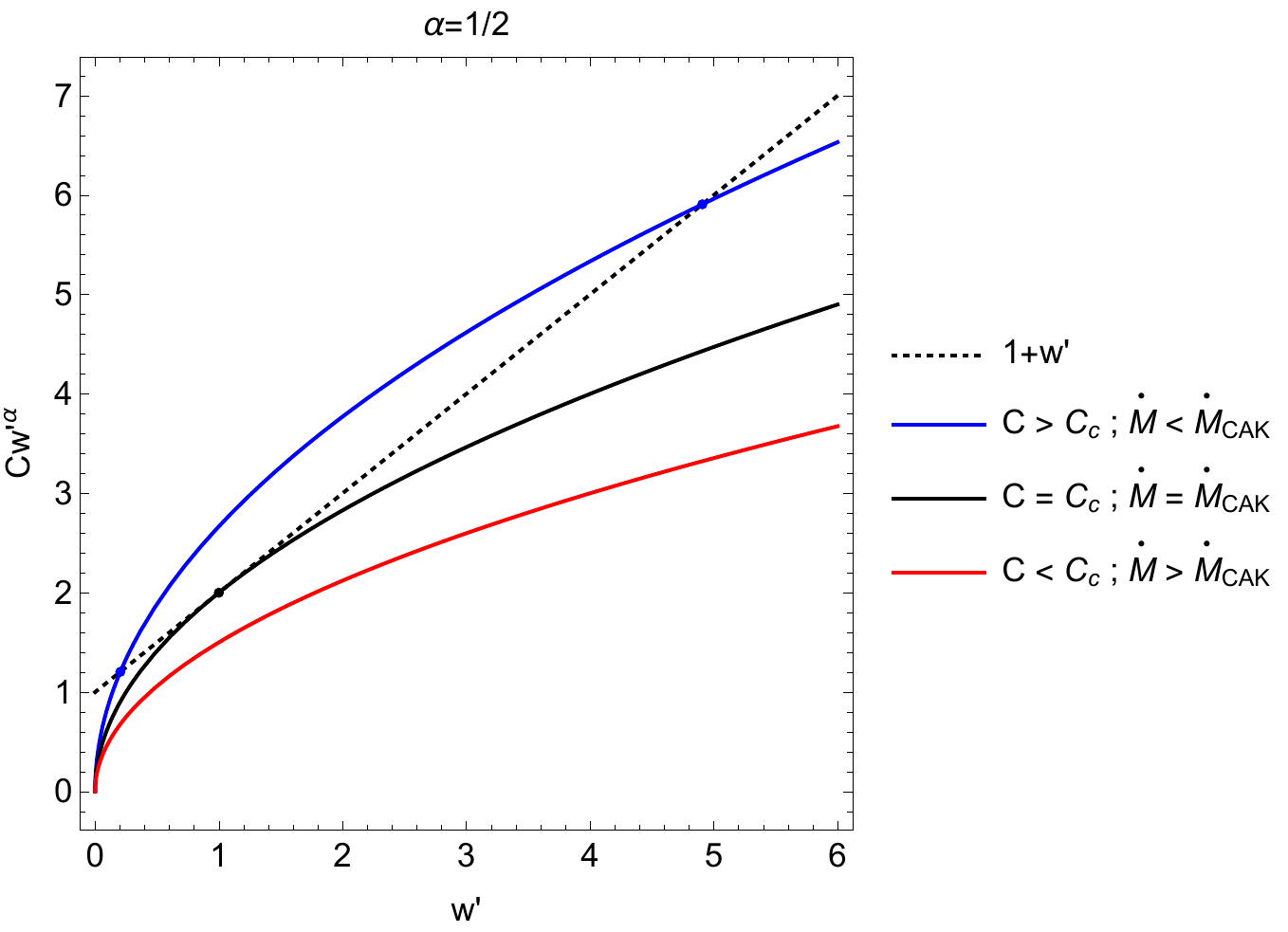}
\caption{
Overplot of the left and right sides of equation \ref{eq:motion_CAK} for different values of C. For C too large two solutions exist while for C too low no solutions exit, such that there is a unique value of C for which only a single solution exists.
}
\label{fig:ccrit}
\end{figure}

Introducing the variable transformations $w=v^2/v_{esc}^2$ (with $v_{esc}^2\equiv GM_*(1-\Gamma_e)/R_*$), $x=1-R_*/r$, and $w'=dw/dx$, equation \ref{eq:sph_sym_eq_mom} reduces to the form
\beq\label{eq:motion_CAK}
Cw'^\alpha = w'+1
\eeq
\beq
C = \frac{1}{1-\alpha} \left(\frac{L_*}{\dot{M} c^2}\right)^\alpha \left(Q\frac{\Gamma_e}{1-\Gamma_e}\right)^{1-\alpha}\, ,
\eeq
where the constant $\dot{M}=4\pi\rho v_r r^2$ is the wind mass loss rate. Figure \ref{fig:ccrit} shows an overplot of the left and right sides of equation \ref{eq:motion_CAK} and demonstrates that, for a given value of $\alpha$, equation \ref{eq:motion_CAK} has zero, one, or two solutions depending on the value of $C\propto\dot{M}^{-\alpha}$.
%For C smaller than this optimal value (i.e. $\dot{M}>\dot{M}_{CAK}$), no solutions exist and any outflow with this mass loss rate will not make it out of the stellar gravitational field. On the other hand, for C larger than this optimal value (i.e. $\dot{M}<\dot{M}_{CAK}$), two solutions exist, one which is slow and dense (the solution with smaller $w'$) and one which is fast and tenuous (the solution with higher $w'$). 
The minimal allowed, or ``critical'', value of $C_c=\alpha^{-\alpha}(1-\alpha)^{\alpha-1}$ is an attractor solution and gives the optimal case for which the star drives the maximum possible mass loss,

\beq\label{eq:mdot_CAK}
\dot{M}_{CAK}=\frac{\alpha}{1-\alpha}\frac{L_*}{ c^2} \left(Q\frac{\Gamma_e}{1-\Gamma_e}\right)^{(1-\alpha)/(\alpha)}\, ,
\eeq
This corresponds to a critical acceleration $w'_c=\alpha/(1-\alpha)$. Since \ref{eq:motion_CAK} has no explicit position dependance, $w'=w'_c$ everywhere, and this can be integrated to give a velocity-law

\beq\label{eq:vel_law}
v_r(r)=\sqrt{\frac{\alpha}{1-\alpha}}v_{esc}\sqrt{1-\frac{R_\ast}{r}}=v_\infty\sqrt{1-\frac{R_\ast}{r}}\, ,
\eeq
where $v_{\infty}$ is the terminal speed of the stellar wind.

Let us now investigate the general scaling of $\dot{M}_{CAK}$ with stellar parameters. For all the stars in this dissertation, and indeed all but the very most massive of stars, the scaling with $(1-\Gamma_e)$ can be ignored leaving

\beq
\dot{M}_{CAK}\propto \frac{L_\ast^{1/\alpha}}{M_\ast^{(1-\alpha)/(\alpha)}}\,.
\eeq
Since $\alpha\sim0.6$ and the scaling of luminosity with mass is approximately $L_\ast\propto M_\ast^3$, mass loss rate is a very steep function of stellar parameters going roughly as $\dot{M}_{CAK}\propto L_\ast^{1.7}\propto M_\ast^{5.0}$.

The preceding analysis ignores the finite cone angle subtended by the star by assuming the star to be a point source. Including this lowers the mass loss rate by about $50\%$ while increasing the wind terminal speed by a comparable factor \citep{FriAbb86,PauPul86}. The derivation also ignores the contribution of pressure forces to accelerating the wind, the inclusion of which increases mass loss rate and decreases the wind terminal speed by around $10\%$ \citep{OwoudD04}. Finally, this analysis also ignores the ionization balance of the stellar wind and how this contributes through the spectral lines available to radiation. This effect is more subtle and is discussed in appendix \ref{app:ionization}.

In general, the inclusion of these same effects tends to flatten the velocity-law. To account for this, we generalize equation \ref{eq:vel_law} to

\beq\label{eq:beta_vel_law}
v_r(r)=v_\infty \left(1-\frac{R_\ast}{r}\right)^{\beta}\,.
\eeq
Fits of this to simulations including the corrections above give $\beta\sim0.8-1.0$.

%\subsection{Inclusion of the effects of a finite stellar cone angle}\label{sec:finite_disk}

%The analysis thus far has all been carried out under the assumption that the star in question could be treated as a point source.\footnote{Several other approximations have also been tacitly made. For a treatment of another of these, see appendix \ref{app:ionization}} However, especially at distances comparable to the stellar radius, this approximation can be improved upon. By not taking the approximation that $\mathbf{\hat{n}}=\mathbf{\hat{r}}$, (immediately following equation \ref{eq:g_line}), the rest of the derivation can still be done with the exception of the derivative over solid angle such that
%\begin{align}
%g_{tot,fd} &= \frac{(\kappa_e Q)^{1-\alpha}}{(1-\alpha)c^{1+\alpha}\rho^\alpha}\oint I_\ast \left(\frac{dv_r}{dr}\right)^\alpha \hat{\mathbf{n}} d\Omega \\
%&\approx f_{fd}\, g_{tot}\, ,
%\end{align}
%where $f_{fd}$ is referred to as the finite disk correction factor. Careful analysis shows that $f_{fd} \approx 1/(1+\alpha)$ \citep{FriAbb86,PauPul86} such that
%\beq
%\dot{M}_{fd}=f_{fd}^{1/\alpha} \dot{M}_{CAK}=\frac{\dot{M}_{CAK}}{(1+\alpha)^{1/\alpha}}
%\eeq

\section{Inclusion of an Exponential Cutoff to the Line Force}

One further adjustment to the CAK line acceleration is the inclusion of an exponential cut-off in the line distribution function, equation \ref{eq:gayley_powlaw}. Since the distribution function in principle allows there to be a line of $q=\infty$, and $g_{thin}\propto q$, formally the line force is allowed to approach infinite strength. This is not a physically meaningful situation which can be remedied by including an exponential cut-off in $q$. \cite{OwoCas88} introduced such a cut-off making the distribution function 

\beq\label{eq:gayley_powlaw_2}
\frac{dN}{dq} = \frac{\bar{Q}}{\Gamma(\alpha) Q_\mathrm{o}^2}\left(\frac{q}{Q_\mathrm{o}}\right)^{\alpha-2}e^{-q/Q_0}\, ,
\eeq
where $Q$ as used before is related to the new $\bar{Q}$ and $Q_\mathrm{o}$ by $Q^{1-\alpha}=\bar{Q}Q_\mathrm{o}^{-\alpha}$. Repeating the analysis done before,

\beq\label{eq:g_tot}
g_{tot}=\frac{g_e\bar{Q}}{(1-\alpha) }\left[ \frac{(1+\tau_{\mathrm{o},r})^{1-\alpha}-1}{\tau_{\mathrm{o},r}} \right]\, ,
\eeq
where $\tau_{\mathrm{o},r}\equiv Q_\mathrm{o}\kappa_e\rho c/(dv_r/dr)=Q_\mathrm{o}t$. In the limit where $\tau_{\mathrm{o},r} \gg 1$, we obtain 
\beq
g_{tot}=\frac{g_e\bar{Q}}{(1-\alpha) Q_\mathrm{o}^\alpha}\frac{1}{t^\alpha}\, ,
\eeq
and, using $Q^{1-\alpha}=\bar{Q}Q_\mathrm{o}^{-\alpha}$, this reduces to $g_{CAK}$.

\section{A Generalized Three-Dimensional Formalism}\label{sec:3d_line_force}

Although the spherically-symmetric case with radially streaming photons is illustrative, for the disk ablation models in this dissertation the non-radial velocity gradient components can also come into play. Therefore, we need a fully three-dimensional (both in flux prescription and components of the velocity gradient) line-acceleration. \cite{CraOwo95} introduced such a formalism. 

While the simplicity of a 1D formalism is emphasized in the prior sections, this is not the key elegance of the CAK formalism. Indeed, without the Sobolev approximation, we would not have been able to obtain any of these results. By realizing that the Sobolev approximation is valid along any line of sight $\mathbf{\hat{n}}$, rather than just along the radial direction $\mathbf{\hat{r}}$, it becomes evident that we can generalize the derivations above by taking

\beq
dv_r/dr\rightarrow\mathbf{\hat{n}}\cdot\nabla(\mathbf{\hat{n}}\cdot\mathbf{v})\, .
\eeq
With this substitution, the preceding derivations which led to equation \ref{eq:g_tot} now generalize to a form with angle integration over source intensity $I_\ast$,

\beq\label{eq:gcak_3d_full}
\mathbf{g}_{CAK,3D} = \frac{(\kappa_e)^{1-\alpha}\bar{Q}}{(1-\alpha)c^{1+\alpha}\rho^\alpha}\oint I_\ast \left[ \frac{(1+\tau_{\mathrm{o},n})^{1-\alpha}-1}{\tau_{\mathrm{o},n}} \right] \hat{\mathbf{n}} d\Omega\, ,
\eeq
where $\tau_{\mathrm{o},n}\equiv Q_\mathrm{o}\kappa_e\rho c/(\mathbf{\hat{n}}\cdot\nabla(\mathbf{\hat{n}}\cdot\mathbf{v}))$ is the maximum optical depth allowed for a line in $dN/dq$.
In parallel to the prior section, for $\tau_{\mathrm{o},n}\gg 1$ this reduces to

\beq\label{eq:gcak_3d}
\mathbf{g}_{CAK,3D} = \frac{(\kappa_e)^{1-\alpha}\bar{Q}}{(1-\alpha)c^{1+\alpha}\rho^\alpha Q_\mathrm{o}^\alpha}\oint I_\ast \left[\mathbf{\hat{n}}\cdot\nabla(\mathbf{\hat{n}}\cdot\mathbf{v})\right]^\alpha \hat{\mathbf{n}} d\Omega\, .
\eeq
As emphasized before, the central scaling of this equation\footnote{The $1/r^2$ scaling cited here is buried in the solid angle integral.}  is with

\beq
g_{CAK,3D} \propto \frac{1}{r^2}\left(\frac{\mathbf{\hat{n}}\cdot\nabla(\mathbf{\hat{n}}\cdot\mathbf{v})}{\rho} \right)^\alpha
\eeq
Like its 1D equivalent, this involves only purely local calculations making its implementation in a finite differencing hydrodynamics code straightforward. 

This dissertation presents the first application of the fully 3D line acceleration to a circumstellar, Keplerian disk. However, prior work has demonstrated the importance of correctly treating these non-radial effects for winds from rotating stars. 
Such effects become particularly pronounced when the star is so rapidly rotating that it becomes ``oblate'' with an equatorial radius that is larger than that over the poles. Since the local net flux is perpendicular to the local surface and the stellar surface is no longer on a sphere, at mid-latitudes it has a poleward component that drives material away from the equator. In addition, since photons take the 
path of minimal optical depth when passing through an optically thick medium, the equator of such rapidly rotating stars becomes ``gravity darkened'' \citep[see, e.g.][]{Cra96,Zei24}. Appendix \ref{app:ob_gravdark} discusses these effects more extensively. 

\begin{figure}[h!]
\includegraphics[width=\textwidth]{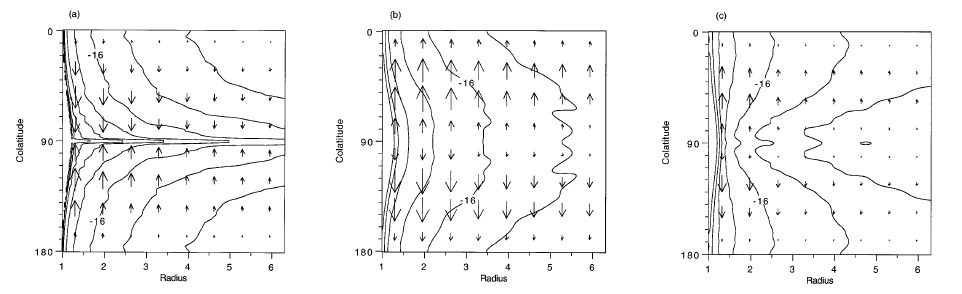}
\caption{
With permission, figure 1 of \cite{OwoCra96}. The panels show density contours and the latitudinal component of velocity. From left to right the panels correspond to a) spherically symmetric radiation, b) inclusion of non-radial line forces, and c) inclusion of both non-radial forces and gravity darkening. 
}
\label{fig:OwoCra96_fig1}
\end{figure}

As discussed in \cite{OwoCra96}, one notable example of the importance of such effects is in their inhibition of the ``wind-compressed disk'' effect, proposed by \cite{BjoCas93} to explain the circumstellar emission from the rapidly-rotating, Be stars\footnote{Such Be stars form a significant part of this dissertation and are discussed in more detail in chapter \ref{chap:disks}.}. Their central idea was that radiatively driven material which overcomes the centrifugally reduced gravity of a rapidly rotating star would, by angular momentum conservation, cross the stellar equator and collide with material from the opposite hemisphere, forming an outflowing wind-compressed disk. Initial simulations of \cite{OwoCra94} that assume pure radial driving, and so ignore oblateness and gravity-darkening, indeed produce such a disk. Subsequent simulations by \cite{OwoCra96} show, however, that when oblateness is included the equatorward flow is reversed, leaving only a weak equatorial over-density associated with the lower net gravity near the equator. Moreover, when equatorial gravity-darkening is also included, the over-density shifts toward the poles, suggesting that rapidly rotating \emph{oblate} stars should actually have \emph{prolate} winds. Figure \ref{fig:OwoCra96_fig1} shows this ``wind-compressed disk inhibition'' in both density contours and latitudinal velocity vectors.

\begin{figure}
\includegraphics[width=\textwidth]{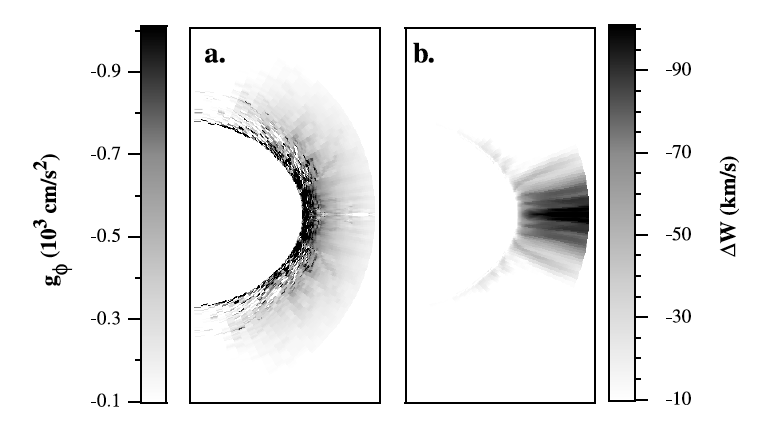}
\caption{
With permission, figure 1 of \cite{GayOwo00}. The panels show, for the region around a rapidly rotating B star, a) the azimuthal force component ($g_\phi$) and b) net change in azimuthal velocity with respect to a model ignoring $g_\phi$.
}
\label{fig:GayOwo00_fig1}
\end{figure}

\begin{figure}
\centering
\includegraphics[width=\textwidth]{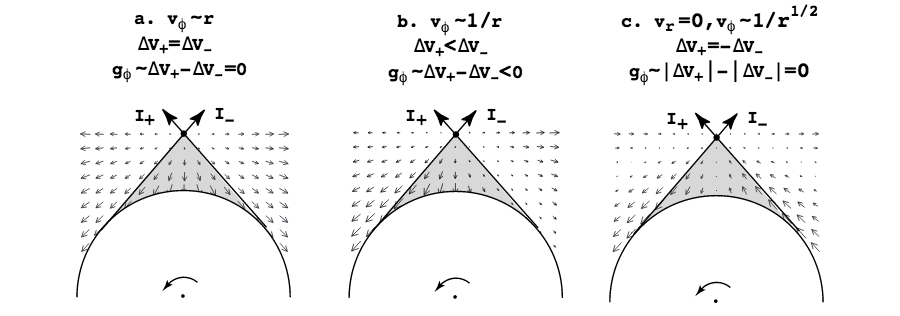}
\caption{
With permission, figure 3 of \cite{GayOwo00}. Each panel shows the velocity as seen by an observer at the intersection between the two rays shown, labeled I$^+$ for the ray from the blue-shifted hemisphere and I$^-$ for the ray from the red-shifted hemisphere. From left to right, the panels show the velocity that would arise for a) rigid rotation and a wind velocity law, b) angular momentum conserving flows with a wind velocity law, and c) a Keplerian disk with no outflow.
}
\label{fig:GayOwo00_fig3}
\end{figure}

Subsequent work by \cite{GayOwo00} showed that the shear arising from the radially declining azimuthal velocity leads to an \emph{azimuthal} line-acceleration that can actually lead to a net ``spin-down'' of the stellar wind (see figure \ref{fig:GayOwo00_fig1}).
Figure \ref{fig:GayOwo00_fig3} shows how the combination of rotation and outflow leads to an asymmetry in the line of sight velocity gradient towards the prograde vs. retrograde hemisphere.
Thus even though there is no net flux in the azimuthal direction, there arises a net azimuthal component of the line-acceleration that acts \emph{against} the direction of stellar rotation, and so causes the wind rotation to ``spin down''.

\begin{figure}
\centering
\includegraphics[width=\textwidth]{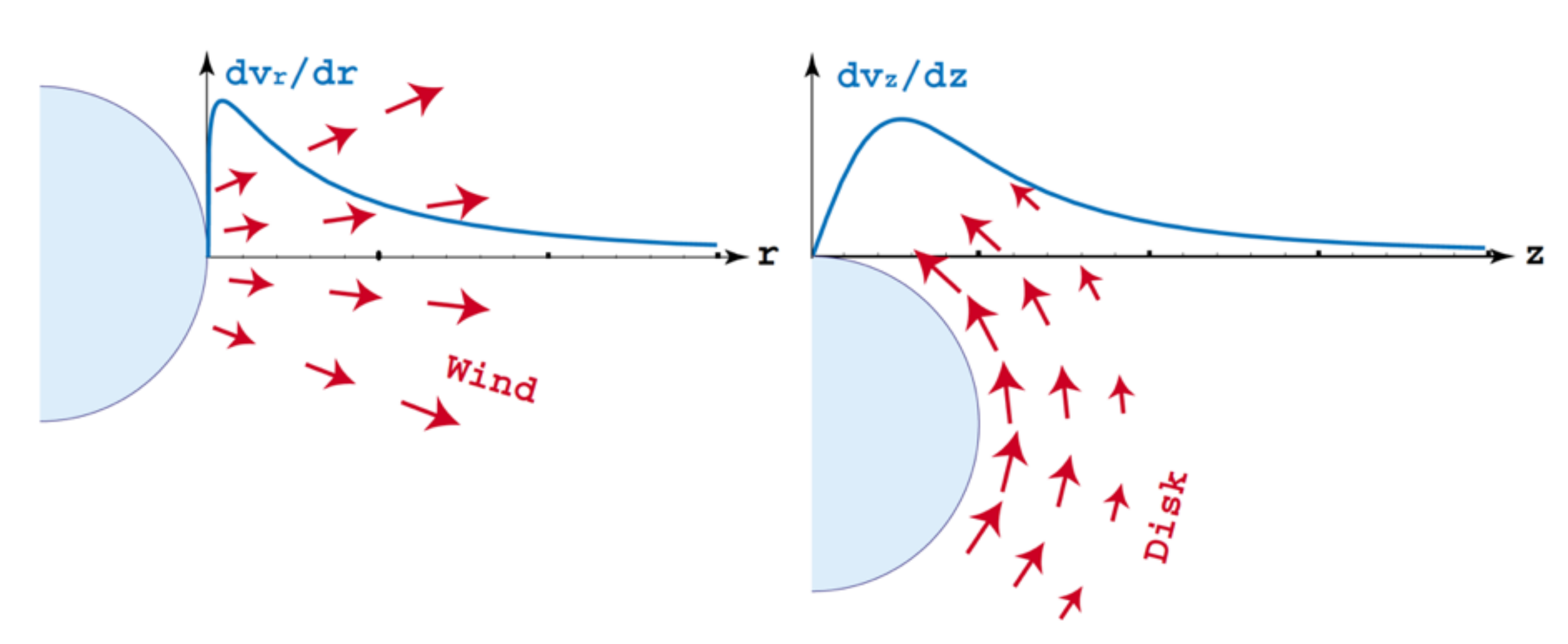}
\caption{
Schematic depiction of the radial velocity gradient, $dv_r/dr$, for a wind (left) and the line of sight velocity gradient, $dv_z/dz$, for a Keplerian disk (right).
}
\label{fig:vel_grad}
\end{figure}

A key motivation for this dissertation is the idea that such non-radial velocity gradients might also play a key role in the forces acting on a circumstellar, Keplerian disk. While part c of figure \ref{fig:GayOwo00_fig3} shows that such a Keplerian disk with no outflow experiences no \emph{azimuthal} acceleration, it is also clear from this figure that there should be a net \emph{radial} acceleration. To reinforce this point, figure \ref{fig:vel_grad} compares the velocity gradient along a ray tangential to the stellar surface, through a Keplerian disk, to the velocity gradient in the radial direction in a spherically symmetric wind. These are comparable in magnitude and qualitatively quite similar in radial variation. This dissertation examines the implications of such shear-enhanced line-accelerations on the ablation of Keplerian disks.
\chapter{Theory and Observations of Circumstellar Disks and Their Host Stars}\label{chap:disks}

Disks are one of the most prominent structures in the universe. From the rings of Saturn to the material accreting onto black holes and even all the way up to the structure of spiral galaxies like the Milky Way, disks are the natural byproduct of rotation and angular-momentum conservation. In effect, disks are the mediators of mass and angular momentum transfer between their host and the surrounding environment.

\section{Formation of Circumstellar Disks by Cloud Collapse}\label{sec:disk_collapse}

Circumstellar disks are a ubiquitous feature of the formation of stars from turbulent gas clouds. Although the turbulence is composed of randomly oriented motions over a range of scales, \cite{BurBod00} showed from simulations that a subset of a cloud undergoing gravitational collapse will generally have a preferred rotation axis with non-zero angular momentum. As the cloud collapses toward a stellar core, angular momentum conservation amplifies the rotational velocities about this axis, while velocities parallel to the axis are cancelled out in collisions.

\begin{figure}
\centering
\includegraphics[width=\textwidth]{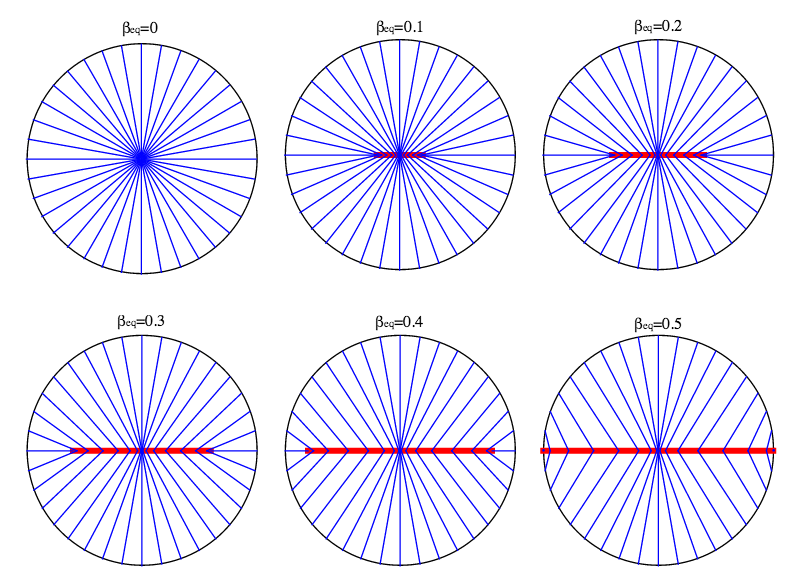}
\caption{A schematic depiction of the collapse of a spherical shell (black circle) around a point mass. The blue lines trace the path taken by gas parcels conserving angular momentum into the final state as a disk, shown in red. The panels vary $\beta_{eq}$ as described in equation \ref{eq:beta_cloud}.}\label{fig:DocO202}
\end{figure}

A simple model illustrating this effect is the collapse of a spherical shell of gas of radius $R$, initially in rigid rotation with angular velocity $\Omega$, about a point mass $M$. Let us define the ratio of rotational kinetic energy $T_{rot}$ of a parcel at the equator, to gravitational potential energy $U_{grav}$,
\beq\label{eq:beta_cloud}
\beta_{eq} = \frac{T_{rot}}{U_{grav}}=\frac{\Omega^2 R^2/2}{G M/R}\, .
\eeq
%For a bound orbit, $T_{rot}=1/2 U_{grav}$ implying that, for $\beta_{eq}\leq0.5$, the cloud will collapse.
Assuming the cloud satisfies the Jean's Criterion, i.e. is cold enough that the internal thermal energy is less than the gravitational potential energy, then it will collapse to form a disk with outer radius $2 \beta_{eq} R$. Parcels at colatitude $\theta$ have $\beta(\theta)=\beta\sin^2\theta$ and, as illustrated by figure \ref{fig:DocO202}, collapse by angular momentum conservation to a Keplerian disk radius $2 \beta_{eq} R \sin^2\theta$. If, instead of a single shell, there are now a large number of concentric shells, each will collapse the same way as before, now under the gravitational attraction of the shells inside it. 

\begin{figure}
\centering

%\begin{subfigure}[b]{0.45\textwidth}
%\includegraphics[width=\textwidth]{figs/rhocol1.jpeg}
%\end{subfigure}

\begin{subfigure}[b]{0.45\textwidth}
\includegraphics[width=\textwidth]{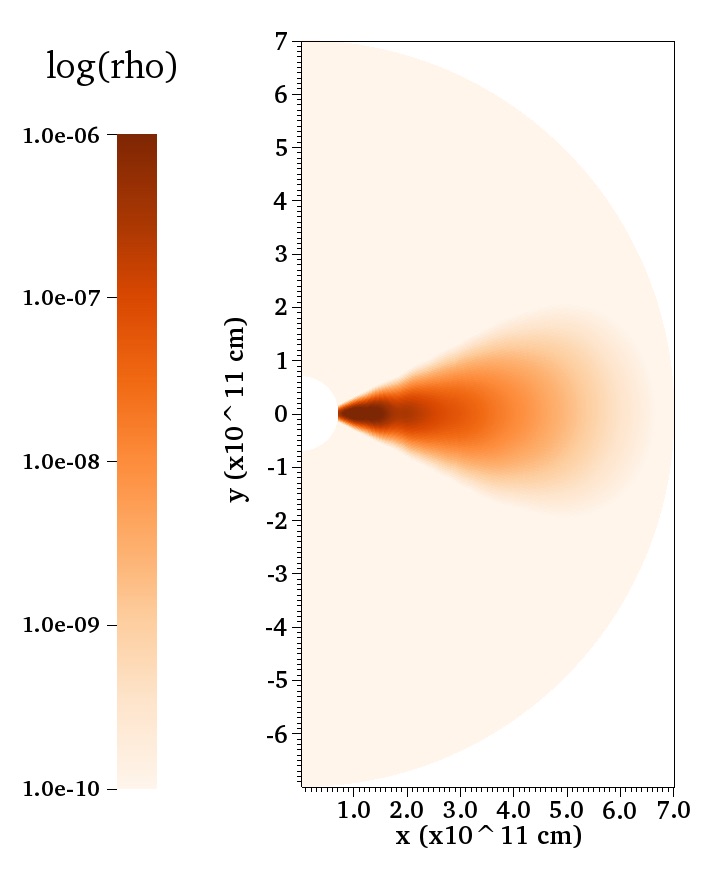}
\end{subfigure}
%\begin{subfigure}[b]{0.45\textwidth}
%\includegraphics[width=\textwidth]{figs/kepcol1.jpeg}
%\end{subfigure}
\begin{subfigure}[b]{0.45\textwidth}
\includegraphics[width=\textwidth]{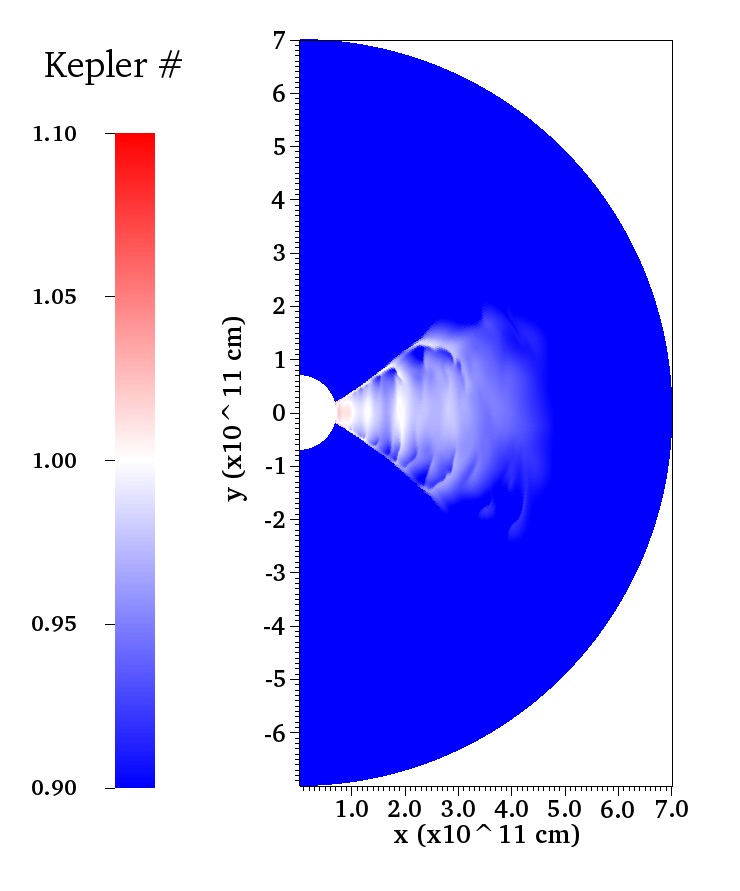}
\end{subfigure}

\caption{
Final state of the collapse of a rigidly rotating cloud of $\beta=0.45$ shown in both density (left) and Kepler number (right).
}
\label{fig:collapse}
\end{figure}

To illustrate this, figure \ref{fig:collapse} shows the final state of a hydrodynamical simulation of the collapse of such a rigidly-rotating, solid sphere of uniform density and fixed isothermal temperature $T$. To avoid the need to compute the self-gravity of the cloud, we retain the assumption of a central point mass $M$ that is much greater than the mass of the cloud. As expected, the final state is a disk with outer radius for this case of $\beta_{eq}=0.45$ near the expected value of $2\beta_{eq}R=0.9R$. However, the left panel of figure \ref{fig:collapse} shows that the finite temperature in the cloud leads to a pressure/density stratification in the vertical direction away from the equatorial plane. The right panel plots ``Kepler Number'', defined as the ratio of azimuthal velocity to local Keplerian orbital velocity
\beq
K\equiv\frac{v_\phi}{\sin\theta\sqrt{GM/r}}\, ,
\eeq
with near unity values around the disk plane demonstrating that this ``flared'' disk is Keplerian.

\section{Disks Around Massive Stars}\label{sec:disk_occurance}

While such protostellar disks are readily observed around stars less than a few solar masses, confirming their presence and studying their properties for luminous, massive stars is observationally more challenging \citep[see, e.g.][]{ZinYor07} because their formation generally occurs deeply embedded in the star's natal cloud.
For O stars this means that, by the time the star can be observed without significant extinction in shorter wavebands than the infrared, the disk has also been largely dissipated by the strong winds and radiative-acceleration discussed in the prior chapter. One goal of this thesis is to quantify this process.

In lower mass B stars, disks can survive to the later pre-main-sequence phase when the shrouding by the natal cloud has been reduced sufficiently to see emission line signatures of the disk, for example in the Herbig Ae and Be stars.
% The rareness of massive stars compounds this issue because, on average, massive star formation will occur at such large distances that resolution in the available wavelengths is poor. Furthermore, despite being rare, massive stars tend to form in groups which causes significant contamination between neighboring stars in observations.
 %\hl{How dense are these disks?} 
 Such Herbig Ae and Be stars provide a good laboratory for the role of disks in massive star formation and providing observational test of the disk dissipation processes discussed in this thesis.

\begin{figure}
\centering
\begin{subfigure}[b]{0.45\textwidth}
\includegraphics[width=\textwidth]{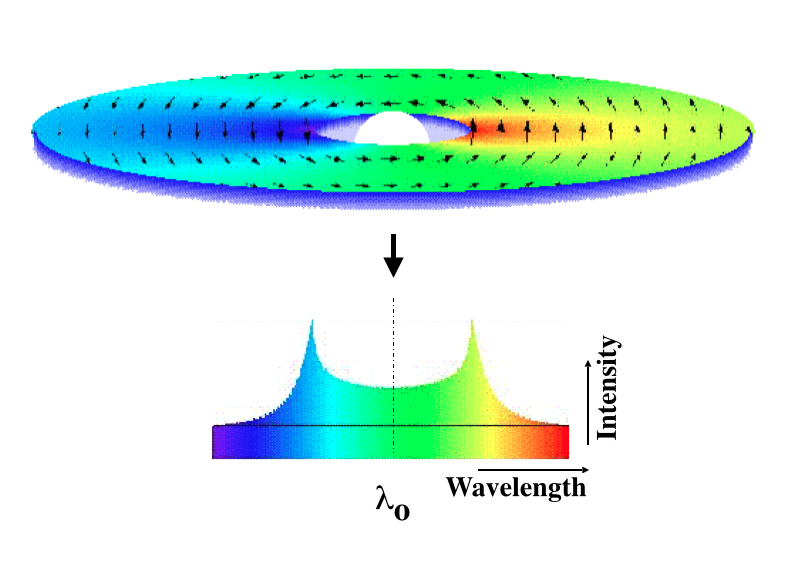}
\end{subfigure}
\begin{subfigure}[b]{0.45\textwidth}
\includegraphics[width=\textwidth]{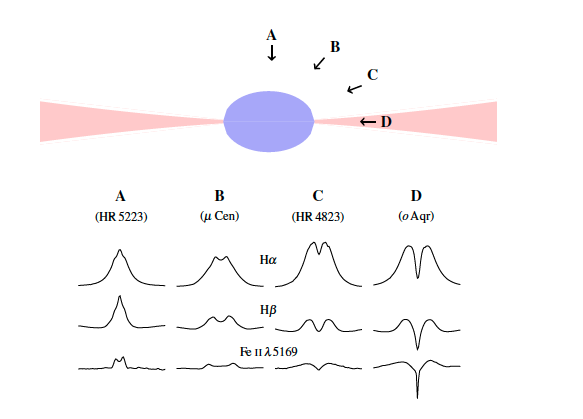}
\end{subfigure}
\caption{By mapping the visible surface area of a Keplerian disk into a histogram by line-of-sight velocity (left), it is easy to see how the majority of disk viewing angles produce double-peaked emission (right panel with permission from \cite{RivCar13}).}
\label{fig:halpha}
\end{figure}

However, massive stars' disks are not confined to the star formation phase. Fully 20\% of \emph{main sequence} B stars show an IR excess as well as H$\alpha$ emission associated with a circumstellar, gaseous disk (see figure \ref{fig:halpha}). Such stars, observed extensively since their discovery nearly 150 years ago by \cite{Sec66}\footnote{In fact, Father Angelo Secchi's 1866 observation of the Be star $\gamma$ Cassiopeiae ($\gamma$ Cas) was the first observation of emission lines in a stellar spectrum.}, are referred to as \emph{Classical} Be stars, and also provide a valuable opportunity to probe the interactions between stars and gaseous disks.
Classical Be stars are far too old to still harbor vestiges of the star formation process, and so are not subject to the extinction that plagues observations of stars still in formation. Additionally, Classical Be star disks come and go on time scales of months to years, allowing for observations that have captured both their growth and decay. This combination of stellar age and transient appearance has led to the consensus that these disks are dynamically generated by the stars themselves, likely as a byproduct of their extreme rotation rates at upwards of $70\%$ of orbital velocity \citep{TowOwo04}. Such a high rotation velocity leads to stellar oblateness and gravity darkening as discussed in appendix \ref{app:ob_gravdark}. Figure \ref{fig:obstar} shows such a star. Chapter \ref{chap:pdome} provides a model for the ejection of material into a disk by rotationally assisted stellar pulsation. Chapters \ref{chap:ab_thin} and \ref{chap:ab_spec_type} discuss how disk destruction by line-driven ablation might explain their transient nature and occasional disappearance. 
For Classical Be stars observed near to edge-on, there is direct observational evidence for enhanced flows along the disk surface \citep[e.g.][]{GraBjo87, GraBjo89} providing a direct test for ablation models. 

Observations show that there is a dependance on spectral type of the fraction of B stars that are observed to be Classical Be stars \citep[see e.g.][]{MarFre06,RivCar13}. On the low mass end, Be stars are thought to transition into a population of A and F disk hosting stars. As these are stars expected to host very weak winds, if indeed they have line-driven winds at all, we do not discuss them any further here. On the high mass end, Be stars seem to cut off fairly abruptly around the B to O spectral transition. Indeed, while there are several possible candidates for O9e stars\footnote{Work by \cite{VinDav09} has shown a systematic mistyping of Oe/Be stars around the B to O transition as earlier spectral types than they actually are due to their strong winds, drawing some question to exactly what is the earliest Oe star.} there are very few claims of more massive Classical Oe stars. One of the central aims of this thesis is to investigate whether this high mass cutoff can be related to the stronger radiative accelerations present in these more massive stars.

\begin{figure}
\centering
\includegraphics[width=0.6\textwidth]{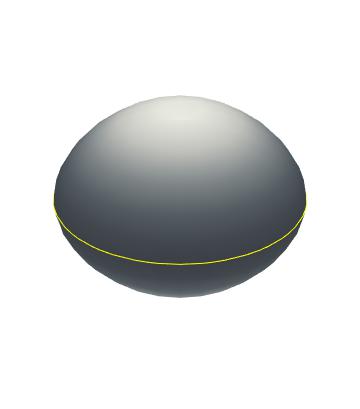}
\caption{A star rotating with equatorial rotational speed 80\% of local orbital speed. Such a star has $R_{equator}/R_{pole}=1.32$ and $F_{equator}/F_{pole}=0.2$. The shading depicts relative surface brightness.}\label{fig:obstar}
\end{figure}

\section{Structure of Circumstellar Disks}\label{sec:disk_struc}

While circumstellar disks occur in a variety of scenarios, there are commonalities to their structures. Solving the radial, $R$, and vertical, $z$, components of the momentum conservation equations yields analytic scalings for the stratification of density $\rho$ in both the radial and vertical directions. For circular orbits with purely azimuthal velocity $v_\phi$ about a star of mass $M_\ast$, taking an isothermal equation of state with pressure $P=\rho a^2$ and sound speed $a$ reduces the vertical component of the momentum to an equation of hydrostatic equilibrium
\begin{align}
\frac{a^2}{\rho}\frac{d\rho}{dz}&=-\frac{G M_\ast z}{\left(R^2+z^2\right)^{3/2}}\\
&\approx -\frac{G M_\ast z}{R^3}\, ,
\end{align}
where the approximate equality assumes that $z\ll R$. The resulting vertical stratification of density then follows a Gaussian
\beq
\rho \propto e^{-1/2\left(z/H\right)^2}\, ,
\eeq
with scale height $H\equiv a^2R^2/(GM)$. Under the same assumptions, the radial component of the equation of momentum conservation is
\beq
\frac{v_\phi^2}{R}=\frac{a^2}{\rho}\frac{d\rho}{dR}+\frac{GMR}{(R^2+z^2)^{3/2}}\,.
\eeq
Since both the radial variation of $v_\phi$ and $\rho$ are unspecified here, and there is no other constraining equation to satisfy, we can freely choose the radial variation of one of them; here we choose to specify $\rho(r)$ to be a power-law of an index $n$ giving the full functional form,
\beq\label{eq:disk_rho}
\rho(R,z)=\rho_\mathrm{o} \left(\frac{R}{R_\ast}\right)^{-n} e^{-\frac{1}{2}(z/H)^2}\, ,
\eeq
with associated velocity
\beq
v_\phi(R,z)=\sqrt{\frac{GM}{\sqrt{R^2+z^2}}}\frac{R}{\sqrt{R^2+z^2}}\left(1-n\frac{H}{R}\right)\,.
\eeq

While $n$ is presented as a free parameter in these equations, prior work has put constraints on its value. For instance, by modeling the H$\alpha$ profile generated by equation \ref{eq:disk_rho} and fitting to the observations of 56 Classical Be stars, \cite{SilJon10} found that $n$ fell in the range $1.5-4.0$ with a statistically significant peak at 3.5. Individual fits for a large number of other stars \citep[e.g.][]{Por99,GieBag07,TycJon08} confirm this result. Following the work of \cite{CarMir06,CarMag07,CarOka09}, the remainder of this dissertation uses the statistically significant peak value $n=3.5$.

\begin{figure}
\centering
\includegraphics[width=\textwidth]{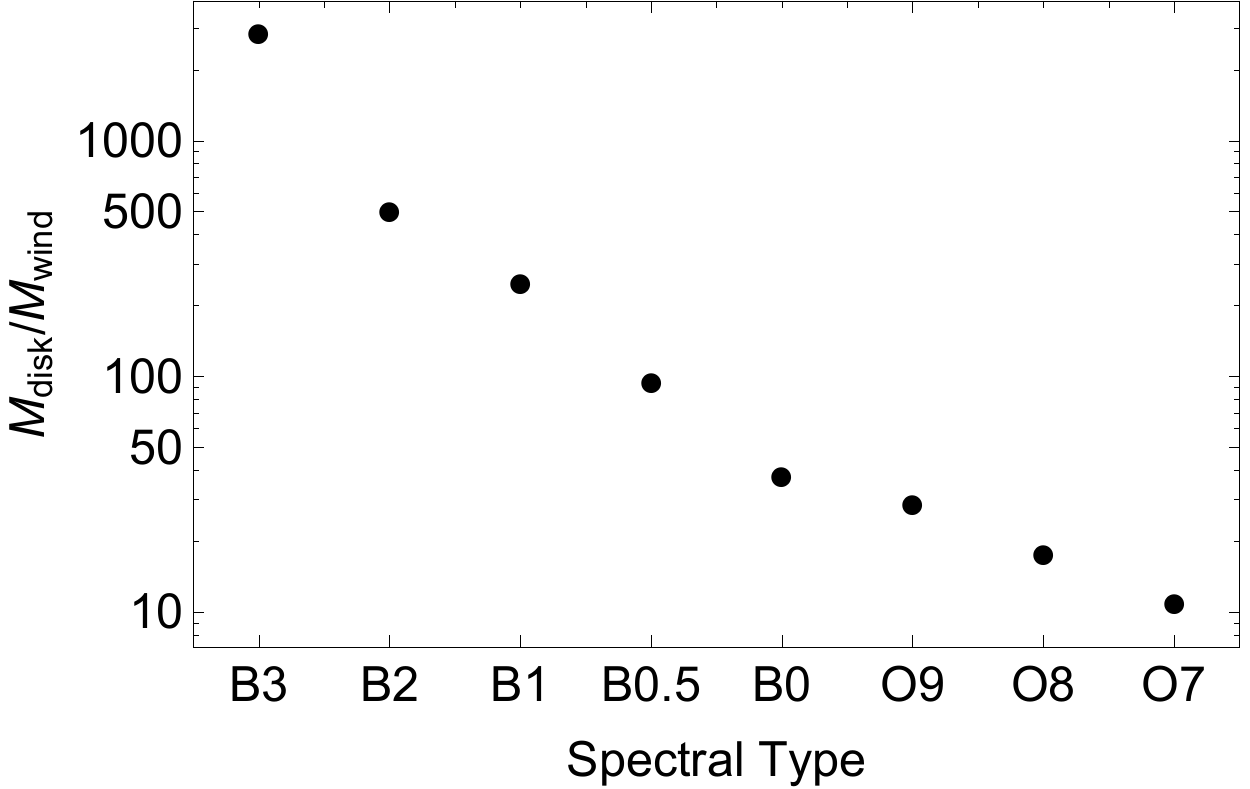}
\caption{The ratio of disk to wind mass for the stars in this dissertation.}\label{fig:mdomw}
\end{figure}

In addition to constraining $n$, observations of Classical Be stars also constrain $\rho_\mathrm{o}$. Since the disks are observed to be marginally thin to electron scattering opacity $\kappa_e$, we here take $\rho_\mathrm{o} = (\kappa_e R_\ast)^{-1}$, yielding a radial optical depth through the equatorial plane of $\tau=1/(n-1)=0.4$. Integrating the disk density profile over the full simulation volume extending from 1 to 10 stellar radii, the total disk mass for power-law index $n=3.5$ depends on the stellar parameters as
\begin{align}
M_{disk}&=\frac{(2\pi)^{3/2}}{\kappa_e} \frac{a}{\sqrt{G M_\ast}} R_\ast^{2.5} \ln(10)\\
&=1.24\times10^{-10} M_\odot \sqrt{\frac{T}{10^4 \,\mathrm{K}}}\sqrt{\frac{10M_\odot}{M}}\left(\frac{R}{5R_\odot}\right)^{2.5}  \, .
\end{align}
For comparison, for a star without a disk but a mass loss rate $\dot{M}$ and a constant outflow, i.e. a $\beta=0$ velocity law, $v_r = v_\infty$, the total wind mass in the same simulation volume is given by
\begin{align}
M_{wind}&=\int \rho_{wind} dV =\frac{9R_\ast\dot{M}}{v_\infty} \\
&= 9.9\times10^{-14} M_\odot \frac{\dot{M}}{10^{-10} M_\odot/yr}\,\frac{R_\ast}{5 R_\odot}\,\frac{10^8 cm/s}{v_\infty}\,.
\end{align}
Figure \ref{fig:mdomw} plots the ratio $M_{disk}/M_{wind}$, demonstrating that the disk makes up the majority of mass in the simulation across all spectral types considered. Accounting for a more realistic, $\beta=1$ velocity-law causes the wind mass to slowly diverge as the lower integration bound approaches $R_\ast$, where $v_r\rightarrow0$ makes $\rho\rightarrow\infty$. However, doing this integral with a lower bound $R_\ast+\epsilon$ only introduces an order unity increase (depending on the lower integration bound), and so the general conclusion that the disk mass is larger than the wind mass does not change.

\section{Viscous Transport}\label{sec:visc}

Modeling of the dynamic evolution of circumstellar disks has previously focused on the role of viscous forces, largely ignoring ablation effects.\footnote{For a detailed review of viscous transport in gaseous disks see \cite{Lod08}.} As concentric rings of gas in the disk shear against each other, angular momentum is transported outwards allowing material to diffuse either inward or outward based on the mass source boundary conditions. In the case of a star in formation, where the dominant mass reservoir is the cloud at the outer edge of the disk, the net material transport is inwards (accretion), while in the case of a Classical Be star, where the rapidly rotating star provides an inner mass reservoir, the net material transport is outwards (decretion\footnote{``Decretion'', literally meaning a decrease, is not a particularly common word in the English language. However, it is vastly preferable to ``excretion'' as was once used.}).

Since viscosity does not determine the vertical structure of a disk with only Keplerian orbital velocity, let us define the vertically integrated surface density $\Sigma\equiv \int_{-\infty}^{\infty} \rho dz$. For a disk with Gaussian vertical stratification, this gives $\Sigma(R)=\sqrt{\pi} H \rho_{eq}(R)$, where $\rho_{eq}(R)$ accounts for the power-law dependence of density on radius. Combining the equations of mass and momentum conservation, the interaction of Keplerian shear with kinematic viscosity $\nu$ causes $\Sigma$ to evolve according to the diffusion equation \citep[see, e.g.][]{Lod08},

\beq
\frac{d\Sigma}{dt}=\frac{3}{R}\frac{\partial}{\partial R}\left[R^{1/2} \frac{\partial}{\partial R}\left(R^{1/2} \nu \Sigma\right)\right]\,.
\eeq

\begin{figure}
\centering
\includegraphics[width=\textwidth]{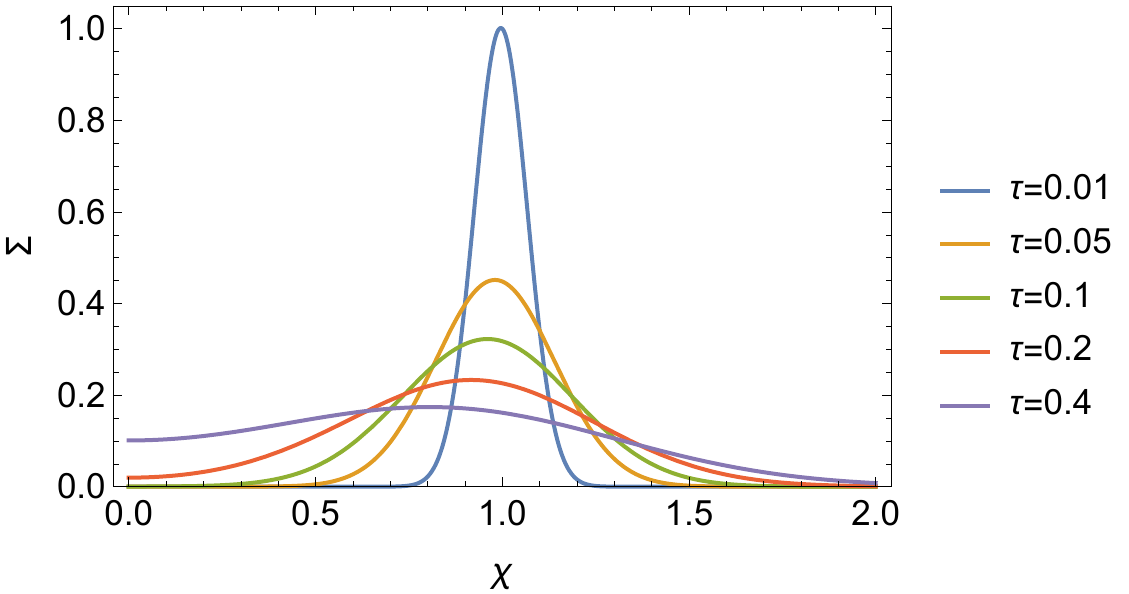}
\caption{The spreading ring Green's function solution at several times demonstrating its behavior in space and time. $\Sigma$ has been normalized such that the peak value of $\Sigma(\chi,0.01)=1$.}\label{fig:spread_ring}
\end{figure}

A simple way to illustrate this diffusive evolution of a disk is to introduce at time $t=0$, a $\delta$-function in surface density at some initial equatorial radius $R_\mathrm{o}$, such that $\Sigma(R,t=0)=m/(2\pi R_\mathrm{o})\,\delta(R-R_\mathrm{o})$ where $m$ is the total mass in the annulus. This material is also given Keplerian orbital velocity $v_\phi=\sqrt{GM/R_\mathrm{o}}$. If $\nu$ is position and density independent, the evolution follows the so called ``spreading ring'' solution, \citep{LynPri74}
\beq
\Sigma(\chi,\tau_v) = \frac{m}{\pi R_\mathrm{o}^2}\frac{\chi^{-1/4}}{\tau_v} e^{-(1+\chi^2)/\tau_v}I_{1/4}\left(\frac{2\chi}{\tau_v}\right)\, ,
\eeq
where $\chi=R/R_\mathrm{o}$, $\tau_v=12\nu t/R_\mathrm{o}^2$, and $I_{1/4}$ is a modified Bessel function of the first kind.
Since $\nu$ is taken to be independent of $\Sigma$, this represents a Green's function solution in space and time to a linear equation. Figure \ref{fig:spread_ring} plots several snapshots of the behavior of this function, showing that the ring spreads both inwards and outwards from its initial position. Using the principles of superposition, this Green's function allows one to solve for the time evolution of disks of arbitrary initial surface density. For a constant rate of mass being added at some source radius $R_\mathrm{o}$, it also allows us to solve for the asymptotic spreading of a disk away from a mass source. If this mass source is at the inner edge of the disk $\Sigma\propto r^{-0.5}$, while $\Sigma$ is independent of radius if the mass source is at the outer edge of the disk. While these are interesting and useful results, they are only valid for a constant $\nu$, which is not expected to be the case.

One particularly common method for accounting for the spatial dependence of $\nu$ is the ``$\alpha$-prescription''. First introduced by \cite{ShaSun76} to treat accretion onto black holes, this method consists of breaking down kinematic viscosity, which has units of a length-squared over time, into the product of a characteristic length, a characteristic velocity, and an efficiency factor $\alpha$ such that 
\beq
\nu = \alpha H a\,.
\eeq
Here the pressure scale height has been chosen as the characteristic length as it is the largest scale over which material can be expected to be coherently transported, and the sound speed has chosen as the characteristic velocity as viscous transport is expected to be subsonic. $\alpha$ then can be alternatively interpreted as the product of the fraction of a scale height over which material is actually transported and the fraction of the sound speed at which it travels.
While this only qualitatively describes the underlying physics of viscous transport, it does provide a simple method for determining the approximate scale of the viscous forces as well as a first order estimate of their spatial dependence. Based on these merits, the $\alpha$-prescription has become one of the most prevalent ways to cast viscosity in astrophysics.

For position dependent kinematic viscosity, as is the case in the $\alpha$-prescription, we no longer have access to the Green's function solution. However, it is still possible to derive an asymptotic surface density distribution as was done for instance in \cite{BjoCar05}. For an $\alpha$-prescription of viscosity and a scale height proportional to $R^{-1.5}$, the asymptotic surface density in an isothermal decretion disk has $\Sigma\propto R^{-2.0}$ and, since $\Sigma\propto \rho H$, volume density varies with radius as $R^{-3.5}$. Comparing this to the results of \cite{SilJon10} shows that this is the statistically significant peak value of $n$, reinforcing the choice to use $n=3.5$ here.

\begin{figure}[b!]
\centering
\includegraphics[width=\textwidth]{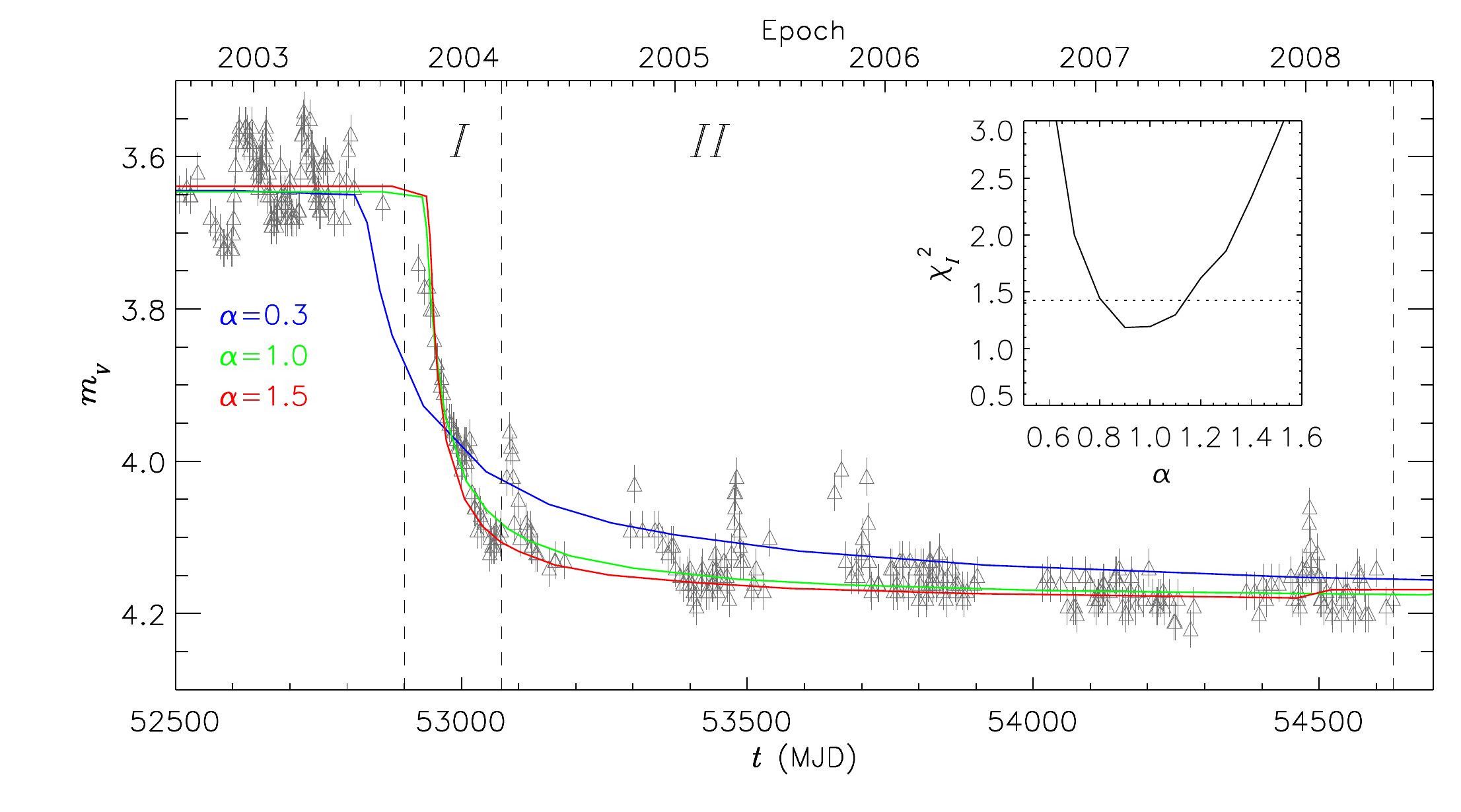}
\caption{With permission, figure 1 of \cite{CarBjo12} showing observations and viscous diffusion fits to the disk decay of 28 CMa.}\label{fig:Car_12}
\end{figure}

With this confirmation that the $\alpha$-prescription is a good model for disks, we can now turn our attention to the scale of $\alpha$. While it is in principle a free parameter of the $\alpha$-prescription, we expect that $\alpha\leq1$ so that material is not transported supersonically or over larger lengths than the disk scale height. Beyond this simple limit, it is also possible to calibrate $\alpha$ by comparing the distribution of gas in a sample of disks of a variety of inferred ages \citep[see, e.g.][]{HarCal98,AndWil07}. Doing so suggests that for star forming disks $\alpha\sim0.01$. This should be taken as a loose order of magnitude estimate only, however, as $\alpha$ itself is not necessarily constant even within a single system as it likely depends on at least temperature, ionization level, density, and radius.

In the study of Classical Be disks, the $\alpha$-prescription has notably been applied to dynamical modeling of 28 CMa \citep{CarBjo12}. Figure \ref{fig:Car_12} shows a series of observations taken between 2003 and 2009 showed over a half a dex dimming of the visual band magnitude of this star, inferred to be associated with the disappearance of a circumstellar disk. To match the observed $\sim$170 day decay time of the disk, an $\alpha$ of 1.0$\pm$0.2 is required\footnote{Re-analysis of the data suggests $\alpha$ closer to 0.4, still large compared to the expected value $0.01\lesssim\alpha\lesssim0.1$. (Carciofi, private communication).}, on the cusp of (if not too large for) the expected range of $\alpha\leq1$. Therefore a much more efficient process for evacuating a circumstellar disk in necessary. This dissertation investigates the possibility that this more efficient evacuation of the disk is related to the non-radial velocity gradient components discussed at the end of chapter \ref{chap:line}.

\chapter{Radiative Ablation of An Optically Thin Disk}\label{chap:ab_thin}

We now have both a formalism for implementing a 3D, vector line-acceleration (chapter \ref{chap:line}) and a description of the disks to which we want to apply it (chapter \ref{chap:disks}).
Before undertaking an investigation of the interaction between disk and radiation over a wide range of stellar parameters, however, let us focus on a single model, here chosen to be a non-rotating main sequence B2 star (section \ref{sec:B2}), with emphasis on resolution and comparison of force implementations.

From physical arguments and prior experience, the spatial resolution is relatively easy to specify.
In contrast, the required number of rays to resolve the stellar radiation and properly model disk ablation is not as clear. We therefore carry out a resolution study in ray quadrature (section \ref{sec:res_tests}).

Another issue regards the importance of non-radial velocity gradients and accelerations found to play such a crucial role in the wind rotation models of chapter \ref{chap:line}. The parameter study here examines their role for disk ablation by comparing a model which has a fully 3D vector acceleration with two simplified models: one with a 1D radial acceleration calculated using $dv_r/dr$ as opposed to $\hat{n}\cdot\nabla(\hat{n}\cdot \mathbf{v})$; and one with a 3D radial acceleration calculated with the non-radial velocity gradients but only implementing the radial component of the acceleration (section \ref{sec:force_imp}).

To wrap up the investigation of a B2 star and allow us to continue on to a parameter study of stellar spectral type, section \ref{sec:w80_rot} presents a discussion of the effects of rotation on ablation. While chapter \ref{chap:disks} points out the ubiquitous observation of rapid rotation in Be stars, including such rotation in a grid-based, numerical hydrodynamics code poses challenges for implementation and adds complexity through the introduction of a new parameter, so we wish here to determine the importance of its inclusion for the ablation rate.
%allow us to determine if the additional complexities associated with its inclusion are necessary for us to handle here. 

\section{Standard Model of a B2V Star}\label{sec:B2}

%For our standard Be star model, we want to be sure to select as prototypical of a Be star as possible in order to constrain numerics and physics for a truly ``standard'' model. 
For our standard model, we choose a B2 star both because of its strong luminosity and also because
this is near the spectral type with the largest fraction of Be stars.
% to peak around $8-10\,M_\odot$ \citep[see e.g.][]{MarFre06,RivCar13}. Therefore, we chose for our standard model a B2 star with $M_\ast=9\,M_\odot$. 
Table \ref{tab:b2_params} provides the full set of stellar and disk parameters as derived from the evolutionary tracks of \cite{GeoEks13} and the effective temperature calibrations of \cite{TruDuf07}. Table \ref{tab:b2_wind} gives the wind parameters as derived by \cite{PulSpr00}.

\begin{table}
\caption{Stellar and Disk Parameters of the B2 Standard Model} \label{tab:b2_params}
\centering
\begin{tabular}{| l | c | c | c | c | c |}
\hline
Sp. type & $T_{eff}$ (kK) & $L_\ast$ ($L_\odot$) & $M_\ast$ ($M_\odot$) & $R_\ast$ ($R_\odot$) & $M_{disk}$ ($M_\odot$)\\
\hline
B2V & 22 & 5.0$\times 10^3$ & 9 & 5.0 & 1.9$\times 10^{-10}$  \\
\hline
\end{tabular}
\end{table}

\begin{table}
\caption{Wind Parameters of the B2 Standard Model} \label{tab:b2_wind}
\centering
\begin{tabular}{| l | c | c | c | c |}
\hline
Sp. type & $\bar{Q}$ & $Q_\mathrm{o}$ & $\alpha$ & $\dot{M}_{wind}$ ($M_\odot/yr$) \\
\hline
B2V & 1800 & 4900 & 0.59 & 7.4$\times 10^{-10}$ \\
\hline
\end{tabular}
\end{table}

\begin{figure}
\centering
\includegraphics[width=0.6\textwidth]{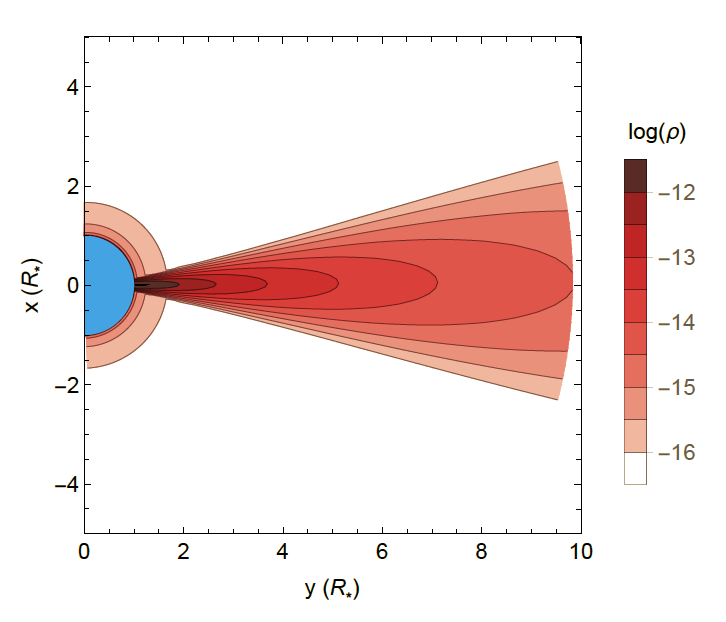}
\caption{
Initial condition of the B2 standard model plotted in $\log(\rho)$ measured in $\log($g/cm$^3)$.
}
\label{fig:b2_init}
\end{figure}

Taking these parameters, we use the hydrodynamics code VH-1 to evolve the time-dependent equations for conservation of mass and momentum,

\begin{align}
&\frac{\partial \rho}{\partial t} + \nabla\cdot(\rho\mathbf{v}) = 0 \\
&\frac{\partial \mathbf{v}}{\partial t} + \mathbf{v}\cdot\nabla\mathbf{v}=-\frac{1}{\rho}\nabla P - \mathbf{g}_{grav} +\mathbf{g}_{rad}\, .
\end{align}
Here we do not need an equation of energy conservation because we assume an isothermal equation of state, $P=\rho a^2$ where $a=\sqrt{kT/\bar{\mu}}$ is the isothermal sound speed for temperature $T$ and mean molecular weight $\bar{\mu}$ ($\sim0.6m_{proton}$). Additionally, gravity only arises from the fixed stellar mass, as the self-gravity of the disk is many orders of magnitude weaker. The initial condition assumes a superposition of a hydrostatic, optically thin, Keplerian disk (as described in chapter \ref{chap:disks}) onto a background spherically symmetric wind (as described in chapter \ref{chap:line}). Each extends from the stellar surface to the outer simulation boundary in its respective portion of the initial conditions. Figure \ref{fig:b2_init} plots $\log(\rho)$ contours for this initial condition.

The outer boundary at $r=10R_\ast$ (see table \ref{tab:grid}) assumes a zero gradient in all quantities; the inner boundary at $r=R_\ast$ assumes a constant density\footnote{This density is $\rho_\ast=5\rho_{sonic}$, with $\rho_{sonic}\equiv\dot{M}_{CAK}/(4\pi R_{sonic}^2 a)$ the density at the sonic radius $R_{sonic}$, where $v_r(R_{sonic})=a$. For this B2 model, $\rho_\ast=5\times10^{-14}$g/cm$^3$.}  and linearly extrapolates the velocity components across the boundary with the constraint that they not be larger than the isothermal sound speed. For this initial model, the stellar radiation field is characterized by a grid of rays, projected onto the stellar disk with $n_{\phi'}=6$ rays in the full circle in azimuthal angle $0<\phi'<2\pi$ about the center, and $n_p=6$ rays in impact parameter $p$. These are distributed and weighted according to a Gauss-Legendre quadrature in $\phi'$ and impact area $y=p^2/R_\ast^2$. For a depiction of this quadrature, see the second panel from the right in figure \ref{fig:ray_quad}.

\begin{figure}
\centering
\begin{subfigure}[b]{0.50\textwidth}
\includegraphics[width=\textwidth]{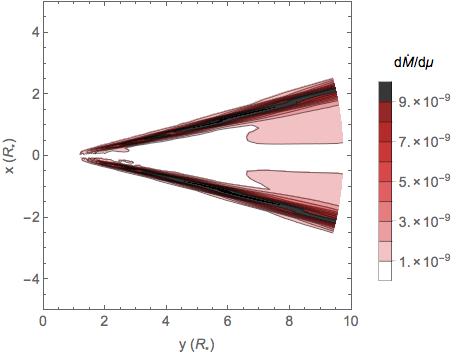}
\end{subfigure}
\begin{subfigure}[b]{0.48\textwidth}
\includegraphics[width=\textwidth]{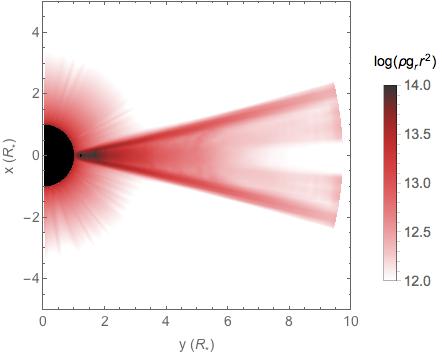}
\end{subfigure}
\caption{
Time averaged mass flux per unit $\cos(\theta)$ of the non-rotating B2 star model, plotted in units of $M_\odot$ yr$^{-1}$ $\cos(\theta)^{-1}$ (left), and force-per-unit-length $\rho g_r r^2$ in cgs units (right). Both been averaged from $10^6$ to $3\times10^6$ s to omit the time dominated by the disk readjusting to the introduction of radiation forces.
}
\label{fig:B2_dmdotdmu}
\end{figure}

In addition to the model parameters and simulation method, we also need to define a few quantities that will be used in this analysis, as well as in the remainder of the dissertation. One particularly important quantity is the mass flux distribution per unit solid angle, for azimuthally symmetric models proportional to,
\beq
\frac{d\dot{M}}{d\mu} \equiv 2 \pi \rho v_{r} r^2\,,
\eeq
where $\mu\equiv\cos\theta$. By plotting the time average of this quantity for the B2 model, figure \ref{fig:B2_dmdotdmu} confirms the expectation that ablation should mostly occur in thin layers along the upper and lower edges of the disk\footnote{The small pink excursions into the central region of the disk are the remnants of small scale oscillatory motions set off by the relaxation of the initial conditions. Elsewhere in the disk, these have averaged out.}. By comparing this with the second panel of figure \ref{fig:B2_dmdotdmu}, which shows force-per-unit-length $\rho g_r r^2$, we can see that the material in these ablating layers is accelerated throughout the full radius of the simulation. However, since the thickness of the ablating layer does not substantially change in solid angle, we interpret this as continuing acceleration of the material that has been dislodged from the disk near the star, rather than an addition of new disk material to the ablation flow. Note that, while this mass loss rate per solid angle is much larger than that of the wind, its much smaller solid angle means that the total mass loss rate can be comparable.

\begin{figure}
\centering
\includegraphics[width=\textwidth]{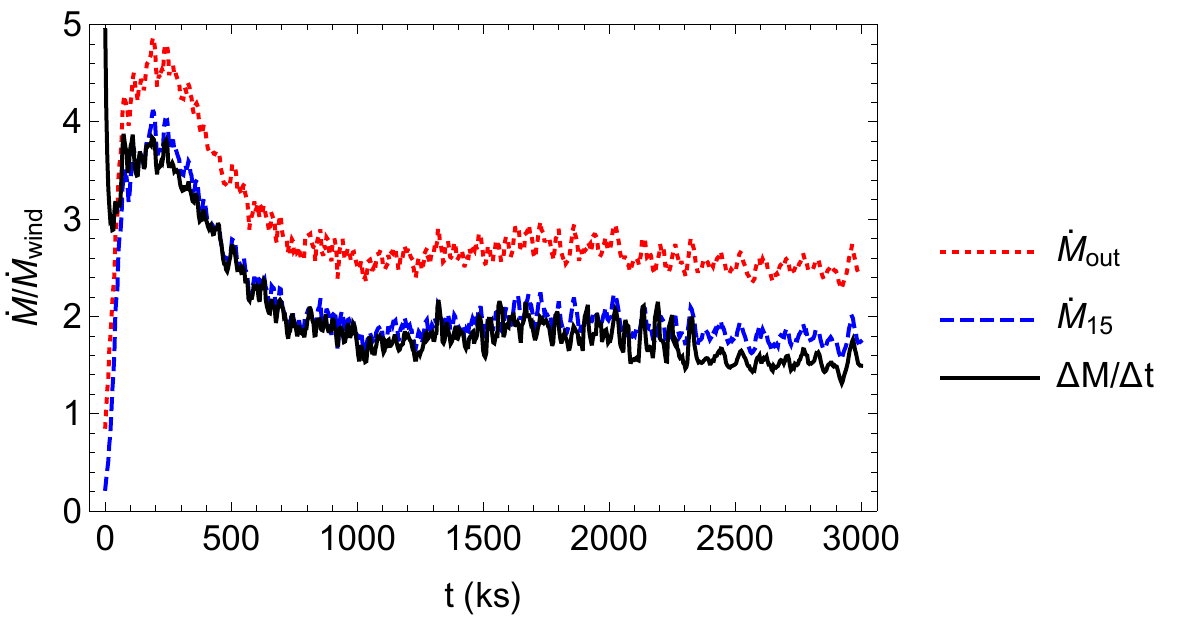}
\caption{
Mass loss rate in units of spherically symmetric mass loss rate for three mass loss metrics of the standard model of a B2 star.
}
\label{fig:B2_mdot}
\end{figure}

To quantify this disk ablation rate, one option is to compute the flux of mass through the outer simulation boundary. By plotting this, figure \ref{fig:B2_mdot} demonstrates that, after an initial adjustment period, the mass loss rate from a model with disk ablation is roughly $3 \dot{M}_{wind}$. However, this includes both the wind and disk mass loss rates. We can attempt to remedy this by choosing only a small angle about the equator. This is also shown in figure \ref{fig:B2_mdot} as ``$\dot{M}_{15}$'', referring to the inclusion of mass loss in a band of 15$^\circ$ above and below the equator
%\hl{If you have extra time change this to $\dot{M}_{\pm15}$}. 
Finally, figure \ref{fig:B2_mdot} also shows the rate of change of mass in the simulation volume $\Delta M/\Delta t$, since this is not affected by the steady state wind. The initial spike in this disk ablation rate is due to accretion of material back onto the star. After a few kiloseconds, however, the ablation rate becomes nearly equal to $\dot{M}_{30}$; the small difference at late times can be accounted for by the shrinking of solid angle of the disk and the contamination of a little bit of wind. Therefore, the remainder of this dissertation uses $\Delta M/\Delta t$ as the disk ablation rate, with the caveat that the initial few ks are dominated by infall. 

The next section investigates how this ablation rate, which is a key outcome of the dynamical simulations, is affected by both resolution and force implementation.
%Thus, any process that is going to build or sustain such a disk against ablation must feed mass in at at least twice the wind mass loss rate, suggesting that the disk is generated by something other than the stellar wind. 
%Note though, that even without such a mechanism feeding mass into the disk, the simulation continues to host a disk for a time scale predicted to be on the order of 100 days (see figures \ref{fig:frac_m_spec} and \ref{fig:t_disk_tvs}) which is within a factor of two of the disk decay time seen in 28 CMa by \cite{CarBjo12}, suggesting that ablation could be a comparably or more important effect than viscosity in Classical Be disk destruction.

%\begin{figure}
%\centering
%\includegraphics[width=\textwidth]{figs/mdot_over_mdotwind_B2.pdf}
%\caption{
%Ablation rate in units of spherically symmetric mass loss rate for the standard model of a B2 star.
%}
%\label{fig:B2_mdot_ab}
%\end{figure}

\section{Resolution tests}\label{sec:res_tests}

%Ideally, we hope to be able to get reliable results in a timeframe which is not prohibitively long. 
%For the spatial grid in $r$ and $\theta$, the necessary resolution is something which is fairly predictable based both on physical arguments and prior experience with radiation hydrodynamic simulations. The necessary resolution in radiation rays over the stellar flux field, however, is not something which is as easily predicted, nor something with which we have such extensive experience. Therefore, while we will be able to select a spatial grid prior to running a simulation as discussed in the first subsection below, the resolution in radiation rays will be tested for, as discussed in the second subsection.

\subsection{Spatial grid}
%\hl{No show stoppers. Wordy}
%The choice of resolution in spatial grid and ray quadrature must balance precision and computational cost.
%The spatial grid resolution requires that the smallest structure of interest not be below the grid scale.
Beginning with the spatial grid, in radius we must resolve the trans-sonic region where the wind goes from sub- to super-sonic. 
To ensure that this condition is met, we compare our grid scale against the radial pressure scale height of the star, given by $a^2/v_{esc}^2\, R_\ast$. 
%This requires high resolution near the stellar surface but not elsewhere
Beyond the trans-sonic region the requirements for resolution are much less stringent, however, so a stretch factor is applied such that each grid cell is 2\% larger than the one immediately interior to it, making the grid resolution lower with increasing distance from the star. 
The first column of data in table \ref{tab:grid} gives the details of the $r$ grid including the minimum grid spacing as well as the pressure scale height for the standard model of a B2V star and for an O7V star which is the most massive model considered. 
In both cases the smallest grid cell is about half the size of the scale height, which we deem sufficient based on prior experience.

\begin{figure}
\centering
\includegraphics[width=\textwidth]{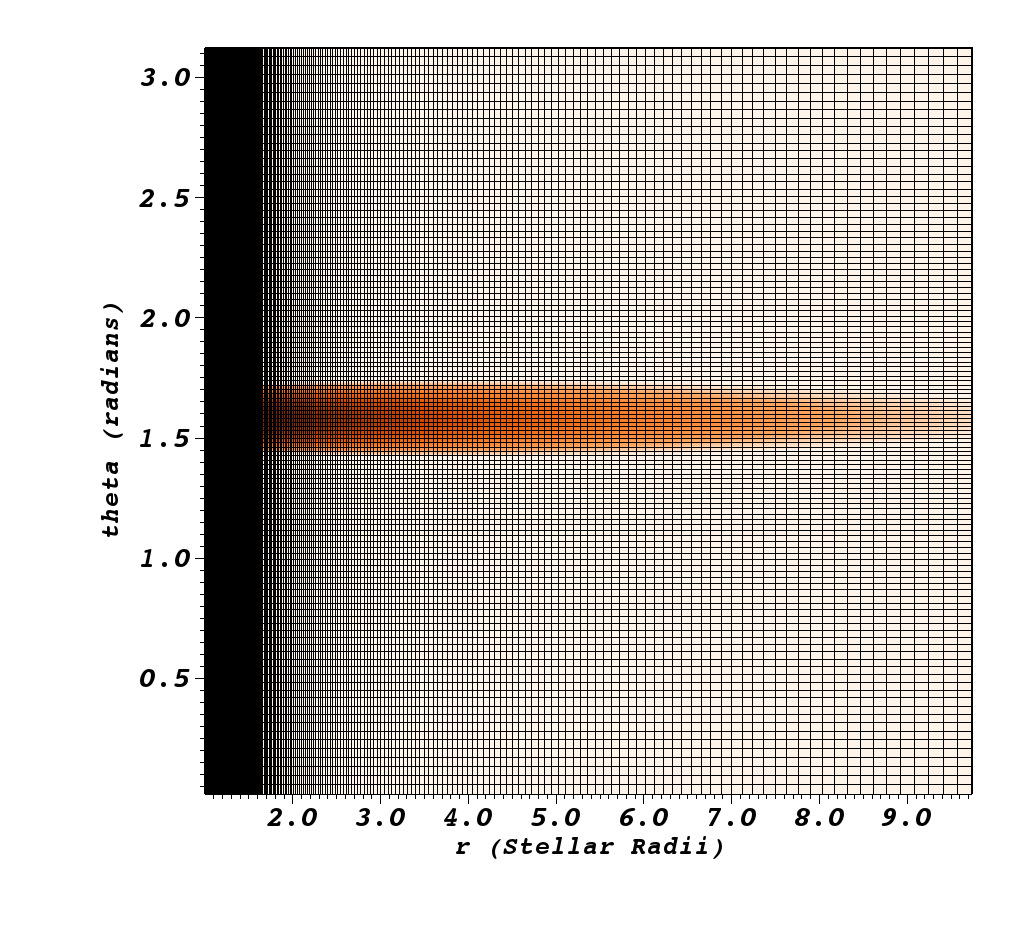}
\caption{
Grid mesh as a function of $r$ (x-axis) and $\theta$ (y-axis) with the initial state in log($\rho$) plotted for comparison.
}
\label{fig:grid_quad}
\end{figure}

In latitude, the scale to be resolved is the vertical pressure scale height of the disk. This now requires higher resolution near the equatorial plane, so again a stretch is applied, this time with the minimum grid spacing at the equator ($\theta=\pi/2$) and the stretch proceeding in both directions toward the poles ($\theta=0$ and $\pi$), making each cell 1.5\% larger than the one immediately closer to the equator. The second column in table \ref{tab:grid} gives details of this $\theta$ grid, including now the minimum grid spacing in $\theta$ and the disk pressure scale height at $R_\ast$ where $H$ has its minimum, again calculated for both the B2V and O7V models. Note that there are again approximately two zones per scale height. Since the scale height grows faster than linear with $r$ (like $r^{1.5}$ near the equator) while the $\theta$ grid spacing scales linearly with $r$, away from stellar surface this resolution of the disk pressure scale height improves. Figure \ref{fig:grid_quad} plots this coordinate mesh in both $r$ and $\theta$.

\begin{table}
\centering
\caption{Grid specifications \label{tab:grid}}
\begin{tabular}{| l | c  c |}
\hline
& $r$ & $\theta$  \\
\hline
min. & $R_\ast$ & 0\\
max. & $10\,R_\ast$ & $\pi$ \\
number of zones & 300 & 120 \\
stretch & 1.02 & 1.015 \\
min. grid spacing & 4.7$\times 10^{-4}\,R_\ast$ & 1.7$\times 10^{-2}\, R_\ast$ \\
B2 scale height & 8.8$\times 10^{-4}\,R_\ast$ & 3.0$\times 10^{-2}\, R_\ast$ \\
O7 scale height & 9.3$\times 10^{-4}\,R_\ast$ & 3.0$\times 10^{-2}\, R_\ast$ \\
\hline
\end{tabular}
\end{table}

\subsection{Ray quadrature}
\begin{figure}
\centering
\begin{subfigure}[b]{0.46\textwidth}
\includegraphics[width=\textwidth]{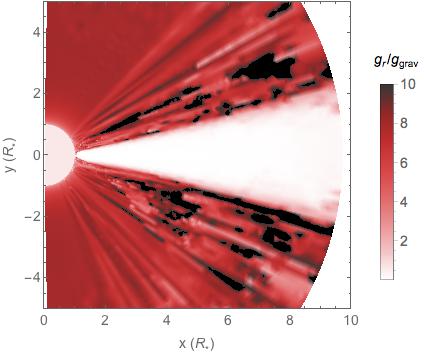}
\subcaption{$g_r$}
\end{subfigure}
\begin{subfigure}[b]{0.48\textwidth}
\includegraphics[width=\textwidth]{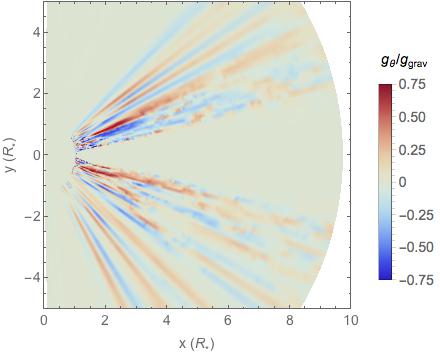}
\subcaption{$g_\theta$}
\end{subfigure}
\begin{subfigure}[b]{0.48\textwidth}
\includegraphics[width=\textwidth]{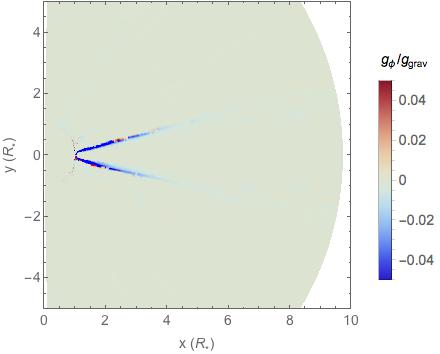}
\subcaption{$g_\phi$}
\end{subfigure}
\caption{Snapshots of $g_r$, $g_\theta$, and $g_\phi$ for a $n_p=n_{\phi'}=6$ ray quadrature at $t=500$ ks}
\label{fig:evo_force}
\end{figure}

Ensuring a sufficient ray quadrature resolution is less obvious than the spatial grid. Predominantly, this is due to the fluctuating small scale structures generated by ablation. Since each viewing direction shows a different velocity gradient, these small scale structures can be of different magnitudes and, in some cases, even different positions with changing ray quadratures. The sharp interface between the disk and wind also contributes to significant differences between velocity gradients for neighboring rays. Figure \ref{fig:evo_force} demonstrates this noisiness by plotting the three components of the line-force for a $n_p=n_{\phi'}=6$ ray quadrature at $t=500$ ks.

\begin{figure}
\centering
\includegraphics[width=0.9\textwidth]{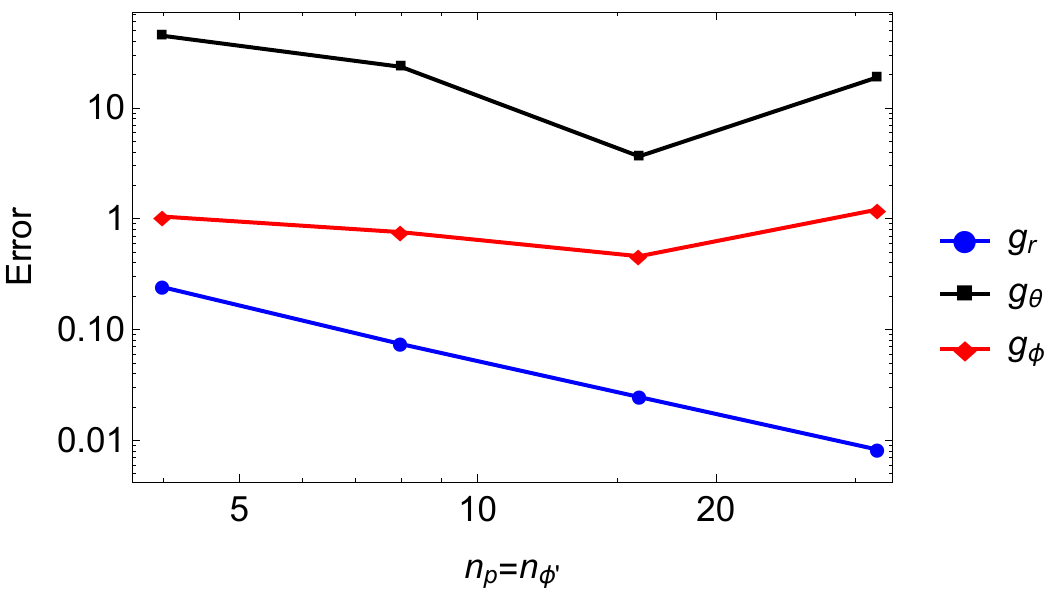}
\caption{
Convergence test for the components of the line-acceleration.
}
\label{fig:quad_conv}
\end{figure}

To quantify whether these structures converge with increasing ray resolution, we define an error measure to be

\beq
Error=\mathrm{Max}\left(1-\left\vert\frac{g_x(n/2)}{g_x(n)}\right\vert\right)\,,
\eeq
where $x$ can be $r$, $\theta$, or $\phi$, and we here fix $n=n_p=n_{\phi'}$. Figure \ref{fig:quad_conv} plots this error versus n for all three components of the line-acceleration\footnote{Points where $g_x(n)=0$ are omitted from this calculation to prevent Error$\rightarrow\infty$.}. While $g_r$ converges to within 10\% error by $n=8$ and continues to converge as $n$ increases, the error in $g_\phi$ nor $g_\theta$ stays high and shows no clear signs of converging up to $n=32$.

\begin{figure}[h!]
\centering
\includegraphics[width=0.8\textwidth]{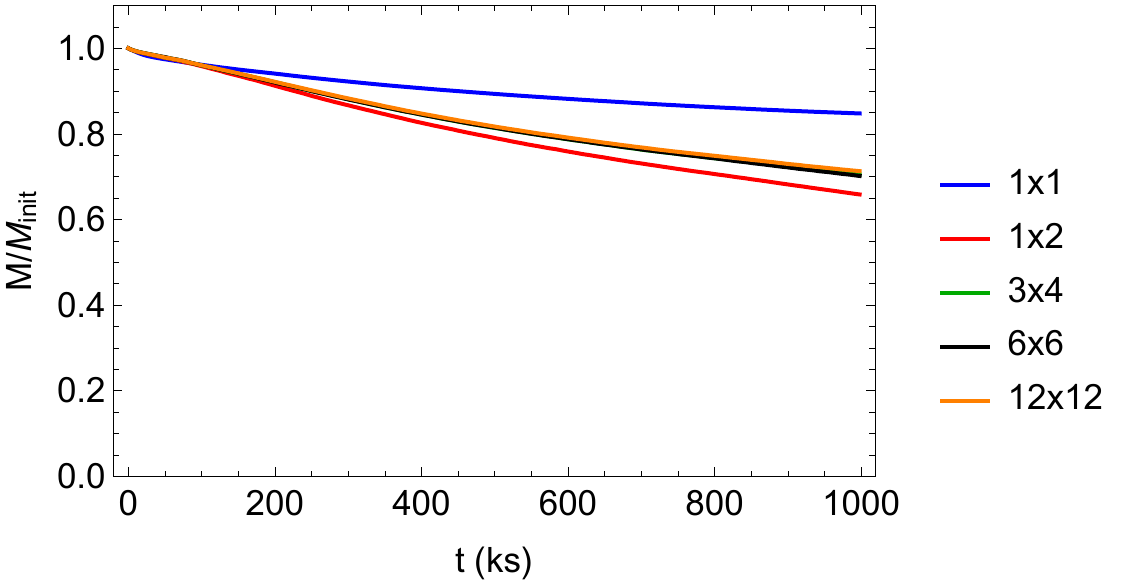}
\caption{
Total mass in the simulation volume as a function of time for several ray quadratures. Each label lists first the number of $y$ points and then the number of $\phi'$ points.
}
\label{fig:res_test_mass}
\end{figure}

\begin{figure}[h!]
\centering
\includegraphics[width=0.8\textwidth]{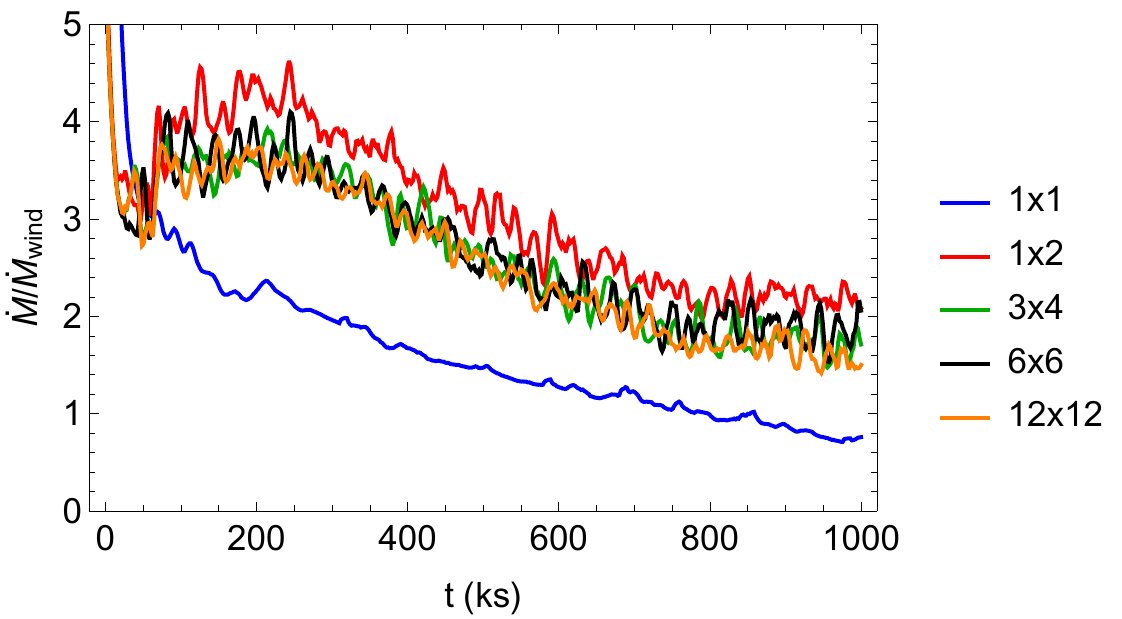}
\caption{
Ablation rate as a function of time for several ray quadratures. Each label lists first the number of $y$ points and then the number of $\phi'$ points.
}
\label{fig:res_test_mdot}
\end{figure}

\begin{figure}[h!]
\centering
\includegraphics[width=0.85\textwidth]{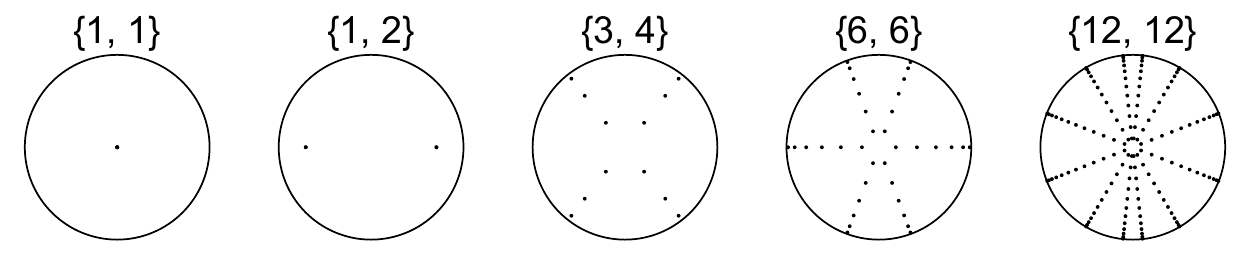}
\caption{
Ray quadratures used for figures \ref{fig:res_test_mass} and \ref{fig:res_test_mdot}. Each label lists first the number of $p$ points and then the number of $\phi'$ points.
}
\label{fig:ray_quad}
\end{figure}

On a grid level, it is thus difficult to identify a quadrature with clear formal convergence, so a more global criterion is needed. As the most important characteristics of the simulations are the ablation rate and the amount of disk material remaining in the simulation, a more meaningful test is to run several models and compare their evolution in these key parameters. Figure \ref{fig:ray_quad} shows the ray quadratures used for this test\footnote{Here the ``1$\times$1'' quadrature is actually a point star model including the finite disk correction factor}. As can be seen in figures \ref{fig:res_test_mass} and \ref{fig:res_test_mdot}, with the exception of the lowest two resolution simulations, the variations between the different quadratures are below the level of the random fluctuation in mass loss rate for a single simulation leading to very similar amounts of mass remaining in these three simulations. 
%We therefore chose to delay the decision between ray quadratures until we have had a chance to compare line-acceleration implementations.

%\newpage

\section{Comparison of Force Implementations}\label{sec:force_imp}

%Now that we have developed an appropriate resolution for the fully 3-Dimensional implementation, we can 
Let us now examine how changes in the line-acceleration implementation affect the ablation rate. To do this, we define three models:
\begin{enumerate}
\item \emph{1D radial}: This assumes $\mathbf{v}(r)=v_r\mathbf{\hat{r}}$ and $v_\theta=v_\phi=\partial v_r/\partial \theta = \partial v_r/\partial \phi=0$ in computing a purely radial\footnote{This is achieved by replacing $\mathbf{\hat{n}}\cdot\nabla(\mathbf{\hat{n}}\cdot\mathbf{v})$ with $\mu^2 dv_r/dr+ (1-\mu^2) v_r/r$ in equation \ref{eq:gcak_3d}.} line-acceleration $\mathbf{g}=g_r\mathbf{\hat{r}}$.
%\item \emph{3D radial}: This allows $\mathbf{v}(r,\theta,\phi)=v_r\mathbf{\hat{r}}+v_\theta\mathbf{\hat{\theta}}+v_\phi\mathbf{\hat{\phi}}$ in computing a purely radial line-acceleration $\mathbf{g}=g_r\mathbf{\hat{r}}$.
\item \emph{3D radial}: This calculates the full integral over the stellar core with $\mathbf{v}(r,\theta,\phi)=v_r\mathbf{\hat{r}}+v_\theta\mathbf{\hat{\theta}}+v_\phi\mathbf{\hat{\phi}}$ in order to calculate a fully 3-dimensional line acceleration. However, only the radial component, $\mathbf{g}=g_r\mathbf{\hat{r}}$, is used in the evolution of the code.
\item \emph{3D vector}: This accounts for the full vector velocity gradients in computing a full vector line-acceleration.
\end{enumerate}
% The first assumes that the velocity is purely radial, such that $\mathbf{v}=v_r\mathbf{\hat{r}}$ and $v_\theta=v_\phi=0$, and also ignores the non-radial velocity gradients $\partial v_r/\partial \theta = \partial v_r/\partial \phi$ in computing a purely radial line-acceleration $\mathbf{g}=g_r\mathbf{\hat{r}}$. We call this the ``1D radial" acceleration implementation. It is worth noting that, for this 1D radial acceleration model, we do not enforce that the non-radial velocity components are in fact zero, nor do we impose that $v_r$ is indeed actually a function of only $r$. Rather, we simply neglect all non-radial velocities and velocity gradients in the calculation of the line-acceleration.

%The other two models relax the assumptions on velocity, using the actual velocity field in the simulation to calculate all 3 components of the velocity gradient dynamically. However, for what we refer to as the ``3D radial'' implementation we continue to assume $\mathbf{g}=g_r\mathbf{\hat{r}}$, whereas the ``3D vector'' implementation uses all 3 components of the line-acceleration and so is the most complete of the three models.

Note that in all these implementations of the line-acceleration, the wind itself \emph{does} have $\mathbf{v}(r)=v_r\mathbf{\hat{r}}$ and, since it is spherically symmetric, $\partial v_r/\partial \theta = \partial v_r/\partial \phi \equiv 0$. Therefore, we expect the wind to be unchanged among the three implementations. This means that what follows can be considered to be an investigation of how disk ablation depends on the force implementation, and that this investigation is not contaminated by variations in the stellar wind.

\subsection{1D radial vs. 3D vector line acceleration}
\begin{figure}
\centering
\includegraphics[width=\textwidth]{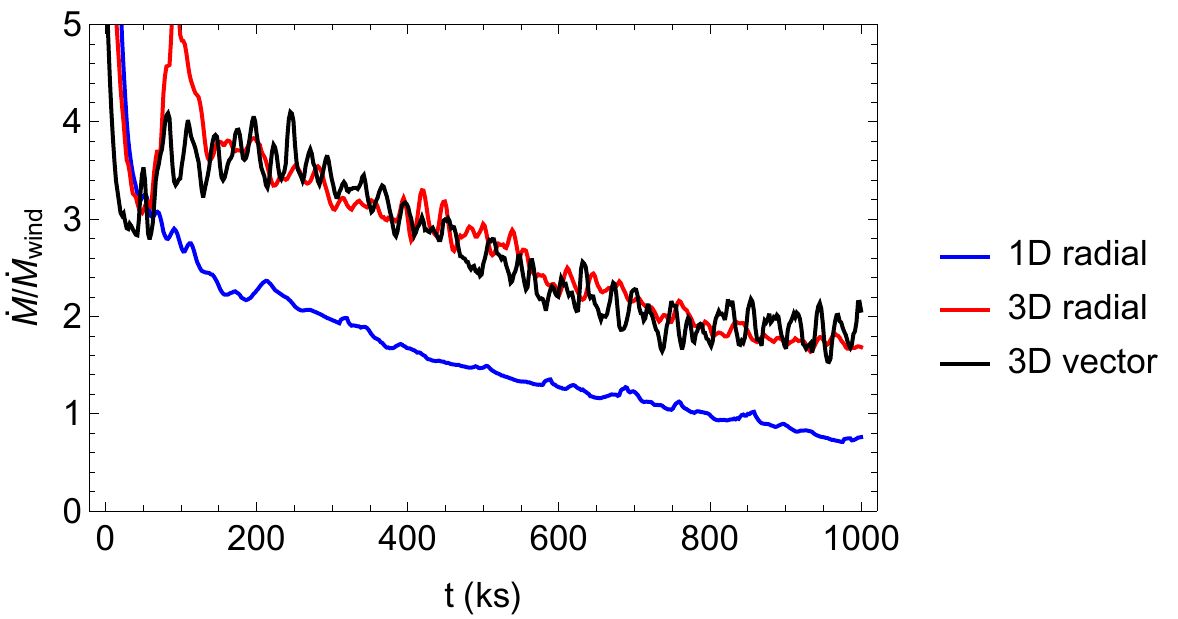}
\caption{
Ablation rate, measured in units of the spherically symmetric mass loss rate, for the three force implementations discussed in the text.
}
\label{fig:1d_v_3d}
\end{figure}

By comparing the 1D radial implementation to the 3D vector implementation, we can test the importance of the non-radial velocity gradient terms in causing ablation. This comparison can also help disentangle what portion of the mass lost from the disk is due to ablation -- by which we mean the removal of material only by direct acceleration of disk material by radiation -- and what fraction is from ``entrainment'', wherein the wind drags low density disk material viscously coupled through a Kelvin-Helmholtz instability. Figure \ref{fig:1d_v_3d} shows that, asymptotically, the 3D vector form yields a disk ablation rate more than twice the 1D radial model. This suggests that the inclusion of full line-acceleration more than doubles any ablation from entrainment alone.

\subsection{3D radial vs. 3D vector line acceleration}
Figure \ref{fig:1d_v_3d} also includes the 3D radial model. While there is a marked difference between including or omitting the non-radial velocity gradients, the non-radial components of the line-acceleration seem to make very little difference to the overall ablation rate. This stands in interesting contrast to the wind-compressed-disk inhibition and wind spin-down discussed at the end of chapter \ref{chap:line}, wherein the angular acceleration components were most important. Note that for both the 3D radial and vector implementations the full 3D line-acceleration is calculated, but for the 3D radial implementation the $\theta$- and $\phi$-components are reset to zero.  Therefore, while it seems that they formally could be omitted here, there is effectively no computational cost difference between including and omitting them, and the full 3D version will be used for the remainder of the dissertation.

In light of these results we can now see that, in terms of ray quadrature resolution, it is the convergence of the radial component of the line-acceleration that is most crucial, since the $\theta$ and $\phi$ components are dynamically unimportant. As the radial component does converge relatively well with increasingly dense ray quadratures (see figure \ref{fig:quad_conv}), we can now understand why there is such good agreement between the simulation mass and ablation rate in figures \ref{fig:res_test_mass} and \ref{fig:res_test_mdot}. Therefore, for the remainder of the dissertation we choose a $n_p=n_{\phi'}=6$ ray quadrature as a balance between ensuring convergence in radial acceleration without having prohibitively expensive simulations.

\section{Effects of Rotation}\label{sec:w80_rot}
 
As discussed in chapter \ref{chap:disks}, Classical Be stars are observed to be rotating at upwards of 70\% of critical, defined here \citep[see][and appendix \ref{app:ob_gravdark}]{RivCar13} by the ratio of their equatorial rotation speed to local Keplerian orbital speed,

\beq
W\equiv\frac{v_{eq}}{\sqrt{G M_\ast/R_\ast}}\,.
\eeq
%\hl{maybe this parameter should be in chapter 3 as well?}
As discussed in chapter \ref{chap:pdome}, it seems that this rapid rotation plays a key role in the feeding of the circumstellar disk. Let us here examine whether this rapid rotation, and the associated stellar oblateness and gravity darkening, has a significant effect on line-driven ablation.

\begin{figure}
\centering
\includegraphics[width=0.6\textwidth]{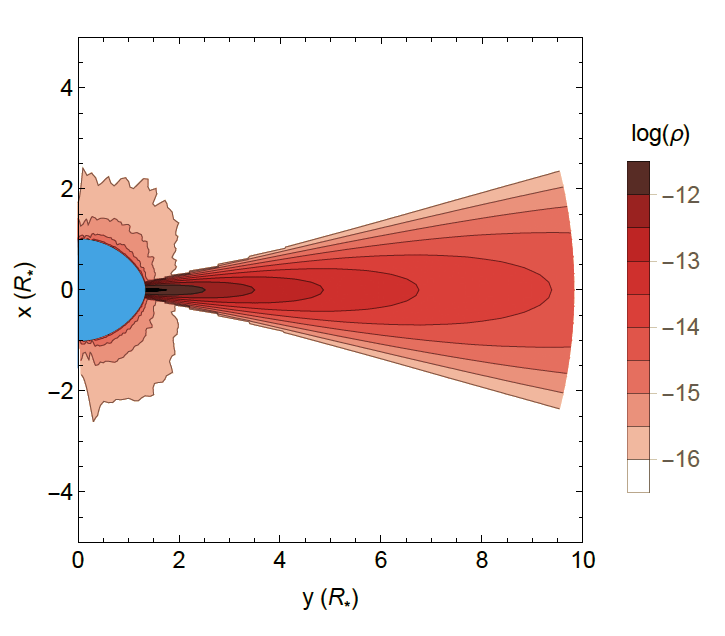}
\caption{
Initial condition of the rotating B2 model plotted in $\log(\rho)$ measured in $\log($g/cm$^3)$.
}
\label{fig:b2_rot_init}
\end{figure}

To test this, we modify our standard B2 model to have $W=0.8$, keeping the equatorial radius constant and allowing the polar radius to shrink\footnote{Such a star may not be viable under stellar evolution models. However, this choice provides the most direct method for comparing ablation effects with and without stellar rotation.}. As done for the non-rotating case, this rotating model is allowed to relax to a steady state wind and then a disk is superimposed on top. Recall from chapter \ref{chap:line} that an oblate star is expected to have a prolate wind, a property that is reproduced in these simulations, as can be seen from the initial conditions plotted in figure \ref{fig:b2_rot_init}.
%. In going from a non-rotating to rotating model, we choose to keep the equatorial radius constant since the disk is mainly in the equatorial plane

\begin{figure}[h!]
\centering
\includegraphics[width=\textwidth]{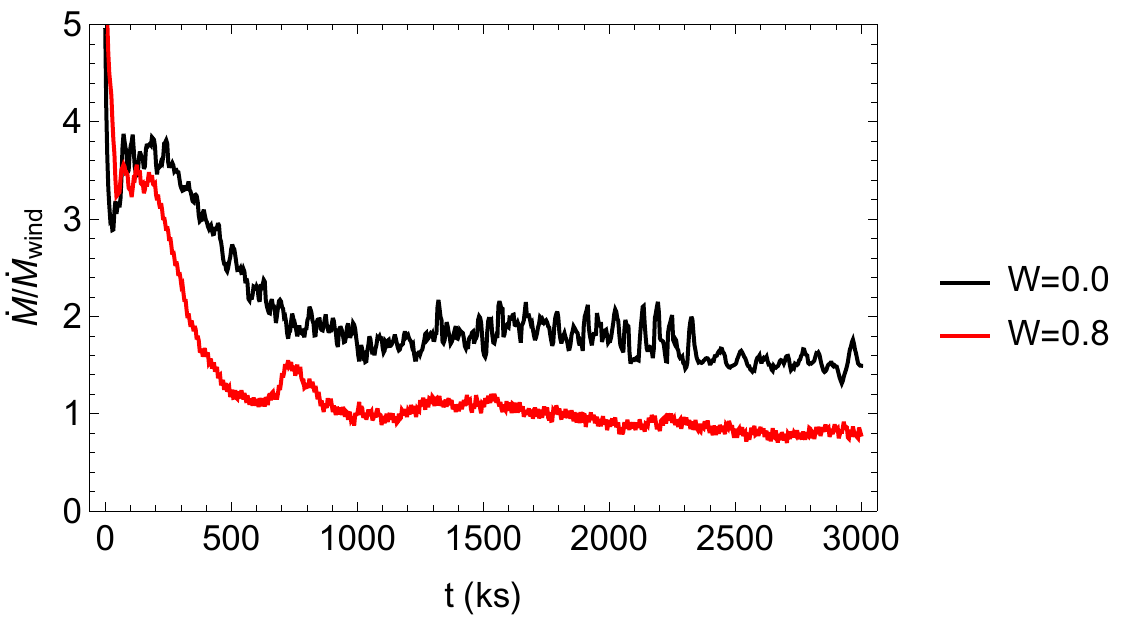}
\caption{
Ablation rate in units of spherically symmetric mass loss rate for both rotating and non-rotating B2 models.
}
\label{fig:rot_v_norot}
\end{figure}

Figure \ref{fig:rot_v_norot} compares the ablation rate of the B2 models with and without rotation, showing that the ablation rate is somewhat reduced, by a maximum factor about two. This is consistent with the comparable factor two equatorial gravity darkening between the non-rotating and rotating models, as discussed in appendix \ref{app:ob_gravdark}, following the review given by \cite{Cra96}. 
%Since stellar luminosities are often uncertain up to a factor two, and the actual form of gravity darkening is also uncertain, the additional complexities of stellar oblateness and gravity darkening outweigh the order unity modification to ablation rate. We therefore choose not to add the additional free parameter for rotation in the spectral type parameter study in chapter \ref{chap:ab_spec_type}.

\chapter{Radiative Ablation of Classical Be-type Disks as a Function of Spectral Type}\label{chap:ab_spec_type}

We now have in hand a standard simulation model describing a B2 star with an optically thin circumstellar disk. Additionally, we can feel confident, in light of the resolution tests performed at the end of the last chapter, that we understand the necessary spatial and ray quadratures for an ablation simulation. Finally, we also now have a more quantitative understanding both of the necessity of the 3D acceleration implementation derived in chapter \ref{chap:line}, and that it is the radial component of this acceleration that is most essential for calculating line-driven ablation.

We can now move forward into a parameter study of the dependance of ablation on stellar spectral type. As pointed out in chapter \ref{chap:disks}, Classical Be stars do not extend significantly into the O star domain. Therefore, a particularly important question is whether such a trend can be explained by the stronger radiative ablation associated with the greater luminosity of such stars. To this end, we first present a model of an O7 star, which should not be expected to host a Be-type disk. By separating this model out from the rest of the spectral type survey, we can discuss its properties in more detail.

Section \ref{sec:spec_type} then presents the complete parameter study of ablation as a function of spectral type. Here there is particular emphasis on the scaling of ablation with stellar and disk parameters. In particular, we present a simple scaling relation that quite well describes the simulation results.

\section{Disk Destruction by an O7V star}\label{sec:o7}

To begin, let us discuss the parameters of an O7 star and its optically thin disk, presented in table \ref{tab:o7_params}, and of its wind, presented in table \ref{tab:o7_wind}. The choice of keeping the disk marginally optically thin leads to a factor 5 increase in disk mass, but the stellar luminosity has increased by a factor of 25, leading to a factor 200 increase in the mass loss rate. Therefore, we should expect much stronger disk ablation.

\begin{table}
\caption{Stellar and Disk Parameters of the O7 Model} \label{tab:o7_params}
\centering
\begin{tabular}{| l | c | c | c | c | c |}
\hline
Sp. type & $T_{eff}$ (kK) & $L_\ast$ ($L_\odot$) & $M_\ast$ ($M_\odot$) & $R_\ast$ ($R_\odot$) & $M_{disk}$ ($M_\odot$)\\
\hline
O7V & 36 & 1.3$\times 10^5$ & 26.5 & 9.4 & 7.0$\times 10^{-10}$  \\
\hline
\end{tabular}
\end{table}

\begin{table}
\caption{Wind Parameters of the O7 Model} \label{tab:o7_wind}
\centering
\begin{tabular}{| l | c | c | c | c |}
\hline
Sp. type & $\bar{Q}$ & $Q_\mathrm{o}$ & $\alpha$ & $\dot{M}_{wind}$ ($M_\odot/yr$) \\
\hline
O7V & 2500 & 2200 & 0.66 & 1.2$\times 10^{-7}$ \\
\hline
\end{tabular}
\end{table}

\begin{figure}
\centering
\begin{subfigure}[b]{0.6\textwidth}
\includegraphics[width=\textwidth]{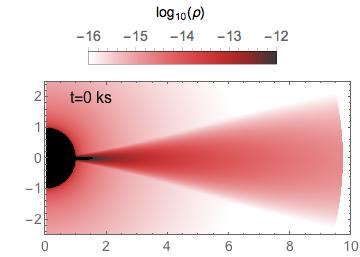}
\end{subfigure}

\begin{subfigure}[b]{0.6\textwidth}
\includegraphics[width=\textwidth]{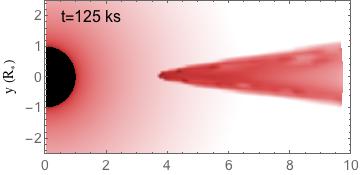}
\end{subfigure}

\begin{subfigure}[b]{0.6\textwidth}
\includegraphics[width=\textwidth]{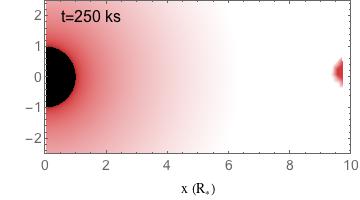}
\end{subfigure}

\caption{
Snapshots of log($\rho$) for the O7 model at 0, 125, and 250 ks.
}
\label{fig:O7_snaps}
\end{figure}

\begin{figure}
\centering
\includegraphics[width=\textwidth]{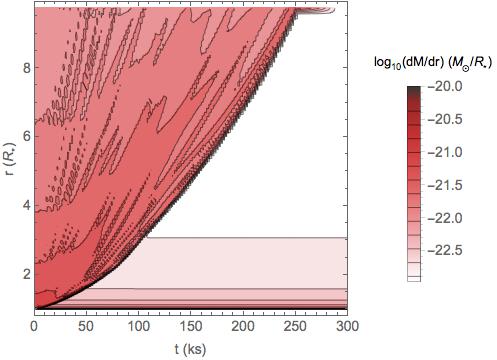}
\caption{
Mass in spherical shells above the stellar surface in units of solar masses per stellar radius.
}
\label{fig:O7_dmdr}
\end{figure}

Indeed, while the model of a B2 star presented in the prior chapter slowly destroys its disk by eating away at it from the edges, the radiative acceleration of an O7 star is strong enough to simply carry the whole disk away in a dynamical timescale as shown by figure \ref{fig:O7_snaps}. This process is rapid enough to preclude producing a comparable figure to \ref{fig:B2_dmdotdmu} for this model. Figure \ref{fig:O7_dmdr} instead plots the spatial and temporal variation of mass in each spherical shell,
\beq
\frac{dM(r,t)}{dr}\equiv \oint \rho r^2 d\Omega\,.
\eeq
Note particularly the rapid removal of material from the inside of the disk outwards, as well as the quite short (less than three day) time it takes for the disk to be completely evacuated.

\begin{figure}
\centering
\includegraphics[width=0.9\textwidth]{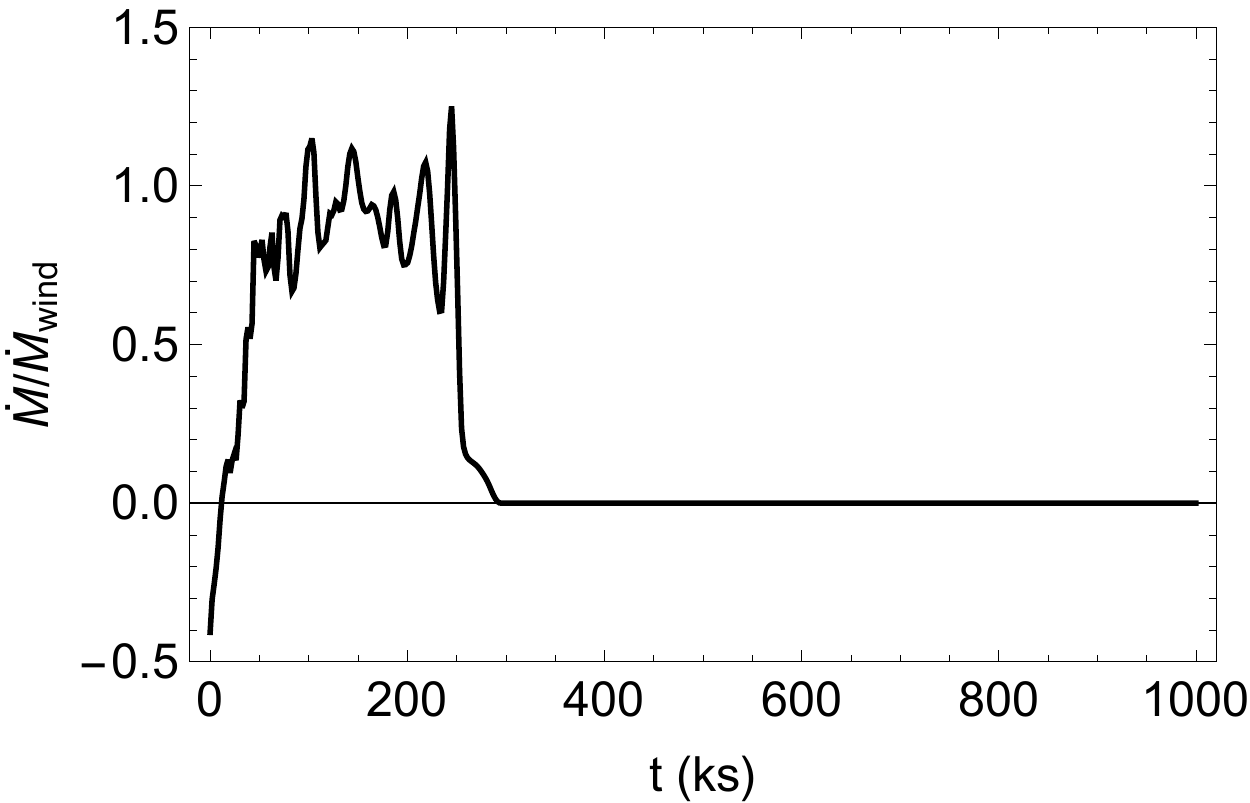}
\caption{
Ablation rate in units of spherically symmetric mass loss rate for an O7 star, the highest mass model considered.
}
\label{fig:O7_mdot}
\end{figure}

Figure \ref{fig:O7_mdot} shows the wind normalized ablation rate for the O7 simulation. Whereas the B2 model slowly relaxed over the simulation duration to a relatively steady ablation rate at about twice the spherically symmetric mass loss rate, the O7 model impulsively ejects the disk material in one large burst and then returns to a steady state wind. This removal of the disk is so sudden that the ablation rate never significantly exceeds the spherically symmetric mass loss rate.

We thus see that an O7 star destroys a pre-existing optically thin disk on a dynamical timescale. In order to counter this destruction, the star would need to be feeding a disk at a rate of over $1.2\times 10^{-7}M_\odot$/yr. Since this rate is so much larger than that needed to maintain a disk for later spectral types, this provides a possible explanation for why there are no O7e stars. \citep[see e.g.][]{MarFre06,RivCar13}.

\section{Variation of Spectral Type}\label{sec:spec_type}

Since the fraction of O and B stars showing the Be phenomena is a function of spectral type, we here investigate whether this may in part be a byproduct of radiative ablation. To address this question, we consider a series of models spanning a spectral range from B3 to O7 with parameters given in tables \ref{tab:spec_stars} and \ref{tab:spec_wind}. The stellar parameters of the O stars are taken from \cite{MarSch05} while, as was done for the B2 star, \cite{TruDuf07} and \cite{GeoEks13} are used to derive the stellar parameters of the B stars. \cite{PulSpr00} is used for all wind parameters.

\begin{figure}
\centering
\includegraphics[width=0.8\textwidth]{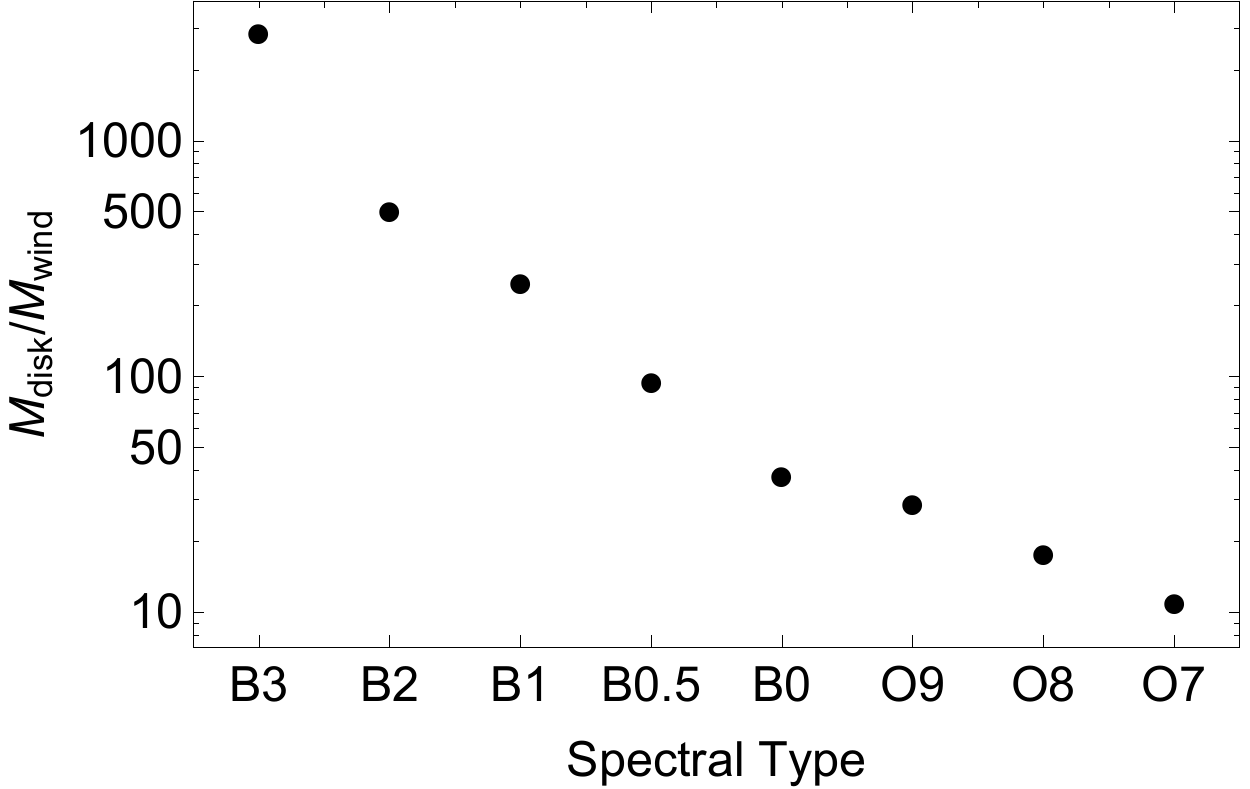}
\caption{
Ratio of disk mass to wind mass as a function of spectral type.
}
\label{fig:disk_to_wind_ratio}
\end{figure}

%\begin{minipage}

\begin{table}
\caption{Stellar and Disk Parameters as a Function of Spectral Type} \label{tab:spec_stars}
\centering
\begin{tabular}{| l | c | c | c | c | c |}
\hline
Sp. type & $T_{eff}$ (kK) & $L_\ast$ ($L_\odot$) & $M_\ast$ ($M_\odot$) & $R_\ast$ ($R_\odot$) & $M_{disk}$ ($M_\odot$) \\
\hline
B3V & 18 & 1.6$\times 10^3$ & 6 & 4.1 & 1.3$\times 10^{-10}$ \\
B2V & 22 & 5.0$\times 10^3$ & 9 & 5.0 & 1.9$\times 10^{-10}$  \\
B1V & 24 & 7.9$\times 10^3$ & 11 & 5.2 & 2.0$\times 10^{-10}$  \\
B0.5V & 28 & 1.6$\times 10^4$ & 13 & 5.5 & 2.3$\times 10^{-10}$  \\
B0V & 31 & 1.6$\times 10^4$ & 16 & 6.0 & 2.7$\times 10^{-10}$  \\
O9V & 32 & 5.0$\times 10^4$ & 18 & 7.7 & 4.8$\times 10^{-10}$  \\
O8V & 33 & 7.9$\times 10^4$ & 22 & 8.5 & 5.7$\times 10^{-10}$  \\
O7V & 36 & 1.3$\times 10^5$ & 26.5 & 9.4 & 7.0$\times 10^{-10}$  \\
\hline
\end{tabular}
%\end{table}
\vspace{20pt}
%\begin{table}
\caption{Wind Parameters as a Function of Spectral Type} \label{tab:spec_wind}
\centering
\begin{tabular}{| l | c | c | c | c |}
\hline
Sp. type & $\bar{Q}$ & $Q_\mathrm{o}$ & $\alpha$ & $\dot{M}_{wind}$ ($M_\odot/yr$) \\
\hline
B3V & 1500 & 7000 & 0.55 & 7.3$\times 10^{-11}$ \\
B2V & 1800 & 4900 & 0.59 & 7.4$\times 10^{-10}$ \\
B1V & 2000 & 4600 & 0.60 & 1.7$\times 10^{-9}$ \\
B0.5V & 2300 & 3900 & 0.63 & 6.0$\times 10^{-9}$ \\
B0V & 2400 & 3400 & 0.64 & 1.9$\times 10^{-8}$ \\
O9V & 2400 & 3300 & 0.65 & 3.4$\times 10^{-8}$ \\
O8V & 2300 & 3100 & 0.65 & 6.3$\times 10^{-8}$ \\
O7V & 2200 & 2500 & 0.66 & 1.2$\times 10^{-7}$ \\
\hline
\end{tabular}
\end{table}

%\end{minipage}

\begin{figure}
\centering
\includegraphics[width=\textwidth]{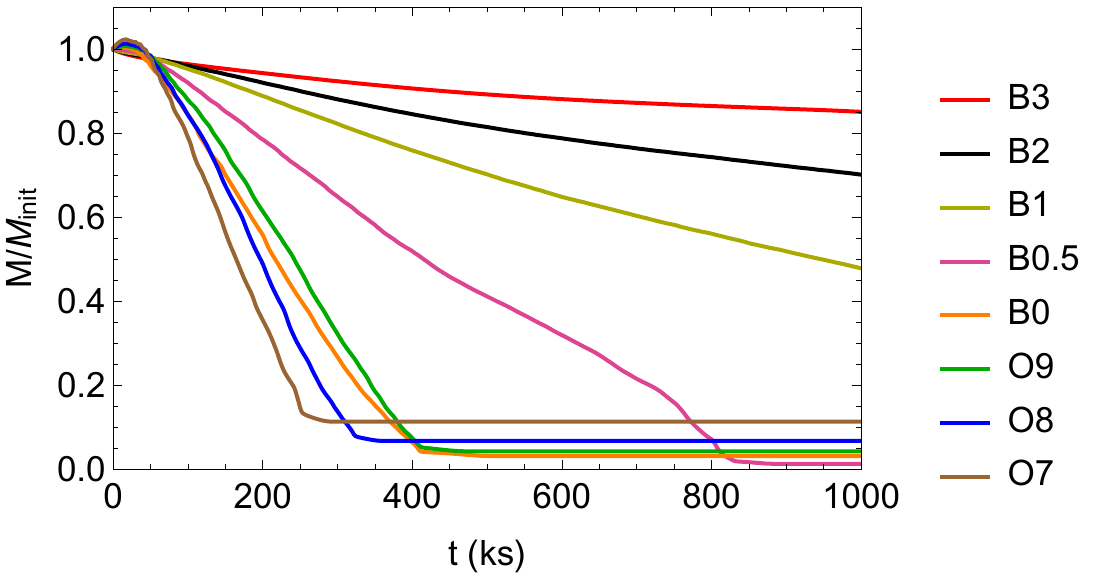}
\caption{
Fractional mass in the simulation volume as a function of time and spectral type
}
\label{fig:frac_m_spec}
\end{figure}

\begin{figure}
\centering
\includegraphics[width=\textwidth]{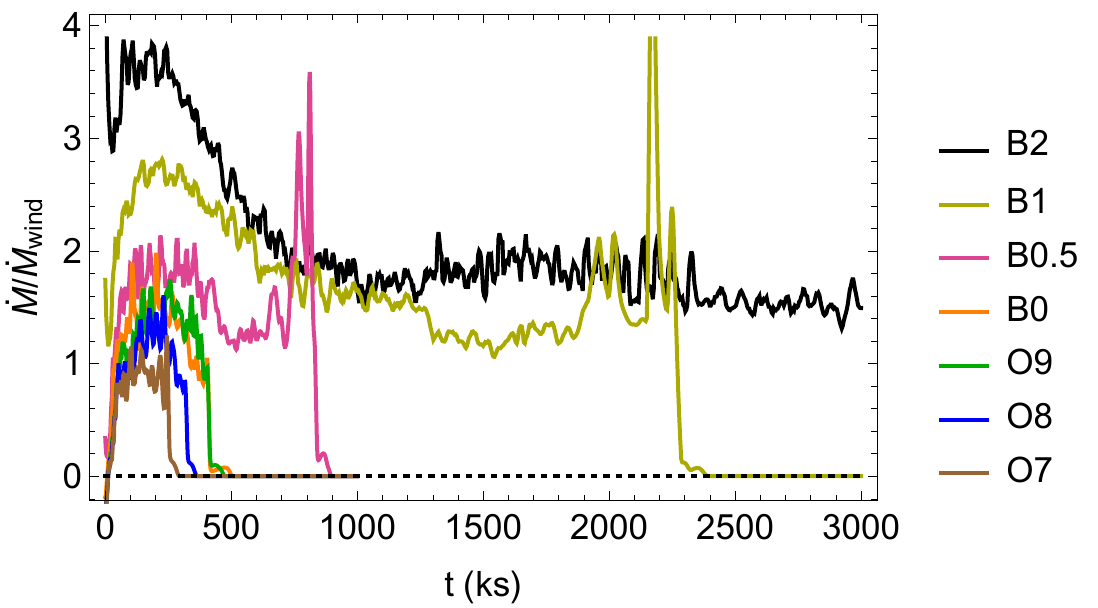}
\caption{
Ablation rate in units of spherically symmetric $\dot{M}_{wind}$ for each spectral type.
}
\label{fig:mdotdisk}
\end{figure}

To facilitate comparison among the models, figure \ref{fig:disk_to_wind_ratio} plots the ratio of disk to wind mass following the scalings introduced in section \ref{sec:disk_struc}.
%To make sure that the models considered here can be compared to one another, the scalings of disk and wind mass introduced in section \ref{sec:disk_struc} need to be discussed as a function of spectral type.
%Figure \ref{fig:disk_to_wind_ratio} plots the ratio of disk mass to wind mass for each of the models considered. 
While this ratio declines fairly significantly going to earlier spectral types, it is never less than 10, and thus the simulations always have their initial mass budget dominated by the disk.
Since this ratio depends on stellar mass, radius, and temperature, the models can most easily be directly compared by focusing on the fraction of initial mass that remains at time t, as shown in figure \ref{fig:frac_m_spec}, or on the ablation rate as shown in figure \ref{fig:mdotdisk}. Both of these metrics highlight two populations, one consisting of stars that behave like the O7 star and dynamically eject their disk, and the other consisting of stars like the B2 star and slowly ablate their disks at a steady rate over a much longer time. We bridge the strong difference between a B0 star (in the first category) and a B1 star (in the second category), with a B0.5 model.

\begin{figure}
\centering
\includegraphics[width=0.8\textwidth]{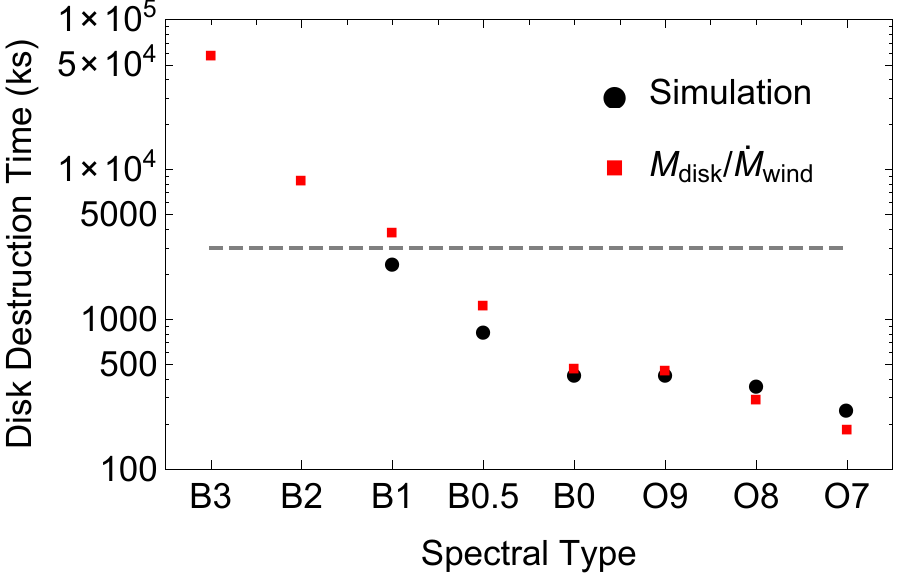}
\caption{
Time to destroy an optically thin disk as predicted by $M_{disk}/\dot{M}_{wind}$ (red squares) compared with the time it actually takes in the simulations (black circles). The dashed line denotes the duration of the longest simulation meaning no simulations have been run long enough to see the disk removal completed for the B2 or B3 case.
}
\label{fig:t_disk}
\end{figure}

While this transition in behavior is very sharp, it is perhaps unsurprising. The ratio between initial disk mass and spherically symmetric mass loss rate, shown by the blue points in figure \ref{fig:t_disk}, reproduces within an order unity factor the actual disk destruction times shown in orange. Therefore, with knowledge of the stellar spectral properties and circumstellar disk mass, one can accurately estimate the destruction time of an optically thin disk.

The ability of this simple scaling to predict simulation results guides its use to predict the behavior of actual stars.
%More striking even than the ability of this simple scaling to describe the simulation results is its ability to describe nature. 
Recall that chapter \ref{chap:disks} reviewed the disk decay inferred observationally by \cite{CarBjo12} for the Be star 28 CMa. In order to reproduce the 170 day disk decay time, these authors invoked strong viscous force with an anomalously high viscous coefficient $\alpha\approx1$ \citep{ShaSun76}. By comparing the inferred stellar parameters of 28 CMa from \cite{MaiRiv03} to the models here, we can see that it is best represented by our B2 model. The simple scaling relation for this case gives a disk destruction time of 97 days, within a factor of two of the observed 170 day disk decay time. 
%This agreement of time scale shows that line-driven ablation can explain Be disk decay without employing anomalously strong viscous transport.

This important result indicates that line-driven ablation can readily explain the relatively short, few-month timescale observed for Be-disk decay, without the need to invoke an anomalously strong viscous diffusion. As discussed further in the conclusions, it also provides a prediction of ablation time with spectral type, which can be readily tested by observations.

\chapter{Pulsational Mass Ejection in Be Star Disks}\label{chap:pdome}

This first appeared as a conference proceeding as \cite{KeeOwo14b}.

\section{Introduction}

Despite the long and extensive history of observations of Classical Be stars, much remains unknown about the physics that forms their circumstellar disks. Of key interest is the mechanism by which such stars are able to launch material into their circumstellar environment with enough angular momentum to form a disk. At this point, it is widely accepted that the universal appearance of rapid rotation in this class of objects plays a central role in this process. Though exact rotation velocities for individual Be stars are not well agreed upon due to the flattening of the relationship between observed line broadening and stellar rotation velocity in the limit of rapid rotation \citep{TowOwo04}, for sub-critical rotation, some additional mechanism(s) are necessary to place material into orbit.

A crucial step toward an understanding of the origins of the circumstellar material came from the wind compressed disk (WCD) model of \cite{BjoCas93}. In this model, the radial velocity contribution of line-driving in the stellar wind, coupled with the tangential velocity component from the rapid rotation of the star, places material in the wind into inclined orbits that eventually pass through the equatorial plane, whereupon collision with material from the opposite hemisphere forms an equatorial WCD. However, the work of \cite{OwoCra96} showed that the combination of non-radial forces and the equatorial gravity darkening present in rapidly rotating, oblate stars combine to reverse the expected equatorward flow into a poleward flow, thereby inhibiting the formation of an equatorial disk.

A possible clue to an alternate mechanism came from observations of line profile variability of $\mu$ Cen by \cite{RivBaa01} and $\omega$ CMa by \cite{MaiRiv03}. In both cases, increases in the amplitude of photometric variability from non-radial pulsations were found to be associated with modulation and growth of emission from the circumstellar disk. Since then, observations have shown that non-radial pulsation in Classical Be stars is a ubiquitous feature, and \cite{RivBaa03} have even gone so far as to suggest that these non-radial pulsations may be causally connected with the Be phenomena.

The Pulsationally Driven Orbital Mass Ejection (PDOME) model here explores the dynamical issues for ejecting material into orbit from non-radial pulsations on a star near critical rotation. Section \ref{sec:model} discusses details of the model and the parameters used, section \ref{sec:results} presents some preliminary findings, and section \ref{conclusions} discusses possible fruitful directions for future work.

\section{Details of the Model}\label{sec:model}

\subsection{Basic Paradigm}

Our approach here is to explore the dynamics of circumstellar material launched within the equatorial plane from perturbations in density and azimuthal velocity on an underlying, rapidly rotating Be star. Before beginning a discussion of the model in full, it is helpful first to define and clarify some terms. The most important among these are concerned with the nature of the surface perturbations intended to mimic non-radial gravity waves ($g$-mode) pulsations.

For computational convenience, the perturbations explored here are always assumed to be $\lvert m\rvert=4$ , where $m$ is the number of nodes around the equator of the star; however, our implementation of lower boundary conditions in density and velocity allows for distinct directions of phase propagation and energy transport. Specifically, in the frame of the star, the \emph{phase} propagation can either be \emph{prograde} ($v_{phase}>0$) or \emph{retrograde} ($v_{phase}<0$) relative to the direction of stellar rotation. Separately, we can also force the \emph{material} velocity at the peak density of the perturbation to be either prograde ($v_{pert}(\rho_{max})>0$) or retrograde ($v_{pert}(\rho_{max})<0$). Since the material motion sets the direction of energy propagation that is normally associated with group velocity, we refer to the four combinations we explore as prograde group/prograde phase (+/+), prograde group/retrograde phase (+/-), etc.

\subsection{Numerical Model Specifications}

Our numerical simulations of the PDOME model use the Piecewise-Parabolic Method \citep{ColWoo84} hydrodynamics code\footnote{ http://wonka.physics.ncsu.edu/pub/VH-1/} VH-1, implemented here in a 2D, spherical equatorial ($r,\phi$) plane with azimuth ranging from 0 to 90 degrees and radius from 1 to 6 $R_*$. For simplicity we use an isothermal approximation without explicit inclusion of viscous terms.

As the scale height of the stellar atmosphere is very small, $H\sim\,R_*/1000$, it is difficult to resolve stellar pulsations within a hydrodynamic simulation focused on the dynamics of a circumstellar disk over several stellar radii. We thus, instead, mimic the effect of pulsations by imposing sinusoidal perturbations in density and azimuthal velocity at the lower boundary,
\begin{align}
\rho(\phi) &= \rho_0 10^{\left(\log\left(\frac{\rho_{max}}{\rho_0}\right)\sin\left(\frac{2 \pi t}{P}+m\phi\right)\right)} \label{eqn:d_pert}\\
v_\phi(\phi) &= v_{rot} + v_{\phi,pert}\sin\left(\frac{2\pi t}{P}+m\phi+\phi_0\right), \label {eqn:w_pert}
\end{align}
where the exponential variation in density reflects the exponential stratification of the stellar atmosphere, with mean density $\rho_0$ and maximum density of the perturbation $\rho_{max}$. Here $v_{\phi,pert}$ is the azimuthal velocity perturbation, $P$ is the perturbation period, and $\lvert m \rvert$ is the number of nodes around the stellar equator, with $m>0$ ($m<0$) giving prograde (retrograde) phase velocity. For $\phi_0 = 0^\circ$, the perturbations in density and velocity are in phase, representing a prograde group velocity; for $\phi_0=180^\circ$ the velocity and density perturbations are in antiphase, signifying retrograde group velocity. Table \ref{tab:vp_and_vg} summarizes the 4 intercombinations explored in the models detailed in section \ref{sec:results} and what each implies for phase and group velocity.

\begin{table}
\begin{center}
\def\arraystretch{1.75}
\begin{tabular}{r|c|c}
~&$m<0$&$m>0$\\
\hline
$\phi_0=0^\circ$&$v_p<0$,$v_g>0$&$v_p>0$,$v_g>0$\\
\hline
$\phi_0=180^\circ$&$v_p<0$,$v_g<0$&$v_p>0$,$v_g<0$
\end{tabular}
\end{center}
\caption{Sense of phase and group velocity as a function of the sign of $m$ and the value of $\phi_0$.}
\label{tab:vp_and_vg}
\end{table}

Apart from the variations noted in table \ref{tab:vp_and_vg}, all models share common parameters (table \ref{tab:params}). Parameters are chosen to be roughly representative of those inferred for a typical pulsating Be star (eg. $\mu$ Cen). For simplicity, $M_*$ and $R_*$ are tuned to give an equatorial surface orbital speed $v_{orb}=500\;\mathrm{km}\,\mathrm{s}^{-1}.$ This fixed value for all models allows for the introduction of the stellar parameter $W\equiv v_{rot}/v_{orb}$, where $v_{rot}$ is the equatorial rotation speed. For the standard set of parameters, $W=0.95$ and $v_{rot}=v_{orb}-c_s$ where $c_s= 25\;\mathrm{km}\,\mathrm{s}^{-1}$ is the sound speed.

\begin{table}[!ht]
\begin{center}
\def\arraystretch{1.75}
\begin{tabular}{|l|c|}
\multicolumn{2}{c}{Stellar Parameters}\\
\hline
$M_*$&9.2$M_\odot$\\
\hline
$R_*$&7$R_\odot$\\
\hline
$v_{rot}$&475\;km\,s$^{-1}$\\
\hline
$c_s$&25\;km\,s$^{-1}$\\
\hline
\end{tabular}
\hspace{10pt}
\begin{tabular}{|l|c|}
\multicolumn{2}{c}{Perturbation Parameters}\\
\hline
$P$&40\;ks\\
\hline
$|m|$&4\\
\hline
\end{tabular}
\end{center}
\caption{Standard simulation parameters.}\label{tab:params}
\end{table}

\section{Results}\label{sec:results}

\subsection{Prograde vs. Retrograde Phase and Group Velocity}

\begin{figure}
\centering
\includegraphics[width=\textwidth]{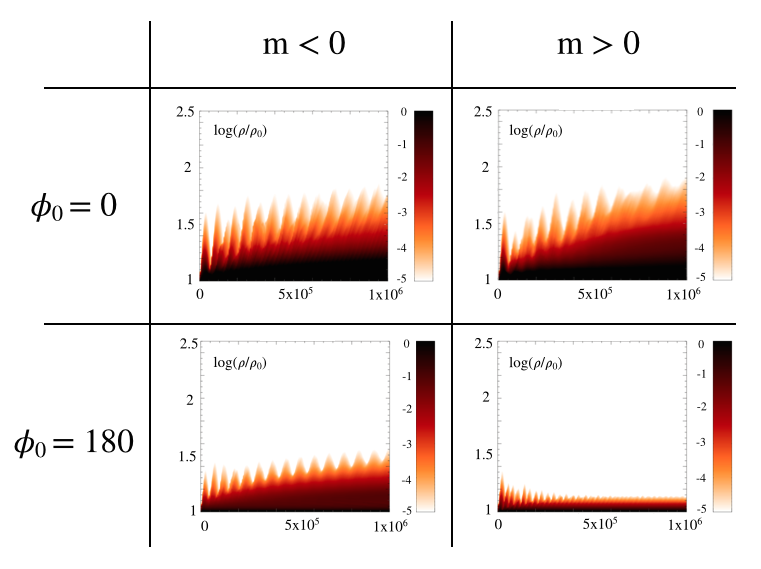}\caption{$\log(\rho)$ in $\mathrm{g}\,\mathrm{cm}^{-3}$ as a function of radius, measured in $R_*$ and time in seconds. While all combinations put some mass into orbit, prograde group velocity simulations do so much more efficiently than those with retrograde group velocity.\label{fig:logrho_comp}}
\end{figure}

\begin{figure}
\centering
\includegraphics[width=\textwidth]{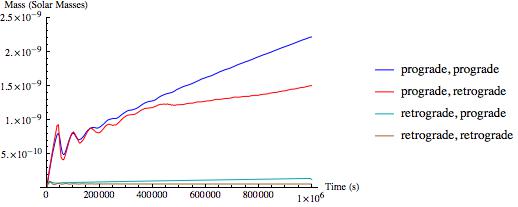}\caption{Mass in the disk in $M_\odot$ as a function of time. Sense of velocity is listed with group first, then phase.\label{fig:mid}}
\end{figure}

In considering results, let us first compare the behavior of the four possible combinations of phase and group velocity listed in table \ref{tab:vp_and_vg}. Figure \ref{fig:logrho_comp} compares results for the log of the azimuthally averaged density, $\log\langle\rho\rangle_\phi$, as a function of radius and time. While the outer boundary of the simulation is at $6R_*$, the figures focus on the most interesting behavior, below $2.5R_*$. Figure \ref{fig:mid} plots the total mass in the disk as a function of time, computed by summing density over the radial direction, while accounting for mass that escapes through the outer boundary. Note that, while all four models do put some material into orbit, prograde group velocity models do so much more effectively. Moreover, observations favor a retrograde phase velocity model \citep{RivBaa03}, thus our favored model is a mixed phase model with prograde group velocity and retrograde phase velocity, henceforth referred to as the +/- model.

\subsection{+/- model}

\begin{figure}
\centering
\includegraphics[width=\textwidth]{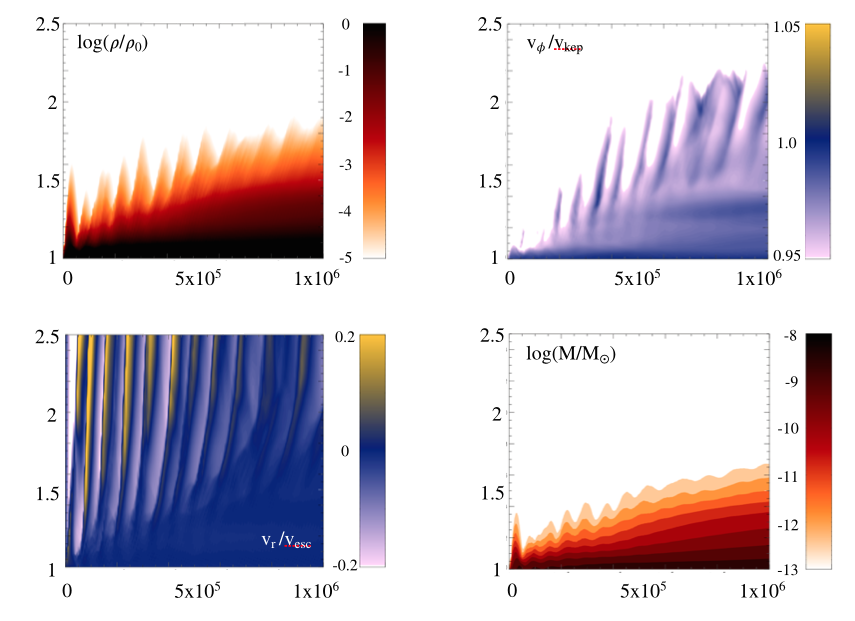}\caption{Clockwise from the upper left, $\log(\rho)$, Kepler number, mass above each radius in $M_\odot$, and radial velocity in units of escape velocity for the +/- model as a function of radius and time.\label{fig:pra}}
\end{figure}

\begin{figure}
\centering
\includegraphics[width=0.8\textwidth]{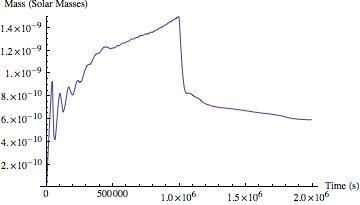}\caption{Mass in the disk in $M_\odot$ as a function of time. Pulsations are turned off at 1 Ms and the disk is allowed to relax.\label{fig:rel}}
\end{figure}

For considering a PDOME model in detail, there are four quantities of particular merit, namely: $\log\langle\rho\rangle_\phi$; mass above each radius; radial velocity; and Kepler number, the ratio of azimuthal velocity to local Keplerian orbital velocity. These are plotted in figure \ref{fig:pra} for the +/- model. The first of these, $\log\langle\rho\rangle_\phi$, has already been shown and discussed for all models in figure \ref{fig:logrho_comp}.  The mass above each radius shows that the mass launched into orbit is comparable to inferred total masses in Classical Be disks.
Radial pressure support allows the disk to be nearly in hydrostatic equilibrium in the radial direction for the inner disk even with Kepler number slightly below unity and, indeed, the lower left panel of figure \ref{fig:pra} shows that radial velocity is significantly below the local escape speed.
Figure \ref{fig:rel} shows that, if the pulsation is turned off, about half of the material begins to settle into a stable disk, while the remainder falls back on to the star. 

\subsection{Variations on the +/- model}

\begin{figure}
\centering
\includegraphics[width=\textwidth]{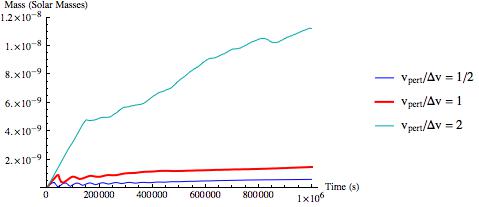}\caption{Mass in the disk in $M_\odot$ as a function of time for $v_\phi/\Delta v$ of 0.5, 1, and 2.\label{fig:vodv}}
\end{figure}

As preliminary investigations into some possible future veins of research, we consider two variations on the +/- model. The first of these is to vary the ratio of the perturbation velocity to $\Delta v \equiv v_{orb}-v_{rot}$. Figure \ref{fig:vodv} shows that decreasing this ratio proportionally decreases disk mass, while increasing it creates a much steeper increase, nearly a factor of ten, in disk mass. Understanding this strong sensitivity requires further study.

\begin{figure}
\centering
\includegraphics[width=\textwidth]{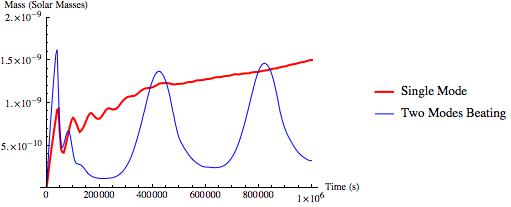}\caption{Mass in the disk in $M_\odot$ as a function of time for a single mode pulsation and two pulsation modes beating against one another.\label{fig:beat}}
\end{figure}

The second variation on the base model, motivated by the complex multiperiodic nature of pulsations on $\mu$ Cen \citep{RivBaa98}, is to consider the effects of two perturbations beating against one another. For this, two perturbations with $v_{pert}=\Delta v/2$ are imposed with 10$\%$ separated periods ($P=40\;\mathrm{ks}\pm2\;\mathrm{ks}$), leading to a $400\;\mathrm{ks}$ beat period. Figure \ref{fig:beat} shows that the disk mass oscillates with this beat period. During constructive interference, the mass peaks at a level comparable to that seen for a single mode mode; during destructive interference much of the material falls back on the star, much as occurred when the perturbations were turned off. Stochastic and impulsive events would be expected to produce a similar behavior, so long as the matter is ejected with sufficient angular momentum and energy to enter orbit.

\section{Conclusions and Future Work}\label{conclusions}

While preliminary, the results here provide a proof of concept for ejection of mass into the circumstellar environment by non-radial pulsations. They also demonstrate how such a disk, once generated, can persist in the absence of pulsations, viscous forces, and radiative forces. However, there is much room for future work. For instance, models should assume more moderate rotation rates, $v_{rot}=0.8-0.85v_{orb}$, which require $9-16$ times more energy to reach orbit. In addition, models should be extended to higher latitudes away from the equatorial plane using either 2D axisymmetry\footnote{Note that such an azimuthally symmetric mode would imply radial pulsation modes which are not observed in Classical Be stars but may still be considered meaningful as a computationally inexpensive first step.} or a full 3D model. Given the success of building and dissipating disks by viscous diffusion \citep{BjoCar05,CarBjo12}, viscous forces should be included in modeling the disk evolution. Given the high luminosity of the central star, one should also consider the effect of line driven forces in inhibiting the build-up of a disk, or ablating its surface layers. Nevertheless, the PDOME models here provide a promising step toward understanding the dynamics of Be disk formation, and a framework for such further investigations.

%\clearpage % To force this stuff to happen by this point in the text, otherwise these will probably end up after the references.

%\acknowledgements We acknowledge funding support received from NASA ATP Grant NNX11AC40G, and helpful feedback from T. Rivinius during initial phases of this investigation.

%\bibliography{kee_london}  % For BibTex

\section{Questions}

\emph{A. Okazaki: You mentioned that no viscosity is included in your simulations. But, your disks extend up to $~1.5\;R_{star}$, where the specific angular momentum is about 20$\%$ higher than at the stellar equatorial surface. What's the mechanism for angular momentum transport in your simulations?}

\vspace{10 pt}

\emph{N. D. Kee: It is correct that viscous forces are not explicitly included in these simulations. However, due to the use of a grid based code, the code is subject to numerical viscosity. This is likely sufficient to provide the radial extent that is seen in the results I have shown. However, confirming that this is the mechanism responsible for the spreading of the disk is still one of the issues to be addressed.}

\chapter{Radiative Ablation of Optically Thick Disks}\label{chap:ab_thick}

Having discussed both the formation and destruction of  the optically thin, decretion disks of Be stars, let us next examine line-driven ablation of the much denser, optically thick, accretion disks that develop during the early star-formation phases of luminous, massive stars.
%Now that we have discussed a possible mechanism for the generation of Classical Be star disks in chapter \ref{chap:pdome}, we can conclude the portion of this dissertation treating Classical Be star disks and move on to investigate the radiative ablation of disks around pre-main sequence objects. 
%As discussed in chapter \ref{chap:disks}, such disks present significant additional challenges, both for observation and modeling. 
For this study, it is this substantial continuum optical depth which is most problematic. In general, fully treating radiative transfer through a multi-dimensional, optically thick medium is computationally prohibitive; thus in section \ref{sec:an_opt_dep}, we develop an approximate method for optically thick, but geometrically thin, isothermal disks in the limit of gray continuum absorption.

Section \ref{sec:ab_thick} then applies this approximate method to radiative ablation of a very optically thick disk (with radial optical depth $\tau=400$ in the equatorial plane; see section \ref{sec:disk_struc}) around an O7 star. 
%\hl{maybe add a sentence on why} 
To understand the effects of continuum absorption, we compare this against a model that ignores continuum optical depth. These two models have the added benefit of bracketing the expected behavior of continuum electron scattering.

\section{Analytic Optical Depth for a Geometrically Thin Disk}\label{sec:an_opt_dep}

To generalize the results of chapter \ref{chap:ab_thin} to disks of arbitrary optical thickness, we need a method for treating continuum optical depth along an arbitrary ray. Here, however, a problem arises. At the temperatures and densities found around high mass stars, the dominant continuum opacity source is electron scattering. Scattering radiative transfer is tricky to calculate as it requires knowledge of a non-local scattering source function at all points along the ray under consideration \citep[see, e.g.][]{MihMih84}. Monte Carlo methods can bypass this problem but are computationally prohibitive to incorporate into a hydrodynamics code\footnote{Recent work by \cite{Har15} suggests that this may be able to be overcome with massively parallel codes, although this method is not yet widespread.}. Thus, for this work, we choose to initially assume a pure absorption model (i.e. no source function) which exaggerates the reduction of continuum flux, and thus leads to a lower limit on net ablation rate. Comparison with no continuum opacity thus brackets the net ablation rate expected from a scattering opacity.

%However, even a pure absorption treatment requires non-local integrations along several rays which, in a grid-based hydrodynamics code such as VH-1, requires extensive interpolation in density and opacity. Such
Unfortunately, even an absorption model is not straightforward in a grid-based hydrodynamics code such as VH-1. The ideal method is to use long characteristics and integrate the optical depth along each of many ray characteristics. For spherical geometry in grid based codes, however, an arbitrary set of rays is not guaranteed to cross any grid vertices where the density is known. Additionally, even for a spherical grid that is designed such that a long-characteristic approach can be used \citep[see][for the description of such a grid]{Owo99,DesOwo05}, the grid would be constrained in such a way to make resolving a disk problematic. Therefore, an alternative method is needed.

\begin{figure}
\centering
\includegraphics[width=\textwidth]{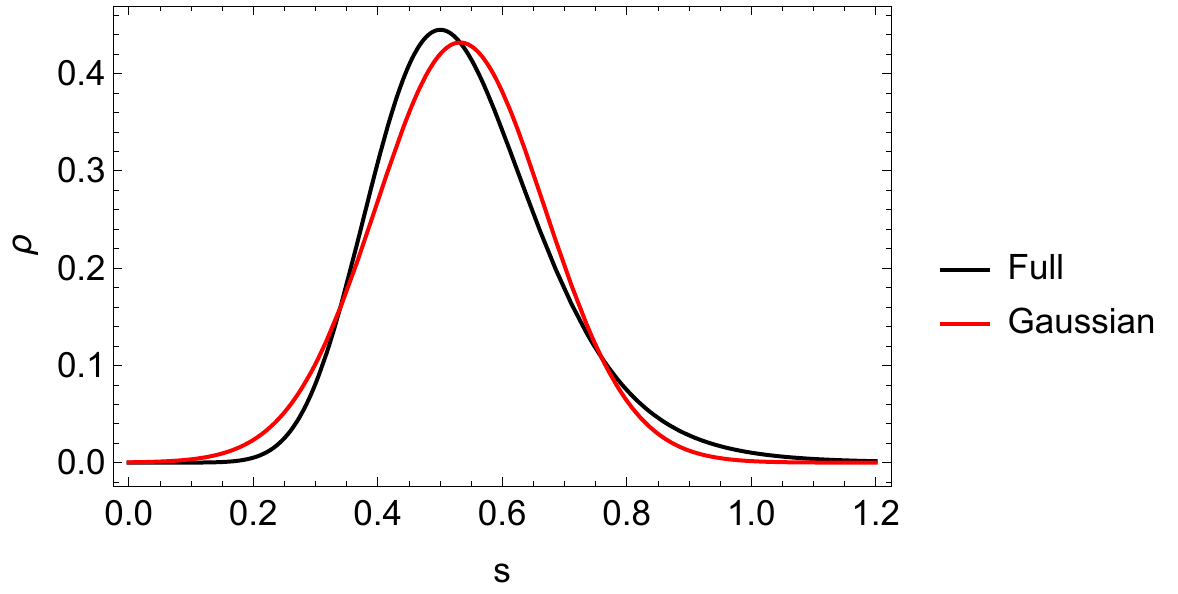}
\caption{\label{fig:rho_comp}
Density profile along an arbitrary ray through the circumstellar disk both along the full, correct path $s$ and using the Gaussian approximate vertical direction.
}
\end{figure}

\begin{figure}
\centering
\begin{subfigure}[b]{0.48\textwidth}
\includegraphics[width=\textwidth]{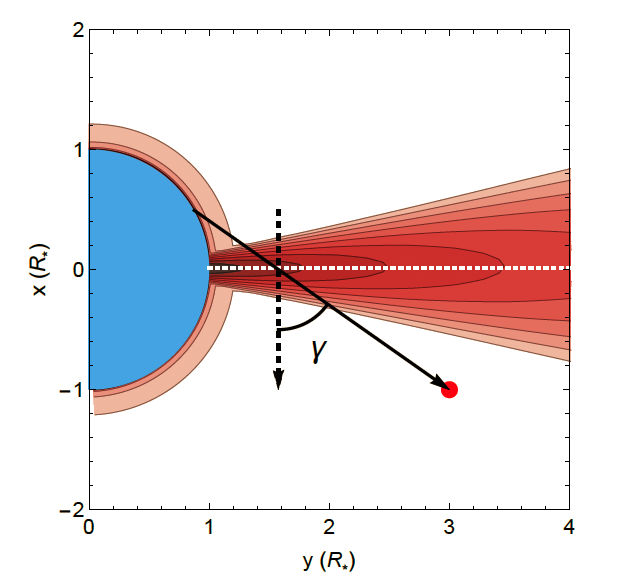}
\label{fig:thin_disk_rho}
\end{subfigure}
\begin{subfigure}[b]{0.48\textwidth}
\includegraphics[width=\textwidth]{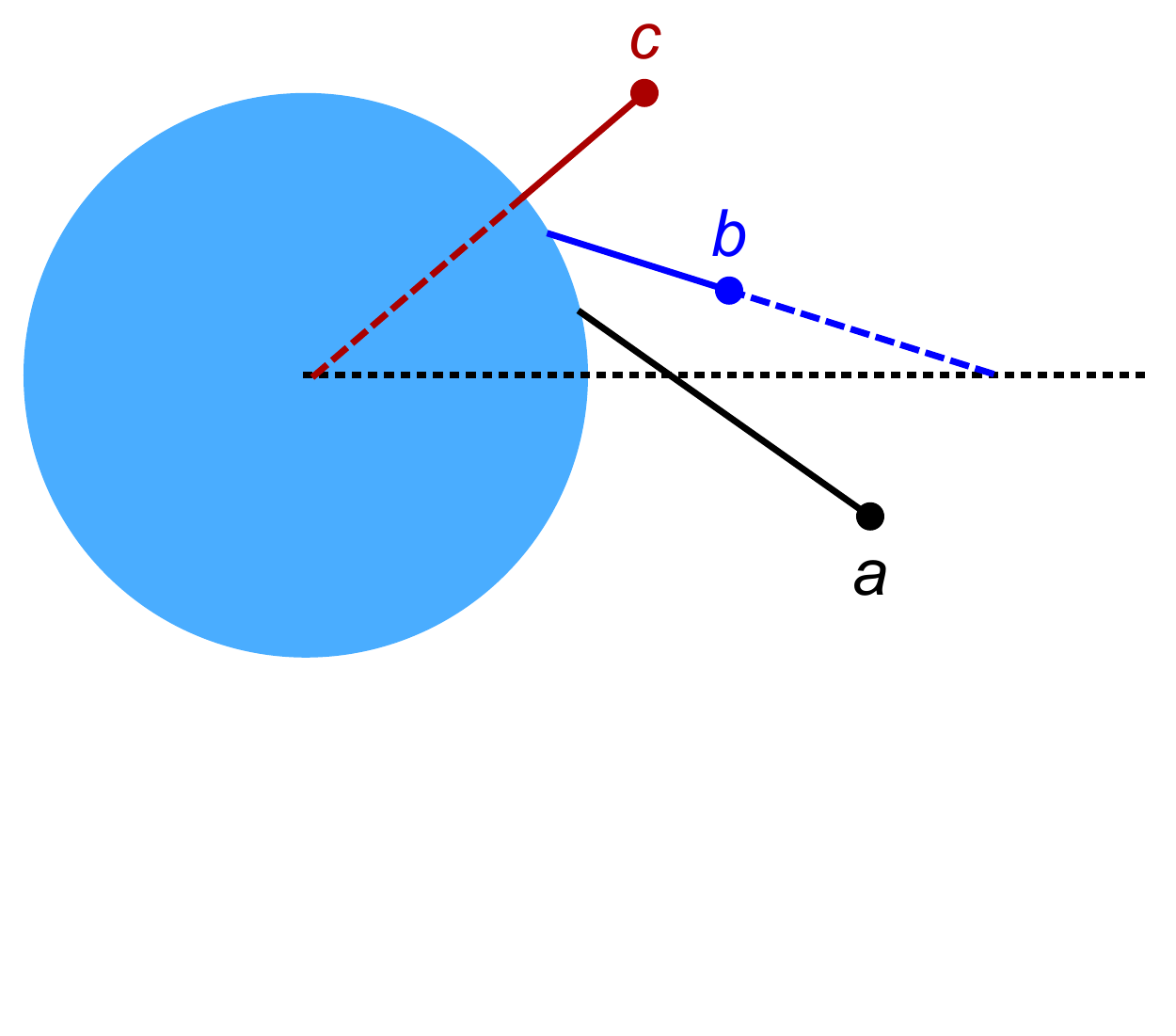}
\label{fig:thin_disk_rays}
\end{subfigure}
\caption{
Optical depth along the solid black arrow is calculated by optical depth along the dashed black arrow in the left panel, corrected by $1/\cos(\gamma)$. Density at the intersection of the black line and the dotted white line is used for $\rho_{eq}$ in the analytic formula. In the right panel, the three possible scenarios discussed in the text for the optical depth calculation are shown.
}
\label{fig:thin_disk_concept}
\end{figure}

Here, we can take advantage of the geometry of the problem. Recalling the general form of optical depth along a path from a point (here on the stellar surface) at $s_\ast$ to a circumstellar point at $s_c$,

\begin{align}\label{eq:long_char}
\tau&=\int_{s_\ast}^{s_c} \kappa(s) \rho(s) ds\\
&=\kappa_e \int_{s_\ast}^{s_c} \rho_{eq}(R(s)) e^{-1/2(z(s)/H(R(s)))^2} ds\,.
\end{align}
For cylindrical coordinates $R(s)$ and $z(s)$ at the local ray position $s$, the latter equality assumes that optical depth arises from continuum electron opacity $\kappa_e$ in a vertically hydrostatic, Keplerian disk (see chapter \ref{chap:disks}) with arbitrary radial dependence of equatorial density $\rho_{eq}$. Under the ``thin-disk'' assumption that the radial variations of density and scale height are much slower than the projected Gaussian vertical variation of density -- i.e. that for a ray at angle $\gamma$  from the disk vertical, $\left|\frac{d\log\rho}{dR}\right|$, $\left|\frac{d\log H}{dR}\right| \ll \frac{\cos\gamma}{H}$ -- the dominant contribution to optical depth will come from the region near the equator at radius $R_{eq}$, where the ray from $s_\ast$ to $s_c$ crosses the equatorial plane. Figure \ref{fig:rho_comp} compares the actual density variation along the full path $s$ to the Gaussian density variation for this ``thin-disk'' approximation. The limited shift in peak value and overall comparable variation of density in the two cases allows for the integral over $ds$ to be approximated by an integral over $dz/\cos(\gamma)$,

\begin{align}
\tau&\approx\frac{\kappa_e \rho_{eq}(R_{eq})}{\cos(\gamma)} \int_{z_\ast}^{z_c} e^{-1/2(z/H(R_{eq}))^2} dz\\
&=\sqrt{\frac{\pi}{2}}\frac{\kappa_e\rho_{eq}(R_{eq}) H(R_{eq})}{\cos(\gamma)} \left| \mathrm{Erf}\left(\frac{z_c}{\sqrt{2}H(R_{eq})}\right) - \mathrm{Erf}\left(\frac{z_\ast}{\sqrt{2}H(R_{eq})}\right) \right| \label{eq:opt_depth}\,,
\end{align}
where the absolute value in the latter form ensures that $\tau$ is always positive. The left panel of figure \ref{fig:thin_disk_concept} graphically illustrates a typical equatorial crossing ray.

In general, not all rays will cross the equator, as illustrated by rays b and c in the right panel of figure \ref{fig:thin_disk_concept}. While case b may seem problematic as $R_{eq}$ does not fall between $s_\ast$ and $s_c$, in practice the difference between error functions for points on the same side of the disk is quite small. Thus, even without handling this case separately, very little attenuation is computed for such a ray, so we can simply continue to use equation \ref{eq:opt_depth}.

For case c, however, $R_{eq}<R_\ast$ and thus we do not have a disk density defined at $R_{eq}$. To handle this case, we simply set $\tau=0$, as we do not expect much optical depth to arise for a ray which only passes through wind and highly stratified disk material.

\begin{figure}
\centering
\begin{subfigure}[b]{0.48\textwidth}
\includegraphics[width=\textwidth]{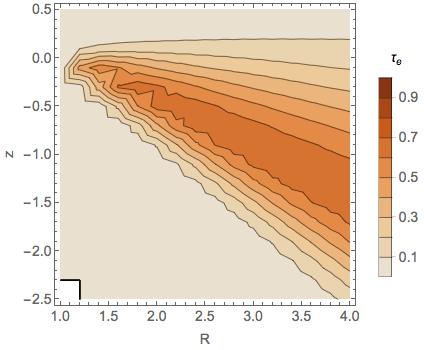}
\caption{
Long Characteristics
}
\label{fig:num_tau_xy}
\end{subfigure}
\begin{subfigure}[b]{0.48\textwidth}
\includegraphics[width=\textwidth]{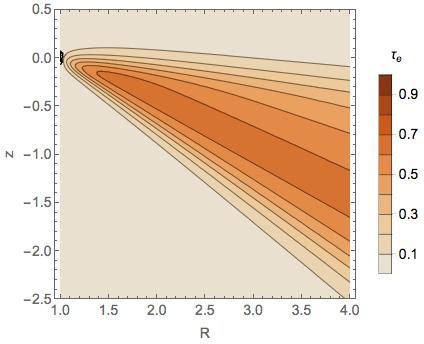}
\caption{
Thin Disk Method
}
\label{fig:an_tau_xy}
\end{subfigure}
\caption{
Normalized optical depth as a function of position around the star for a ray of impact parameter, $p=1/\sqrt{2}$ and radiation azimuthal position $\phi'=45^\circ$. 
%\hl{Make x,y -> R,z}
}
\label{fig:tau_xy}
\end{figure}

\begin{figure}
\centering
\begin{subfigure}[b]{0.48\textwidth}
\includegraphics[width=\textwidth]{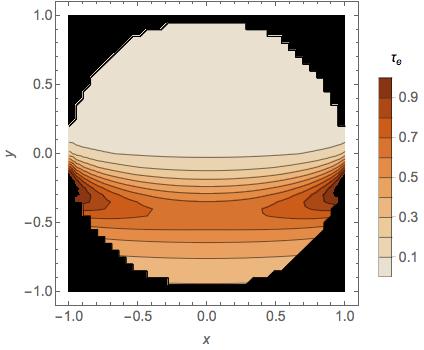}
\caption{
Long-Characteristics
}
\label{fig:num_tau_ypp}
\end{subfigure}
\begin{subfigure}[b]{0.48\textwidth}
\includegraphics[width=\textwidth]{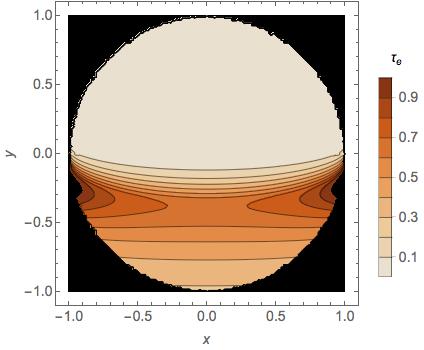}
\caption{
Thin-Disk Method
}
\label{fig:an_tau_ypp}
\end{subfigure}
\caption{
Normalized optical depth from a single point above the star at $r=2R_\ast$ and $10^\circ$ above the equator.
}
\label{fig:tau_ypp}
\end{figure}

For the initial analytic condition of a stratified disk, we can calculate optical depths using both the full long-characteristic integral (equation \ref{eq:long_char}), and this thin-disk approximation (equation \ref{eq:opt_depth}), allowing for a discussion to test the accuracy of this method. Normalizing by the radial optical depth in the equatorial plane, figure \ref{fig:tau_xy} compares optical depths from the two methods as a function of circumstellar position $R$ and $z$, for a fixed impact parameter $p=1/2$ and projected azimuthal position $\phi'=45^\circ$ on the star; figure \ref{fig:tau_ypp} compares the same normalized optical depth as seen from a fixed point at $2R_\ast$ and $10^\circ$ above the equator looking back at various $p$ and $\phi'$ on the stellar surface. Generally, the agreement between the two methods is quite striking, with the modest discrepancies occurring in the regions where the ray does not directly impact the disk so that the optical depth is already small. This thin-disk method thus provides an accurate and much more computationally efficient method for computing optical depths in optically thick, geometrically thin disks.

\section{Initial Tests of Ablation on a Disk of Star Forming Density}\label{sec:ab_thick}
Let us now use this method to study ablation of much denser, optically thick disks, such as those found around stars still in the process of forming. As discussed in chapter \ref{chap:line}, at the temperatures found near luminous, massive stars, continuum opacity is dominated by electron scattering. The thin-disk model calculates an absorption optical depth, which systematically underestimates the local flux expected from scattering and, therefore, the line-accelerations. On the other hand, omitting continuum optical depth effects completely systematically overestimates the flux and acceleration everywhere. Since scattering optical depths are not easy to compute, we use these two methods of continuum absorption and no continuum optical depth to bracket the expected behavior of scattering. 

As the O7 model (see section \ref{sec:o7}) has the strongest radiation field considered here, we use it for this study, now with an equatorial plane optical depth of $\tau=400$ (see chapter \ref{chap:disks}). Here we continue to use the density power-law index 3.5 that is expected for decretion. For an accretion disk we would generally expect a power-law index closer to 1.5. By leaving the power-law index unchanged, however, we can investigate the effects of disk optical depth separately from the effects of changing this power-law, which we leave for a future investigation.

\begin{figure}
\centering
\includegraphics[width=0.7\textwidth]{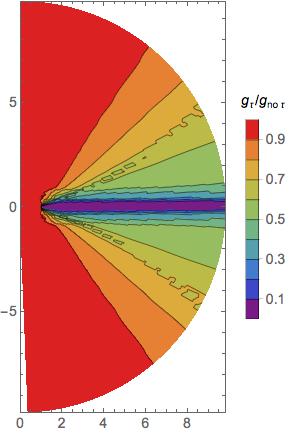}
\caption{
Reduction of the radial line-acceleration component due to continuum optical depth with respect to the same acceleration ignoring continuum absorption. Here, the disk optical depth in the equatorial plane is $\tau=400$.
}
\label{fig:gx_rat}
\end{figure}

Figure \ref{fig:gx_rat} shows the ratio of the radial acceleration component computed with continuum absorption to the same component computed with continuum optical depth omitted. Note that the largest reduction occurs near the equatorial plane, where large portions of the star are occulted by the disk, with the differences tapering off toward the poles where the full star is visible. For points not inside the disk, the peak reduction is about 50\% as the disk occults at most half of the star as seen from such positions.

\begin{figure}
\centering
\includegraphics[width=\textwidth]{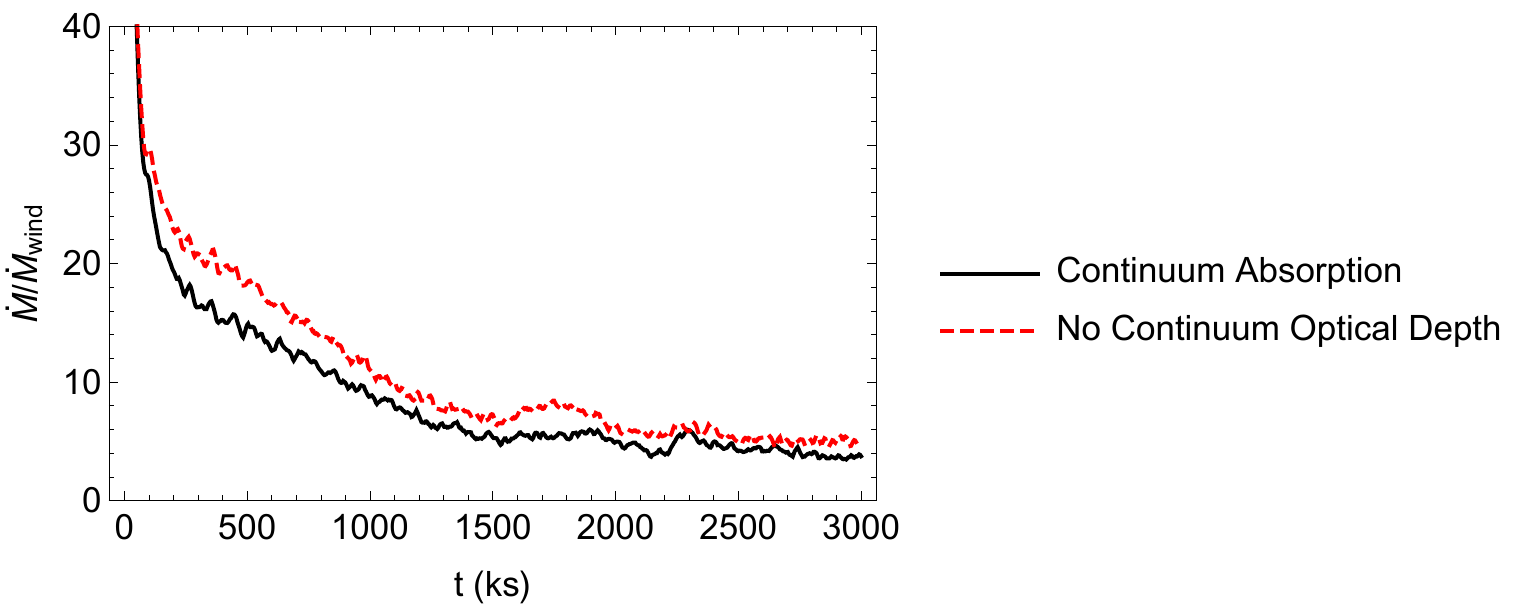}
\caption{
Ablation rate at time $t$ for models with and continuum absorption and no continuum optical depth.
}
\label{fig:mdot_tau}
\end{figure}

Figure \ref{fig:mdot_tau} confirms the expectation of a smaller disk ablation rate for the case with continuum absorption with respect to the case with no continuum attenuation. The difference between the two is modest, however, with only about a 50\% reduction in ablation rate. Thus, these two methods provide an even tighter than expected constraint on the ablation from an intermediate case with electron scattering rather than absorption.

Here we find an extremely high ablation rate in the early phases of the simulation, tens of times the spherically symmetric wind mass loss. This is further evidence of the general incompatibility of a disk with a wind and an intense radiation field. Where previously the radiation of an O7 star dynamically destroyed a disk, however, now there is sufficient disk material to require a protracted adjustment period before the radiation and disk settle into an asymptotic configuration. Note that the O7 star ablates its optically thick disk at perhaps 4-5 times its wind associated mass loss rate, as opposed to the factor of 2 found for the B1 and B2 stars. Nevertheless, this modest increase in disk ablation rate allows the continued use of the simple $M_{disk}/\dot{M}_{wind}$ scaling introduced in chapter \ref{chap:ab_spec_type} to approximate the disk destruction time around stars with optically thick disks. 

Recall that this comes with the caveat that here we have used the radial power-law exponent of 3.5. However, the analysis in chapter \ref{chap:ab_thin} suggests that material is mainly ablated from the disk near the stellar surface, and here we recover an ablation rate that is within about a factor of two of the steady state disk ablation found in chapter \ref{chap:ab_spec_type}, even though here the disk is 1000 times more massive. Thus, we do not anticipate that changing the disk density profile away from $R_\ast$ will significantly impact the ablation rate, but we leave it to future work to confirm this expectation and more stringently test the applicability of the $M_{disk}/\dot{M}_{wind}$ scaling for star-forming accretion disks.

\chapter{Suppression of X-rays from Radiative Shocks by their Thin-shell Instability}\label{chap:thin_shell}

This is a pre-copyedited, author-produced PDF of an article accepted for publication in Monthly Notices of the Royal Astronomical Society following peer review. The version of record \cite{KeeOwo14a} is available online at: \url{http://mnras.oxfordjournals.org/cgi/content/full/stt2475?ijkey=aylHC8lTpOPICXD&keytype=ref}.

\section{Introduction}
\label{sec:intro}

Shocks that arise from collision between highly supersonic flows are a common source of X-ray emission from astrophysical plasmas. 
A prominent example is the case of colliding wind binaries (CWB's), wherein the collision is between strong, highly supersonic  stellar winds from the individual components of a massive-star binary system.
In relatively wide, long-period binaries, wind material that is shock-heated by the collision cools gradually by adiabatic expansion, leading to a spatially extended region of hot post-shock flow, with X-ray emission that is readily computed from the local density-squared emission measure.
This gives an overall X-ray luminosity that scales with the square of the mass loss rate ($L_X \sim \dot{M}^2$) of the source stellar wind, and with the inverse of the distance $d$ from the star to the interaction front.
Numerical hydrodynamics simulations of such wide CWB systems have thus been quite successful in modeling both the level and, in the case of eccentric systems, the orbital phase variation, of observed X-rays from long-period CWB's such as $\eta$~Carinae and WR~140 
\citep{ParPit09, ParPit11, Rus13}.

In closer, short-period binaries, the higher density at the interaction front means that cooling in the post-shock region can become dominated by  radiative emission, with an associated radiative cooling length $\ell_c \ll d$ that can be much smaller than the star to interaction-front distance $d$ that characterizes the scale for adiabatic expansion cooling \citep{SteBlo92, Pit09}.
Since the X-ray emission over the cooling layer is limited by the incoming kinetic energy flux, the X-ray luminosity from such radiative shocks is expected now to scale {\em linearly} with the mass loss rate, $L_X \sim \dot{M}$
\citep{OwoSun13}.

However, analytic stability analyses \citep{Vis94} show that the narrowness of such radiatively cooled shock layers makes them subject to a non-linear {\em thin-shell instability}, wherein lateral perturbation of the interaction front causes material to be diverted from convex to concave regions, converting the direct compression from the oppositely directed flows into multiple elongated regions of strong shear.
Beginning with the pioneering work by \citet{SteBlo92}, all numerical hydrodynamics simulations of flow collisions \citep[e.g.,][]{WalFol98, Pit09, ParPit10} 
in this limit of radiatively cooled shocks indeed show the interaction front to be dominated by highly complex regions of dense, cooled gas, with little high-temperature material to emit X-rays.

While it is clear that such thin-shell instability structure is likely to reduce the X-ray emission from what is expected from a simple laminar compression analysis, so far there have been only limited attempts to quantify the level of this reduction, and how it might scale with physical, and even numerical, parameters.
\citet{ParPit10} have emphasized the potential role of ``numerical conduction'', and other numerical effects associated with limited grid resolution, in lowering  the temperature of shock-heated regions, and so reducing the X-ray emission.
\citet{LamFro11} also assess the spatial grid needed to resolve the instability, and identify a related ``transverse acceleration instability'' that can further contribute to flow structure.
Numerical resolution is certainly a challenge for simulating the small-scale structure that arises with a small cooling length $\ell_c$, especially when carried out over the much larger separation scale $d$; but even in simple planar slab collision models with grids set to well resolve this cooling length (see \S \ref{sec:advect_rhoT}), there is a substantial reduction in X-ray emission.
While numerical diffusion and other artifacts may play a part in this, it seems that much or even most of this reduction stems from a robust {\em physical} effect, namely the conversion of direct compressive shocks to highly oblique shocks along the elongated ``fingers'' of shear from the oppositely directed flow.

The simulations and analysis here aim to quantify such effects of thin-shell instability-generated structure in reducing X-ray emission from radiative shocks.
To focus on thin-shell structure, we ignore the added complexity of fully 3D models explored by other authors \citep[e.g.,][]{vanKep11, ParGos11} within the specific context of CWB's, using instead the simplest possible 2D simulations of equal colliding flows that still allow full development of thin-shell structure.
Beyond CWB's, this has relevance for interpreting X-rays from shocks in other dense stellar outflows, for example the ``embedded wind shocks'' that originate from instabilities in the line-driving of hot-star winds. To explain the observed linear $L_X \sim L_{bol}$ relation between X-ray and stellar bolometric luminosity of single O-type stars,  \citet{OwoSun13} proposed that thin-shell mixing in these radiative shocks reduces their X-ray emission in proportion to some power -- dubbed the mixing exponent -- of their mass loss rate.
The complexity of treating the nonlocal radiation transport makes it difficult to develop multi-D simulations of the structure arising from this line-driving instability, and so a general goal here is to use a study of direct shocks from opposing supersonic flows as a first test of this proposed scaling for thin-shell-mixing effects.

\begin{figure}
\includegraphics[width=\textwidth]{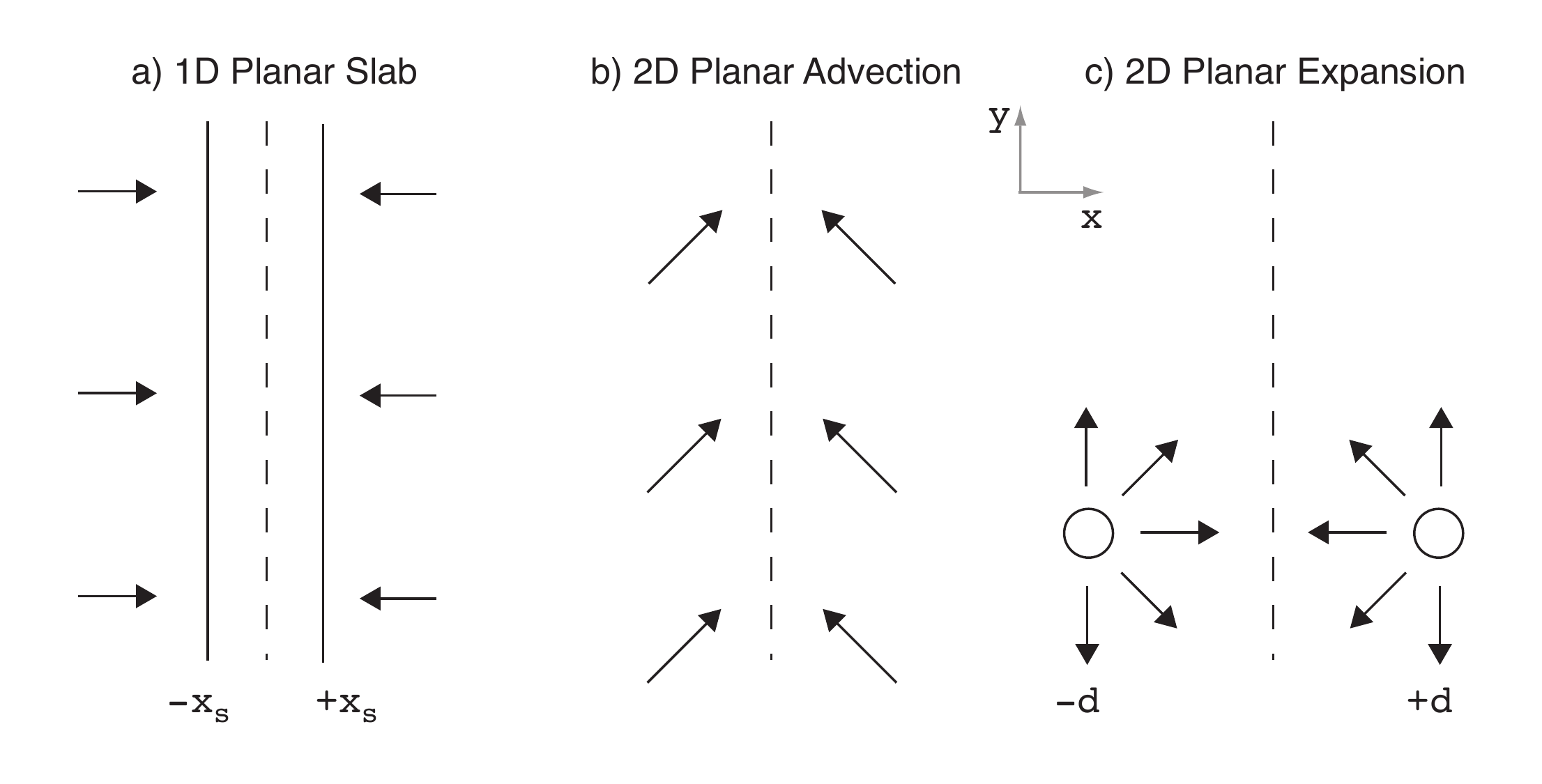}
\caption{
Schematic illustration of the 3 types of flow collision modeled here, as denoted by the figure labels. The vectors represent flow velocity, and the dashed lines represent the mean contact interface for momentum balance.  The solid lines in the 1D slab model {\bf a} represent the shock pair located at offsets $\pm x_s$ from the interface.
The tilted vectors in the 2D planar advection model {\bf b} indicate vertical advection at a speed equal to that of the horizontal compression.
The circles in the 2D planar expansion model {\bf c} indicate stellar wind mass sources at locations $\pm d$ from the interface, with constant outward expansion speed.
}
\label{fig:3configs}
\end{figure}

To provide a firm physical basis, we do this through 3 tiers of simulation, based on the 3 configurations of flow collision illustrated in figure \ref{fig:3configs}.
For the simple case of a 1D planar slab with collision between equal and opposite flows (figure \ref{fig:3configs}a), we first derive analytic scalings for the temperature variation and X-ray emission (\S \ref{sec:cool_anal}) within a cooling length $\ell_c$ on each side of the interaction front, under the idealization of steady-state, standing shocks.
We next (\S \ref{sec:cool_osc_rhoT}) use time-dependent numerical simulations to illustrate the cooling oscillation \citep{CheIma82} of such 1D slab collision, along with associated variation in X-ray emission.
\S \ref{sec:advect_rhoT} extends this planar collision model to include vertical advection in 2D (figure \ref{fig:3configs}b), showing how the initial cooling oscillation breaks up into extensive shear structure along the vertical advection, with an associated factor $\sim1/50$ reduction in the X-ray emission.
For 2D CWB-like models of collision between mass sources with planar expansion (figure \ref{fig:3configs}c), \S \ref{sec:source} carries out a systematic parameter study of how the structure formation, and X-ray reduction, depend on the mass source rate and an associated 2D cooling parameter $\chi_{2D} \sim 1/\dot{M}$.
A key result is that $L_X$ follows the expected analytic scalings for the adiabatic regime $\chi_{2D} \gtrsim 1$, but is reduced by a nearly fixed factor $\sim1/50$ from the linear $L_X \sim \dot{M}$ scaling expected for the strongly radiative limit $\chi_{2D} \ll 1$.
The final section (\S \ref{sec:summary}) discusses open issues from these simulations and outlines directions for future work.

\section{Cooling Analysis}
\label{sec:cool_anal}

\subsection{General equations of hydrodynamics}
\label{sec:genhydro}

All  flow models in this chapter assume no gravity or other external forces, so that the total advective acceleration in velocity ${\bf v }$ stems only from gradients in the gas pressure $p$,
\beq
 \frac{D\mathbf{v}}{Dt} \equiv  \frac{\partial \mathbf{v}}{\partial t} +  \mathbf{v} \cdot \nabla  \mathbf{v}  =-\frac{\nabla p }{\rho}
 \, .
\label{eqn:pcons}
\eeq
Here the mass density $\rho$ satisfies the conservation condition,
\beq
 \frac{\partial \rho}{\partial t} +  \nabla  \cdot (\rho \mathbf{v} ) = 0 \, .
 \label{eqn:mcons}
 \eeq
The internal energy density, given by $e = (3/2) p$ for a monatomic ideal gas, follows a similar conservation form,
but now with non-zero terms on the right-hand-side to account for the sources and sinks of energy,
\beq
 \frac{\partial e}{\partial t} +  \nabla  \cdot (e \mathbf{v} ) = 
- p  \nabla\cdot \mathbf{v} - C_{rad}
  \, .
  \label{eqn:econs}
 \eeq
 Here the pressure term represents the effect of compressive heating ($\nabla \cdot {\bf v} < 0$) or expansive cooling ($\nabla \cdot {\bf v} >  0$), and the $C_{rad}$ term accounts radiative cooling\footnote{For the shock-heated flows considered here, we do not explicitly include any external heating term; but to mimic the effect of stellar photoionization heating in keeping circumstellar material from falling below a typical hot-star effective temperature \citep{Dre89}, we do impose a `floor' temperature $T=30,000$~K.}.
This volume cooling rate has the scaling,
\beq
C_{rad} = n_e n_p \Lambda (T)  =   
\rho^2 \Lambda_m (T)
\, ,
\label{eq:qrad}
\eeq
where $\Lambda (T)$ is the optically thin radiative loss function
\citep{CooChe89,SchKos09},
 and the latter equality defines a mass-weighted form $\Lambda_m \equiv \Lambda/\mu_e \mu_p$.  
For a fully ionized plasma the proton and electron number densities $n_p$ and $n_e$ are related to the mass density $\rho$ through the associated hydrogen mass fraction $X= m_p/\mu_p = m_p n_p/\rho  $ and mean mass per electron  $\mu_e = \rho/n_e = 2 m_p/ (1+ X)$. We assume here the standard solar hydrogen abundance $X=0.72$.
For all numerical simulations below, we use the radiative loss tabulation from \citet{CooChe89}, implemented within the exact integration scheme from \citet{Tow09}. 

Using the ideal gas law $p = \rho kT/\bar{\mu}$ (with $k$ the Boltzmann constant and $\bar{\mu} = 0.62 m_p$ the mean atomic weight), we can combine eqns.\ (\ref{eqn:mcons}) and (\ref{eqn:econs}) to derive a general equation for the total advective variation of the temperature,
\beq
\frac{1}{T}  \frac{DT}{Dt}  = - \frac{2}{3} \, \nabla \cdot \mathbf{v} - \frac{2}{3} \frac{\bar{\mu} \rho \Lambda_m(T)}{kT}
\, .
\label{eqn:DTDt}
\eeq

\subsection{Planar steady shock with cooling length $\ell_c$}
\label{sec:lcool}

To provide a basis for interpreting the time-dependent numerical models below, let us first consider the  idealized case
in which two highly supersonic, {\em planar} flows with the same fixed density $\rho_o$ and equal but opposite speeds $v_o$ along the $x$-direction collide at a fixed interface position $x=0$, resulting in a pair of standing, {\em steady-state} shocks at  fixed positions $x=\pm x_s$ on each side of the interface (see figure \ref{fig:3configs}a).
Since  $\rho v$ is constant,  
eqn.\ (\ref{eqn:DTDt}) for temperature variation in the post-shock layers within $|x| < x_s$  takes the form
\begin{equation}
-\frac{v}{T} \frac{dT}{dx}=\frac{2}{3}\frac{dv}{dx}+\frac{2}{3}\frac{\bar{\mu}\rho\Lambda_m(T)}{kT} \, .
\label{eqn:1D_steady_dT}
\end{equation}
Because the post-shock flow is subsonic, the final deceleration to zero speed at the interface can be achieved with only a mild gradient in pressure; thus, with only a minor fractional correction in the energy balance ($\sim 1/16$; see \citet{AntOwo04} and \S \ref{sec:fx1dslab} below), the cooling layer can be approximated as {\em isobaric}.
Together with constancy of the mass flux $\rho v$, this means $p \sim \rho T  \sim T/v $ are all constant, allowing us to eliminate density and velocity variations in terms of fixed post-shock values,
\begin{equation}
-T^2\frac{dT}{dx}=\frac{2}{5}\frac{\bar{\mu}(\rho_s T_s)^2\Lambda_m}{k(\rho_s v_s)}
\, ,
\label{eqn:dTdx}
\end{equation}
where these post-shock (subscript ``$s$") values are set by the
strong-shock jump conditions\footnote{These jump conditions are most generally cast in terms of the incoming speed {\em relative to the shock}, but in the present idealization of a steady {\em standing} shock this is just set by the speed $v_o$ measured relative to the fixed interface.},
\begin{align}
\label{eqn:RHjump}
\begin{split}
v_s&=\frac{v_o}{4} \\ 
\rho_s &= 4\rho_o  \\
T_s&=\frac{3}{16} \frac{\bar{\mu}}{k} \, v_o^2
= 14 \, {\rm MK} \, 
v_8^2 
 \,.
\end{split}
\end{align}
The latter evaluation for post-shock temperature again assumes a fully ionized gas with solar metallicity, 
with 
$v_8 \equiv v_o/10^8$cm\,s$^{-1}$.

For such typical post-shock temperatures $T_s \gtrsim 10$\,MK, $\Lambda_m$ is a quite weak function of temperature.
If we thus make the further simplification that this $\Lambda_m$ is strictly constant, 
direct integration of eqn.\ (\ref{eqn:dTdx}) yields an analytic solution for the decline in temperature from the post-shock value $T(x_s)=T_s$ to a negligibly small value at the interface ($x=0$),
\begin{equation}
\left ( \frac{T(x)}{T_s} \right )^3 = \frac{|x|}{\ell_c} ~~~ ; ~~~ |x| \le x_s=\ell_c 
\, ,
\label{eqn:Tofx}
\end{equation}
where the total cooling length $\ell_c = x_s$  from the shock to the contact interface has the scaling,
\begin{equation}
\ell_c \equiv \frac{5}{6}\frac{k v_s T_s}{\bar{\mu} \rho_s \Lambda_m}
= \frac{5}{512}\frac{v_o^3}{\rho_o \Lambda_m}
\approx 1.4 \times 10^{11} {\rm cm} \, \frac{v_8^3}{\rho_{-14}}
\, .
\label{eqn:cooling_length}
\end{equation}
At temperatures of order $10$\, MK, the radiative loss function has a value $\Lambda = \mu_e \mu_p \Lambda_m \approx 3 \times 10^{-23}$\,erg\,cm$^3$\,s$^{-1}$ \citep{CooChe89,SchKos09}, giving the numerical scalings in the last equality,  with $\rho_{-14} \equiv \rho_o/10^{-14}$\,g\,cm$^{-3}$.

We can also define a characteristic post-shock cooling time as the post-shock flow speed through this cooling length,
\beq
t_c  \equiv \frac{\ell_c}{v_s} =  1.2 \times 10^4  {\rm s} \, \frac{v_8^2}{\rho_{-14}}
\, .
\label{eqn:tauc}
\eeq
The values for $\ell_c$ and $t_c$ provide convenient reference scales for the cooling length and time in the more general numerical models below.

\begin{figure*}
\includegraphics[width=1\textwidth]{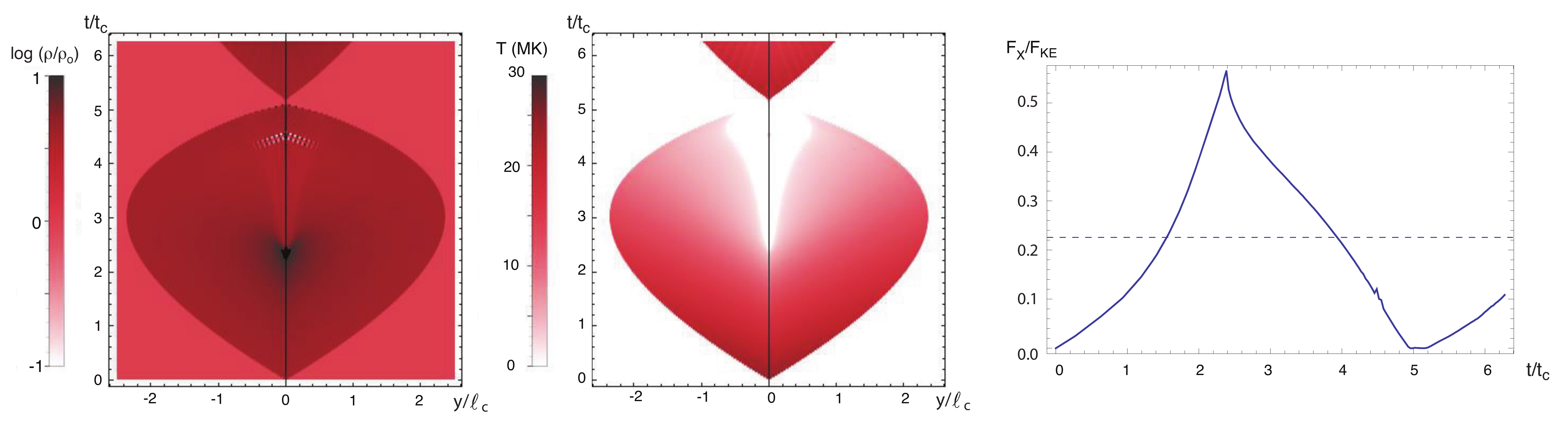}
\caption{The 1D cooling oscillation in  density $\log( \rho (x,t)/\rho_o)$ (left panel) and temperature $T(x,t)$ (in MK; middle panel), plotted vs.\ horizontal position $x$ (in units of the cooling length $\ell_c$) and time $t$ (in units of the cooling time $t_c$). The right panel shows the resulting time variation of the X-ray flux $F_X$, in units of the total flow energy flux $F_{KE} = \rho_o v_o^3$, with the horizontal dashed line showing the time-averaged value over a cooling oscillation cycle.
}
\label{fig:1D_cool_osc}
\end{figure*}

\subsection{X-ray flux from planar, steady shock}
\label{sec:fx1dslab}

This simple model of steady shock in a 1D slab also provides a useful illustration for the scaling of the X-ray emission. The volume emissivity for radiation of energy $E$ depends on the local density and temperature,
\beq
\eta(E,x) = \frac{\rho^2 (x) }{\mu_e \mu_p} \Lambda(E,T(x)) 
\, ,
\label{eqn:etaE}
\eeq
where the energy-dependent radiative loss function  $\Lambda(E,T)$ (erg\,cm$^3$\,s$^{-1}$\,keV$^{-1}$) is derived from APEC thermal equilibrium emission models \citep{SmiBri01}, with the total radiative loss function $\Lambda(T) = \int_0^\infty \Lambda(E,T) \, dE$.
The associated radiative flux comes from integration over both shock cooling layers,
\begin{align}
F(E) &=& \int_{-x_s}^{x_s} \frac{\rho^2 (x) }{\mu_e \mu_p}  \Lambda(E,T(x)) \, dx 
\label{eqn:FEint}
\\
&=& \frac{15}{16} \, 
\rho_o v_o^3
\int_0^{T_s} \frac{\Lambda(E,T)}{\Lambda(T)} \, \frac{dT}{T_s}
\, ,
\label{eqn:FE}
\end{align}
where the latter equality comes from using  eqn.\ (\ref{eqn:dTdx}) to change integration variable to temperature, and the factor $15/16$ reflects the above-mentioned 1/16 loss due to neglect of compressive work within the isobaric cooling model
\citep{AntOwo04}.
The total flux above some representative X-ray threshold, taken here to be $E_X = 0.3$\,keV, thus just scales with the kinetic energy flux $F_{KE} \equiv \rho_o v_o^3$ from both sides,
\beq
F_X = \int_{E_X}^\infty F(E) \, dE 
\approx 
F_{KE} \, f_X 
\, ,
\label{eqn:FXslab}
\eeq
where  
$f_X$ characterizes the fraction of total radiative loss that is emitted in the X-ray bandpass above $E_X$. Note that for  $E_X =0$, we have $f_X=1$, so that, apart from the 1/16 loss,  the bolometric radiative flux nearly equals the total incoming kinetic energy flux.

An important point to emphasize here is that, even though the local volume emissivity (\ref{eqn:etaE}) scales with {\em density-squared emission measure}, the integrated fluxes (\ref{eqn:FE}) and (\ref{eqn:FXslab}) over such a radiative cooling layer scale only {\em linearly} with the inflow density $\rho_o$.

\section{Numerical simulations for laminar flow collisions}
\label{sec:lamflow}

\subsection{The 1D cooling oscillation}
\label{sec:cool_osc_rhoT}

While the above analytic solution for a presumed steady-state shock provides a simple overall characterization of  the shock cooling layer, the linear stability analysis of \cite{CheIma82} shows that, even for a 1D steady laminar incoming flow, a post-shock layer undergoing such radiative cooling is generally {\em not} steady-state, but is {\em unstable to cooling oscillation} modes.

As a basis for 2D models below, let us first examine 1D numerical simulations of this cooling oscillation for this case of direct collision of two equal and opposite laminar flows.
Using the PPM \citep[Piecewise Parabolic Method;][]{ColWoo84} numerical hydrodynamics code 
VH-1\footnote{http://wonka.physics.ncsu.edu/pub/VH-1/},
we solve the time-dependent conservation equations (\ref{eqn:pcons}) --
(\ref{eqn:econs}) on a fixed 1-D planar grid of $n_x = 1000$ zones extending $\pm 2.5 \ell_c$ on each side of central contact discontinuity (again set at $x=0$) between the two flows; the uniform spatial zones of size $\Delta x= 0.005 \, \ell_c$ thus very well resolve the cooling region.
The left and right boundaries at $x \approx \pm 2.5 \ell_c$ assume highly supersonic inflow at speeds $v_o = 1000$\,km\,s$^{-1}$ of  gas with a typical hot-star temperature $T=30,000$\,K and thus sound speed $a \equiv \sqrt{kT/\bar{\mu}} \approx 20$\,km\,s$^{-1} = v_o/50$.
The initial condition at $t=0$ extends these inflow conditions to direct collision at the $x=0$ contact surface.
Since in practice stellar photoionization heating tends to keep gas from cooling much below the stellar effective temperature, we also use $T_o$ as a ``floor'' temperature for both pre- and post-shock gas \citep{Dre89}.

The color scale plots in figure \ref{fig:1D_cool_osc} show the time evolution (plotted along the vertical axis in units of  the cooling time $t_c$) of the density  and temperature within the cooling layers on each side of the contact discontinuity (vertical dark line), bounded on the outside edges by the oscillating shock discontinuity.  
Initially, the limited factor 4 compression means, from mass continuity, that the shock location propagates back from  the contact at a speed $v_o/3$.
However, as the gas near the contact cools, the lower pressure allows compression to higher density, which slows and then reverses this back propagation, but with a roughly factor two {\em overshoot} of the equilibrium cooling length.  As the gas near the contact continues to cool, the cooling  layer contraction leads to direct collapse of the shock onto the contact, whereupon the cycle repeats, leaving just a buildup of cold dense material at the contact discontinuity.  
The overall period of the oscillation is a few ($\sim$5) cooling times, with an amplitude of a couple cooling lengths. For this model, this corresponds to a time of $\sim0.7$ days.

During the initial phase of the oscillation, the back-propagation of the shock means the net velocity jump in the shock frame actually exceeds, by a factor 4/3, that in the steady flow model, and so this leads to an initial post-shock temperature that is $(4/3)^2 = 16/9$ higher than given in the steady shock scalings of eqns.\ (\ref{eqn:RHjump}).
On the other hand, during the contraction phase, the shock velocity jumps are weaker, leading to lower post-shock temperatures.

This 1D model of the cooling oscillation provides another sample case for deriving  X-ray emission from such shock compressions.
For each time $t$ of a simulation,  we carry out the 1D integration (\ref{eqn:FEint}) to obtain the now time-dependent flux at a selected energy, $F(E,t)$.
Integration of $F(E,t)$ over an X-ray energy bandpass $E_X > 0.3$\,keV yields the associated X-ray flux $F_X (t)$. 
For this cooling oscillation model, the rightmost panel of figure \ref{fig:1D_cool_osc} plots the time variation of $F_X$, normalized by the total
kinetic energy flux from both sides of the inflow, $F_{KE} = \rho_o v_o^3$.
Note that, due to a accumulation of shock heated material, $F_X$ increases up to the time $t \approx 3 t_c$ when the cooling region reaches its maximum size, with a peak that
reaches about half the kinetic energy input rate.
But as noted above, during the subsequent cooling zone compression the weaker shocks give lower temperatures. This abruptly reverses the $F_X$ into a decline, making it nearly vanish near the minimum.

\begin{figure}
\centering
\includegraphics[width=0.7\textwidth]{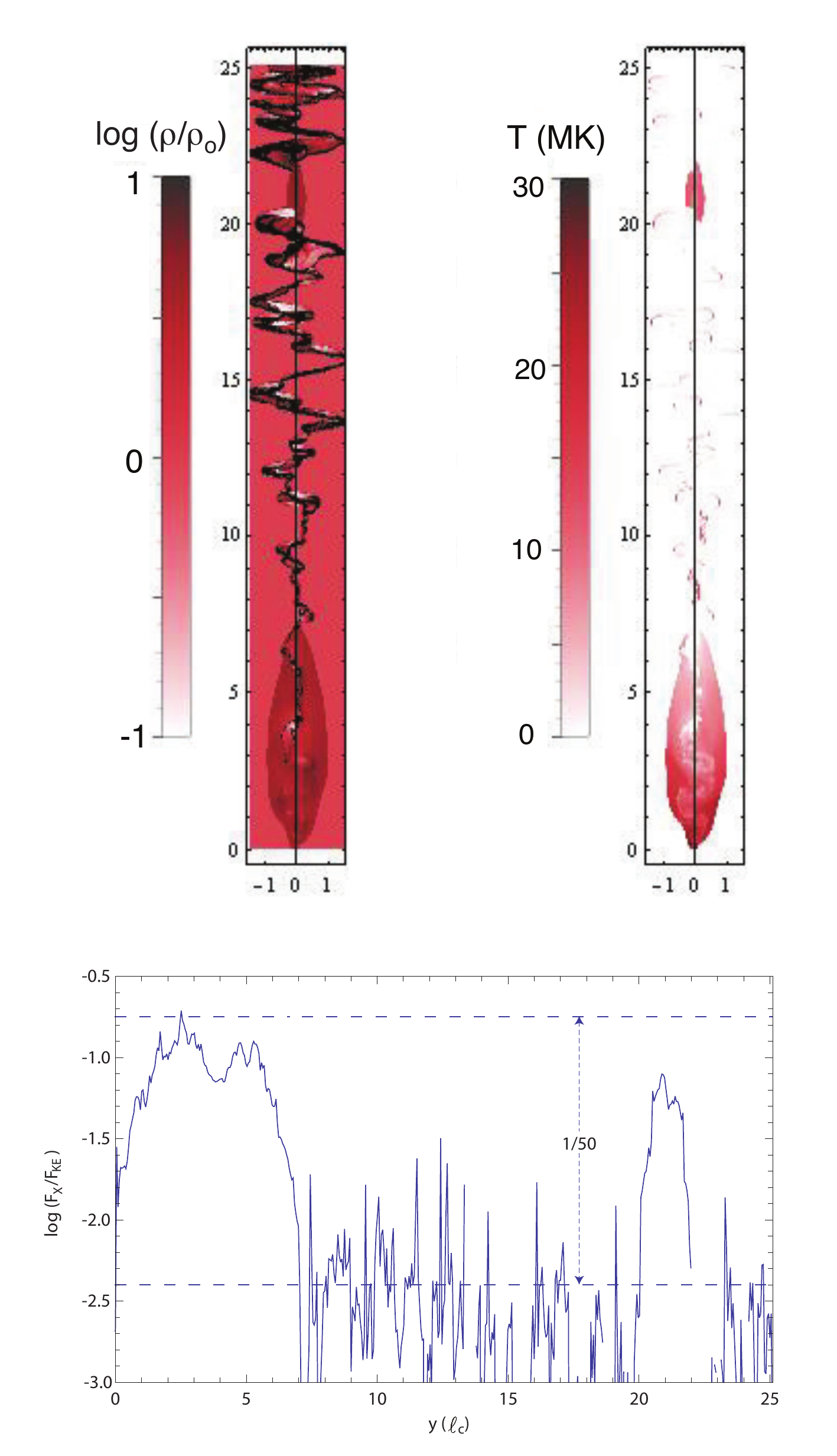}
\caption{Evolved-time snapshots of the 2D advection model showing spatial $(x,y)$ variations of log density (upper left) and  temperature (upper right), along with vertical {\bf $(y)$} variation of $x$-integrated X-ray emission $F_X (y)$ (bottom: now on a log scale, but again normalized by the total kinetic energy flux $F_{KE} = \rho_o v_o^3$). 
The horizontal dashed lines compare the $F_X$ time-averaged compressive value from the 1D cooling oscillation (upper)  and the final  shear-dominated state (lower), showing a roughly factor 1/50 reduction in the X-ray emission.}
\label{fig:advectFrames}
\end{figure}

\subsection{2D spatial breakup of density and temperature structure}
\label{sec:advect_rhoT}

Let us next consider a 2D simulation of this same basic model of direct collision between two equal and opposite planar flows.
In addition to the temporal variation from the cooling oscillation, the addition of a second, {\it transverse} (y) direction now  allows the possibility of a {\em spatial} break-up of  the post-shock cooling layer.
To illustrate this within a single time snapshot, it is convenient to introduce a constant {\em vertical advection} speed $v_y$, set here to be the same as the horizontal (x) inflow compression speed $v_o$, so that material introduced at all boundaries enters the computational domain at a fixed angles of $\pm 45^\circ$ on each side of the central axis $x=0$.
At the lower boundary $y=0$, this leads to a direct interaction at this $x=0$ interface. The flow also makes a 45$^\circ$ angle with this interface, but because the shock normal speed is kept at the same $v_o=v_x=1000$~km/s as in the 1D model, the shock strength is the same. The advection thus represents a simple Galilean transformation of what would occur in a 2D extension of the 1D direct collision model. Moreover, away from the lower boundary, it provides a convenient way to visualize the temporal evolution of the structure through a single time snapshot.

%{\bf Away from the lower boundary, this advection velocity provides a convenient way to illustrate the temporal evolution of the structure through a single time snapshot; but since the velocity perpendicular to the shock interface remains unchanged, it otherwise does not fundamentally alter the physics of the interaction.}

Figure \ref{fig:advectFrames} plots such a snapshot of the density (upper left) and temperature (upper right) structure at a time $t \approx 12 t_c$, corresponding to twice the advection time {\bf$t_{adv}=y_{max}/v_0$} through the vertical extent {\bf$y_{max}$}.
Both spatial directions are now scaled by the cooling length $\ell_c$, but for the vertical axis the simple advection at a fixed speed means that each scaled length unit can be readily translated to  time in cooling times via 
$y/\ell_c = (4 v_y/v_o) (t/t_c) = 4 (t/t_c)$; this thus allows for direct comparison with the 1D space + time variation plot in figure \ref{fig:1D_cool_osc}.
The computational grid now contains $n_x=600$ horizontal zones over the range $-1.5 \ell_c < x  < +1.5 \ell_c$, and $n_y=5000$ vertical zone over the range $0 < y < 25  \ell_c$. This agains corresponds to a uniform mesh size  $\Delta x = \Delta y  = 0.005 \, \ell_c = \ell_c/200$ that  very well resolves the 1D cooling length.

Note that the cooling oscillation still appears in the initial flow interaction near the lower boundary, but the dense cooled material near the central contact  quickly breaks up into a complex structure. The enhanced cooling associated with mixing of this dense structure reduces both the amplitude and period of the oscillation. Indeed, after just one initial  cycle, the 1D compressive cooling oscillation is now effectively overridden by an extensive {\em shear} structure, which grows with increasing distance (or advection time) from the lower boundary, forming complex ``fingers'' of cool, dense gas that bound regions of oppositely directed flow.

The net result is to transform the strong shock compression of the 1D collision into a complex of extended {\em shear }layers, along which any shocks are very oblique and thus very weak, with direct shocks limited to very narrow regions at the (convex) ``tips''  and (concave) ``troughs'' of the fingers.
As shown in upper right panel of figure \ref{fig:advectFrames}, this leads to a corresponding reduction in the spatial extent of high-temperature gas in the downstream flow. The bottom panel of figure \ref{fig:advectFrames} shows that, after an initial sharp rise due to the accumulation of strong compressive shock heating near the lower boundary, the associated horizontally integrated X-ray emission $F_x$ drops abruptly after a single cooling oscillation extending over a few cooling lengths. It  then varies greatly with y-position, but apart from a limited segment of more direct compression around $y \approx 20 \ell_c$, the overall level after the initial oscillation cycle is reduced by about a factor 1/50 
(as denoted by the range between the horizontal dashed lines).

\begin{figure}
\centering
\includegraphics[width=0.65\textwidth]{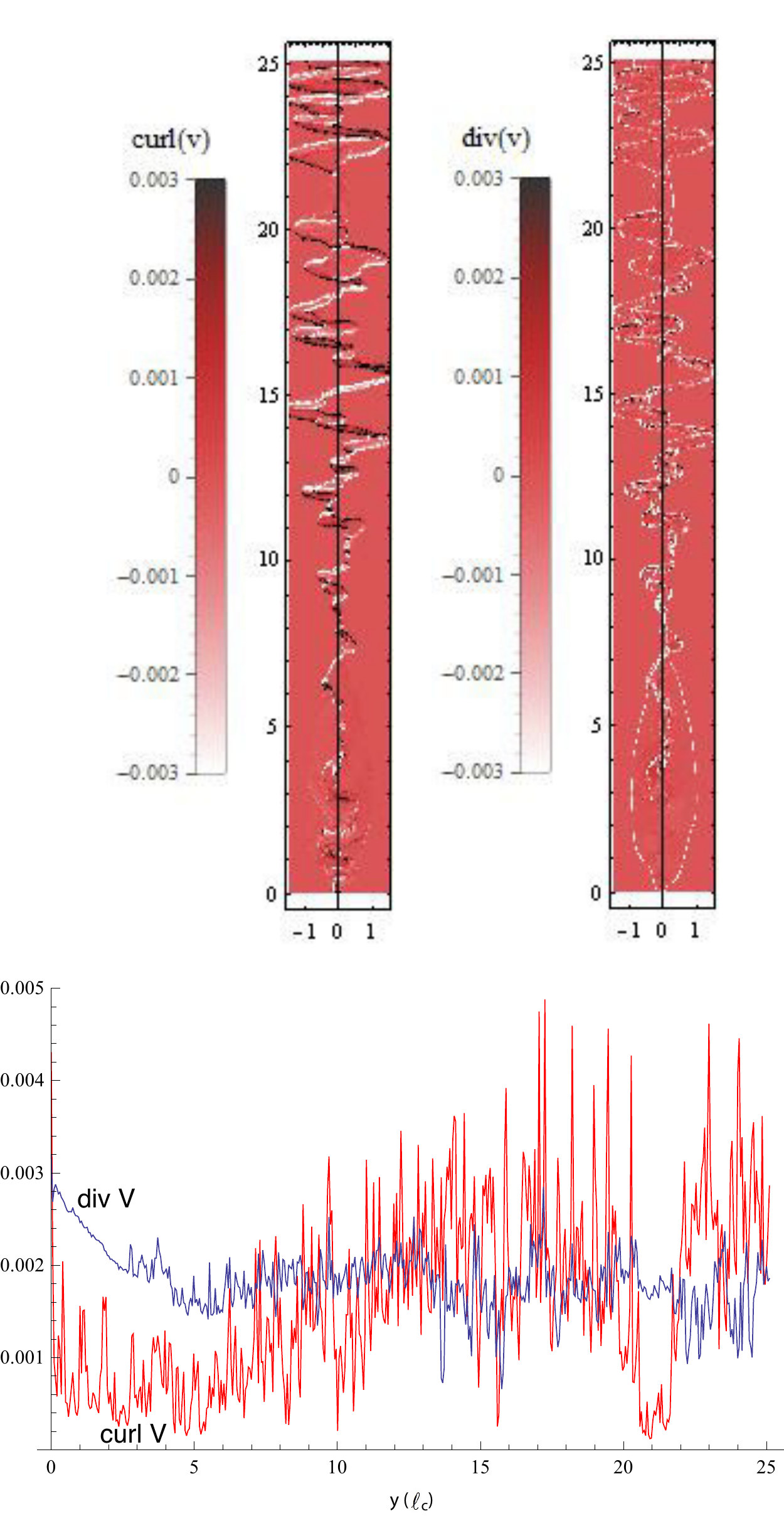}
\caption{{ Snapshots of the 2D advection model at the same time as figure \ref{fig:advectFrames}, showing spatial $(x,y)$ variations of velocity divergence (upper left) and curl (upper right). The bottom panel plots vertical evolution of r.m.s. horizontal averages of these quantities. Note that from the initial condition at the lower boundary (y=0) the divergence decreases while the curl increases.}}
\label{fig:div_curl}
\end{figure}

Note that an important aspect of this thin-shell-instability reduction in X-rays is the transformation of the flow compression, characterized by a velocity divergence $\nabla \cdot {\bf v}$, into strong velocity shear, characterized a flow vorticity $\nabla \times {\bf v}$. The spatial distribution of these quantities, as well as their relative vertical evolution, is illustrated in figure \ref{fig:div_curl}.

\section{Collision between Expanding Outflows}
\label{sec:source}

In practice, shock collisions in astrophysics often occur in {\em expanding} outflows, for example in the collision between two spherically expanding stellar winds in a binary system.
Such expansion tends to give the velocity divergence on the right side of the temperature equation (\ref{eqn:DTDt}) a positive value, contributing then to an {\em adiabatic expansion cooling} that competes with the radiative cooling term 
\cite[see, e.g.,][]{SteBlo92}.
To examine how this competition affects radiative cooling instabilities and their associated reduction in shock temperature and X-ray emission, let us now generalize the above flow collision models to allow for expansion within the 2D plane.

\begin{figure*}
\includegraphics[width=1\textwidth]{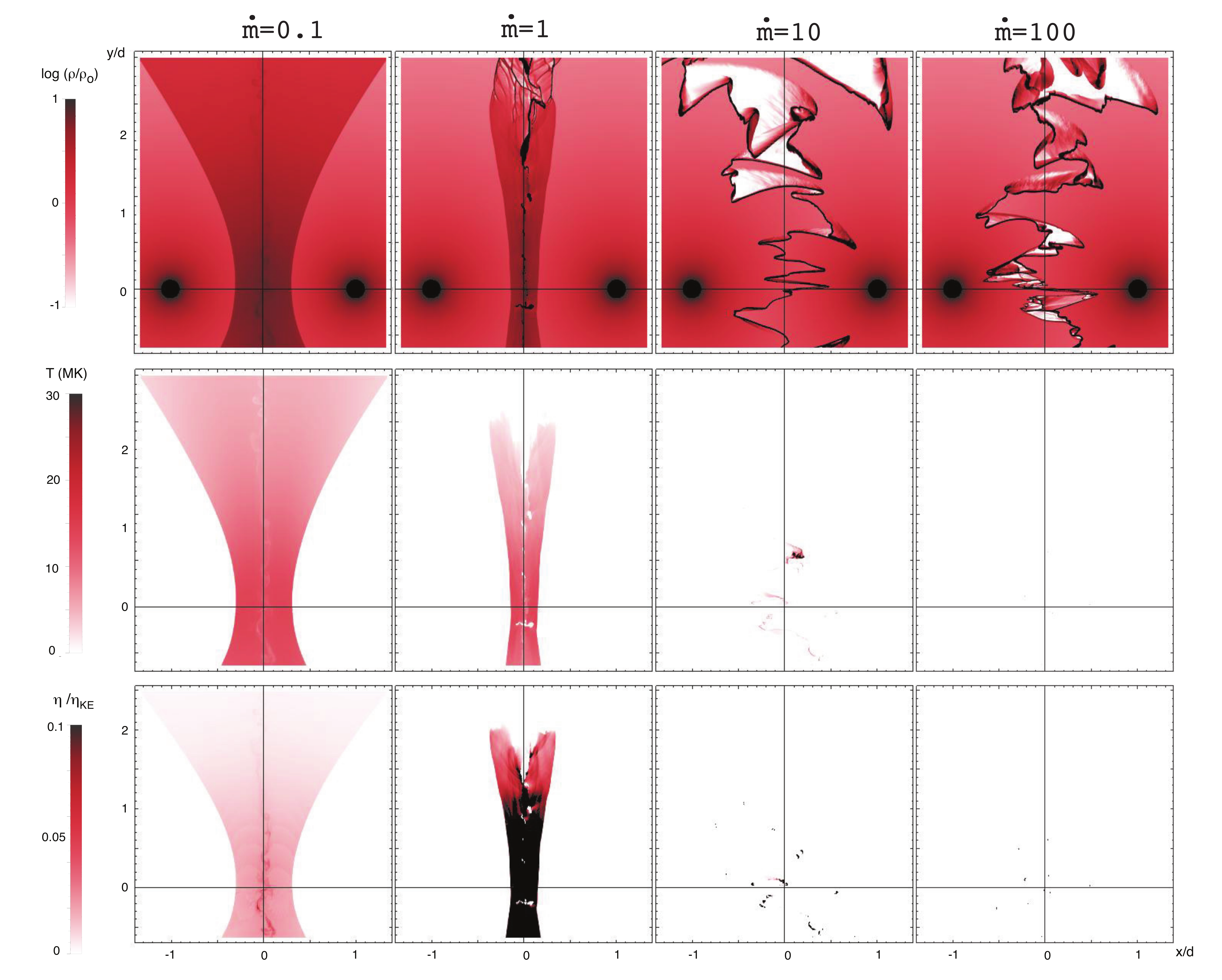}
\caption{
Final time ($t_f=32 d/v_o$) snapshots for models with mass cooling parameter $\dot{m} =$ 0.1, 1, 10, and 100, arranged in columns from left to right, with rows showing spatial $(x,y)$ variation of density $\log \rho$ (top),  temperature $T$ (middle), and X-ray emissivity $\eta_X$ (bottom).
The temperature is in MK, but the density is scaled by $\rho_o(d) = \dot{M}/(2 \pi v_o d)$ and the emissivity by an associated kinetic power density,
 $\eta_{KE} \equiv \dot{M} v_o^2/(2 \pi d^2)$. The axes for horizontal (x) and vertical (y) position are  in units of the star-interface distance $d$.
}
\label{fig:sourceFrames}
\end{figure*}

\subsection{Cooling parameter $\chi_{2D}$ for 2D models with planar expansion}
\label{sec:chi2D}

Specifically, instead of two opposing laminar flows, let us now assume the configuration in figure \ref{fig:3configs}c, namely
a 2D {\em planar expansion} from two distinct  localized mass sources at positions $\{x_m,y_m\}=\{\pm d, 0\}$, with equal constant outflow speed $v_o$ and equal mass ejection rate $\dot{M}$.
To facilitate connection to 3D outflows characterized by a volume density $\rho$, we can formally assume a {\em cylindrical} expansion, with $\dot{M}$ representing the mass source rate {\em per unit length} along an arbitrarily extended $z$-axis perpendicular to the 2D (D) computational plane.  Ahead of any interaction, at a distance $r = \sqrt{(x-x_m)^2+y^2}$ from either source, mass conservation for steady-state, constant-speed, cylindrical expansion implies a mass density  $\rho = \dot{M}/2 \pi v_o r$.
The equality of the two mass sources means that their interaction will again center on the bisector symmetry line at $x=0$.

Along the $x$-axis line between the sources at $x_m = \pm d$, this means that the mass flux within the cooling layer now varies as {\bf $\rho v = \dot{M}/2 \pi r$}.
Applying this for the isobaric post-shock cooling layer model of \S \ref{sec:lcool}, the temperature eqn.\ (\ref{eqn:dTdx}) can now be 
written in a generalized, scaled form that accounts for this $r$-dependence of the mass flux,
\begin{equation}
T^2 \frac{\partial T}{\partial r}=-\frac{2 T_s^3 }{3 \chi_{2D}} \, \frac{r}{r_s^2}
\, , 
\label{eqn:dTdrs}
\end{equation}
which applies along the ($y=0$) $x$-axis, with $r=d-|x|$.
Here the shock distance from the source is set by
\beq
r_s = \frac{d}{\sqrt{1+\chi_{2D}}}
\, ,
\label{eqn:rsanal}
\eeq
where the dimensionless cooling parameter,
\begin{equation}
\chi_{2D} \equiv \frac{5\pi v_o^4}{128\Lambda_m \dot{M}}
= \frac{2 \ell_c}{d}
\, ,
\label{eqn:chi_2D}
\end{equation}
with $\ell_c$ defined here by (\ref{eqn:cooling_length}) using the inflow density at the interface, $\rho_o (d)=\dot{M}/2\pi v_o d$.
Integration of (\ref{eqn:dTdrs}) with the requirements that $T(r_s)=T_s$ and $T(d)=0$ now gives for the temperature variation within the cooling zone along the $x$-axis,
 \beq
 \left ( \frac{T(r)}{T_s} \right )^3 = \frac{1}{\chi_{2D}} \, \left ( \frac{d^2-r^2}{r_s^2} \right )
 \, ; ~~~ r_s < r < d ~~ , ~~ y=0
 \, .
 \label{eqn:Trcyl}
 \eeq
 For $\chi_{2D} \ll 1$, $r_s/d \approx 1-\chi_{2D}/2 = 1 - \ell_c/d$, and since $r/d = 1- |x|/d$, we find from first-order expansion in $|x|/d \ll 1$, that (\ref{eqn:Trcyl}) recovers the laminar collision scaling (\ref{eqn:Tofx}).
 
The dimensionless parameter $\chi_{2D}$ characterizes the ratio of length scales for cooling vs. expansion, set by $\chi_{2D} = 2 \ell_c/d$.
It serves as a 2D analog to the commonly quoted, standard cooling parameter $\chi$, defined by  \cite{SteBlo92} in terms of the ratio of {\em time} scales for cooling vs.\ expansion in the full 3D case of colliding stellar winds.
Both have identical scalings with the fourth power of the flow speed and inverse of the mass source rate; but note that, unlike the full 3D case,  the 2D scaling here has no dependence on the separation distance $d$.  Because planar expansion has one less dimension than the full spherical case, this 3D  scaling with distance in the numerator becomes replaced with division by a mass source  rate $\dot{M}$ {\em per unit length} from cylindrical sources that formally extend perpendicularly from the 2D plane.

Nonetheless, for a fixed flow speed $v_o$, which through eqn.\ (\ref{eqn:RHjump}) sets the post-shock temperature $T_s$, we can still readily examine the effects of varying the relative importance of radiative vs.\ adiabatic cooling by adjusting this mass source rate $\dot{M}$ to change the parameter $\chi_{2D}$.
Indeed, since $\chi_{2D} \sim 1/\dot{M}$, it is convenient to characterize the radiative cooling efficiency in terms of $\dot{m} \equiv \dot{M}/\dot{M}_1$,  where $\dot{M}_1\equiv 5 \pi v_o^4/(128 \Lambda_m)$ is the mass source rate for which $\chi_{2D} = 1$.
This gives $\dot{m} = 1/\chi_{2D}$.

In the strong radiative cooling limit $\dot{m} \gg 1$, the cooling region in this steady shock model is confined to narrow layer with $|x| \le \ell_c = d/2\dot{m}$ on each side of the interaction front at $x=0$. 
Following the planar scaling (\ref{eqn:FXslab}), the X-ray emission in this case should increase linearly with the mass flux, $\dot{m}$.
But the very narrowness of this layer makes it subject to the thin-shell instabilities \citep{Vis94} that give rise to extensive spatial structure seen in the above 2D laminar collision models, with associated reduction in shock temperature and X-ray emission.

In contrast, in the limit of inefficient radiative emission $\dot{m} \ll 1$, the cooling is instead by adiabatic expansion, with a much thicker offset from the interface. The X-ray emission integrated over this extended interface now depends on the density-squared emission measure, implying a total emission that scales with $\dot{m}^2$.   Moreover, this extended layer  can now effectively suppress the thin-shell instability, allowing the post-shock gas to retain higher temperatures, and so an extended X-ray emission.

In the absence of instability-generated structure, \citet{OwoSun13} proposed a simple scaling law that ``bridges" the adiabatic vs.\ radiative limits, which in the current notation can be expressed in units of twice the X-ray luminosity for the transition case $\dot{m}=1$,
\beq
L_X \approx \frac{\dot{m}^2}{1+\dot{m}}
\, .
\label{eqn:bridgelaw}
\eeq
A general goal here is to use numerical simulations to test how this scaling is modified by the effects of the thin-shell instability for high-density flows with $\dot{m} > 1$.

\subsection{Numerical simulation settings}
\label{sec:source_rhoT}

Let us now quantify these expectations with a full numerical simulation parameter study that examines how shock structure and X-ray emission in this 2D model of shock collision and planar expansion depends on this cooling efficiency parameter $\dot{m}$.
The  numerical model again assumes a constant source flow speed $v_o=1000$\,km\,s$^{-1}$, but now with the outflow mass sources at $\{x_m,y_m\}=\{\pm d,0\}$ {\em embedded} in a uniform spatial grid with $n_x = 1024$ zones ranging over $-1.33 d < x < + 1.33 d$ and $n_y = 1204$ zones over $-0.63 d < y < 2.5 d$. This implies fixed zone sizes $\Delta x = \Delta y = 0.0026\,d = 0.0052 \, \dot{m} \ell$.

The separation between the mass sources is set to a typical stellar binary separation $2d=3.1 \times 10^{12}$\,cm = 0.21\,au. All models are run to a final time $t_{f}=$500\,ks, corresponding to nearly 32 characteristic flow times $t_d=d/v_o=$15.75\,ks from the sources to the interface; this is sufficient for even material that initially collides along the source axis to flow out through either the bottom or top boundary, following its vertical pressure acceleration away from this $x$-axis. The simulation uses simple supersonic outflow boundary conditions along both horizontal and vertical edges of the computational domain. 

\begin{figure*}
\includegraphics[width=1\textwidth]{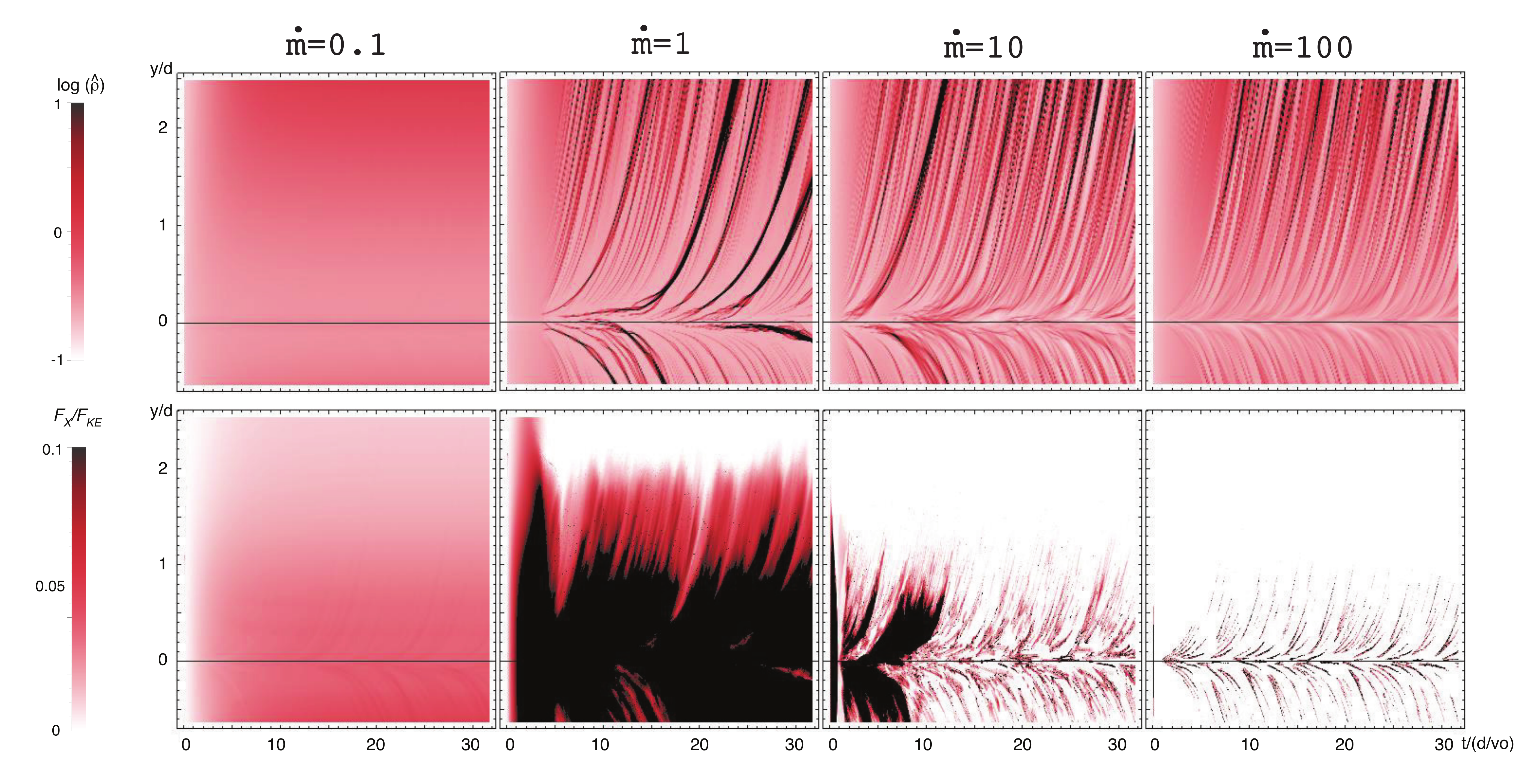}

\caption{
Time-height (t,y) variation of lateral (x) integrations of density ${\hat \rho} (y,t)$ (top row; on log scale),
%normalized by ${\hat \rho}_o = \dot{M}/2 \pi v_o$) and X-ray flux $F_X (y,t)$ (bottom row), 
for models with mass cooling parameter $\dot{m} =$ 0.1,  1,  10, and 100, again arranged in columns from left to right.
The X-ray flux is normalized by the total kinetic energy flux at the interaction front along the axis between the stars, $F_{KE} = \dot{M} v_o^2/(2 \pi d)$.
The vertical (y) spatial axis is in units of star-interaction distance $d$, and the horizontal time axis (t) is in units of the characteristic flow time $d/v_o$.
}
\label{fig:sourceSpaceTime}
\end{figure*}

\subsection{Structure snapshots at final, evolved time}
\label{sec:source_rhoT_snap}

For the final, well-evolved time ($t = 32 d/v_o$),  figure \ref{fig:sourceFrames} compares results for simulations with $\dot{m} = $ 0.1, 1, 10 and 100, arranged along columns from left to right.
The upper two rows give color plots of the log density (top) and temperature (middle); the bottom row shows the associated X-ray volume emission $\eta_X$ for energies $E>E_X=3$\,keV.

The results quite vividly illustrate the trends anticipated above.
For the lowest-density case with $\dot{m}=0.1$ (leftmost column), the relative inefficiency of radiative vs. adiabatic cooling  leads to a distinct standoff of the shock from the interface, with only very weak instability structure forming along the interface.
Along the direct collision between the sources, the high post-shock pressure drives material away from this axis, allowing a shock standoff  radius $r_s/d \approx 0.68 $ that is larger than the $r_s/d = 1/\sqrt{1+4} = 0.45$ predicted by 
eqn.\ (\ref{eqn:rsanal}) (which accounts  for planar expansion, but not such vertical pressure acceleration).
As the standoff distance increases at larger $|y|$, there develops an extended region of high post-shock temperature, with an associated extended region of X-ray emission.

For the factor-ten higher-density model with $\dot{m} =1$ (left central column), the cooling layers become narrower, with now quite notable instability near the interface that becomes strongly developed at large $y$. The high temperature region is accordingly narrower, with vertical extent terminated at the location that instabilities become strong.  The X-ray emission occurs over a similar spatial extent, but is now stronger in the high density regions near the $x$-axis interface.

For a further factor-ten higher-density model with $\dot{m} = 10$ (right central column), the cooling layers now become completely unstable, with density showing extensive finger-like structure similar to the above laminar models. The high-temperature gas, and associated X-ray emission, is now limited to small regions, again associated with the tips of the fingers and the troughs between them.

Finally, in the highest-density model with $\dot{m} = 100$  (rightmost column), the cooling instability leads to even more extended fingering, with the few zones of high-temperature and X-ray emission now hardly noticeable. However, as quantified below, the high density of the few remaining hot regions can still lead to significant overall X-ray luminosity.

\subsection{Time-height evolution of flow structure}
\label{sec:source_rho_tvar}

To complement such single-time snapshots of the flow structure, let us next examine its time evolution.
Specifically, to illustrate the flow evolution in time $t$ and height $y$, let us define horizontal $x$-integrations of the density and X-ray emission.
For the density, to compensate for the $1/r$ decline, we weight the integration by $r$, normalized by the associated integration through the unperturbed mass source,
\beq
{\hat \rho} (y,t) \equiv   \frac{3 \pi v_o}{4 \dot{M} d}  \int_{-4d/3}^{4d/3} \rho (x,y,t) \,r \, dx
\, .
\label{eqn:dyt}
\eeq
We can similarly define a horizontally integrated X-ray flux,
\beq
F_X (y,t) \equiv   \, \int_{-4d/3}^{4d/3} \eta_X (x,y,t) \, dx
\, .
\label{eqn:FXyt}
\eeq

Figure  \ref{fig:sourceSpaceTime} shows color plots of the time-height evolution of ${\hat \rho} (y,t)$ (upper row) and  $F_X (y,t)$ (lower row), again for models with $\dot{m} = $ 0.1, 1,  10, and 100, arranged along columns from left to right.

For the lowest-density model with $\dot{m} =0.1$, both density and X-ray emission quickly adjust to a time-independent steady-state with smooth distribution in $y$. But for all higher density models, there develops clear structure, with density compressions that diverge away from the mass-source $x$-axis, along with associated structure in X-ray emission.
In the $\dot{m} = 1$ case, and in the initial evolution of the $\dot{m}=10$ case,  the overall level of X-ray emission is increased in proportion to the higher source mass rate $\dot{m}$.

But at a time around $t \approx 13 d/v_o$, the latter case shows a sharp decline in filling fraction of X-ray emitting structures, reflecting the formation of strong fingerlike structures from the thin-shell instability, and its associated limitation of X-ray emission to tips and troughs of the fingers.
For the highest-density case $\dot{m}=100$ this X-ray volume reduction starts near the initial time,  and is even more pronounced.
This lower filling factor significantly reduces the X-ray luminosity, though this can be partly compensated by the more intense local emission from dense, strongly emitting regions. 

\begin{figure}
	\centering
	\includegraphics[width=\textwidth]{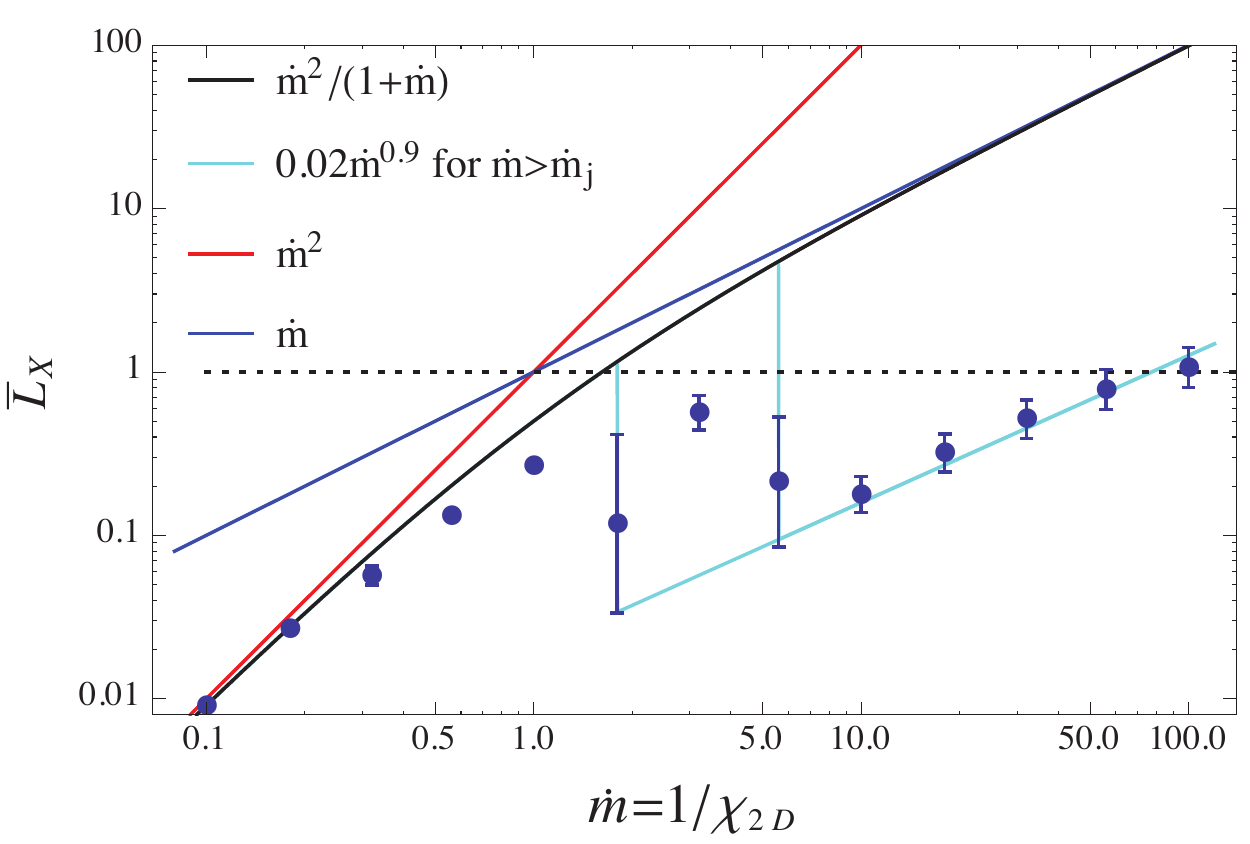}
\caption{
Time-averaged X-ray luminosity ${\bar L}_X$ vs.\ mass-loss-scaled cooling efficiency $\dot{m} = 1/\chi_{2D}$. 
The legend identifies curve styles for  various analytic scaling formulae. 
The points for simulation results are scaled by anchoring ${\bar L}_X$ for the lowest-density case $\dot{m} = 0.1$ to the bridging form (\ref{eqn:bridgelaw}),
 with error bars representing $\pm 1\sigma$ time variability.  
In the nearly adiabatic regime $\dot{m} \lesssim 1$, the simulations nearly follow the black curve for the scaling (\ref{eqn:bridgelaw}).
In the strongly radiative limit $\dot{m} \gg 1$, they are (almost) linear with $\dot{m}$, but with strong reduction below the analytic form (\ref{eqn:bridgelaw}).
The transition regime $2 < \dot{m} <  6 $  has large variations from switching between high and low ``bi-stable'' states. The cyan curves show factor 1/50 jumps at values $\dot{m}_j$ bracketing this transition regime,
with slightly sublinear ($\sim\dot{m}^{0.9}$) scaling above the jump.
}
\label{fig:masterPlot}
\end{figure}

\subsection{Scaling of X-ray luminosity with $\dot{m}$}
\label{sec:source_master}
Let us now quantify the overall scaling of the X-ray luminosity with the cooling efficiency $\dot{m} = 1/\chi_{2D}$. 
For this we derive the asymptotic time-average  X-ray luminosity ${\bar L}_X$ as the integral of $F_X (y,t)$ over positive $y$, with then a  time average over the time $t > 13 d/v_o$ after transition to its asymptotic state.  
In addition to the 4 models detailed above that differ in 1 dex increments of $\dot{m}$, we also compute 9 additional models to give a denser parameter grid of 13 models in 0.25 dex increments over the full range from $\dot{m}=0.1$ to 100.

Figure \ref{fig:masterPlot} plots (on a log-log scale) the resulting data points for $ {\bar L}_X$ vs. $ \dot{m}$, along with error bars to indicate the level of 1$\sigma$ temporal variation.
The black curve compares the \citet{OwoSun13} bridging law scaling (\ref{eqn:bridgelaw}), along with the linear $L_X \sim \dot{m}$ scaling (blue) and quadratic $L_X \sim \dot{m}^2$ scaling  (red) expected respectively in the high-density ($\dot{m} \gg 1$), radiative shock limit, and  in the low-density ($\dot{m} \ll 1$) adiabatic shock limit.
We normalize all the simulation data by anchoring the smooth emission from the lowest-density ($\dot{m} = 0.1$), nearly adiabatic case to exactly fit this bridging law, with 
 ${\bar L}_X = \dot{m}^2/(1+\dot{m}) = $0.009.

In figure \ref{fig:masterPlot} the first 5 data points with lowest $\dot{m}$, ranging from 0.1 to 1, do indeed nearly follow this simple bridging law, with small temporal variability indicated by error bars that are less than the point sizes. This is consistent with the relatively extended shock compression and X-ray emission shown in the two leftmost columns of figures \ref{fig:sourceFrames} and \ref{fig:sourceSpaceTime} for models with $\dot{m}$= 0.1 and $\dot{m}=$1. 

In contrast, for the 5 highest-density models, with $\dot{m} \ge 10$, the ${\bar L}_X$ all fall roughly a fixed factor 1/50 below the linear increase that applies in the $\dot{m} \gg 1$ limit of this bridging law, with somewhat larger, but still modest errors bars indicating only a moderate level of time variability. This is similar to the factor 1/50 reduction in X-ray emission of the 2D advection models once the initial compressive oscillation is transformed to the strong shear flow and finger structure.  It is also consistent with the extensive shear fingers and reduced X-ray emission shown in the two rightmost columns of figures \ref{fig:sourceFrames} and \ref{fig:sourceSpaceTime} for models with $\dot{m}$= 10 and $\dot{m}=$100. 

\begin{figure*}
\includegraphics[width=1\textwidth]{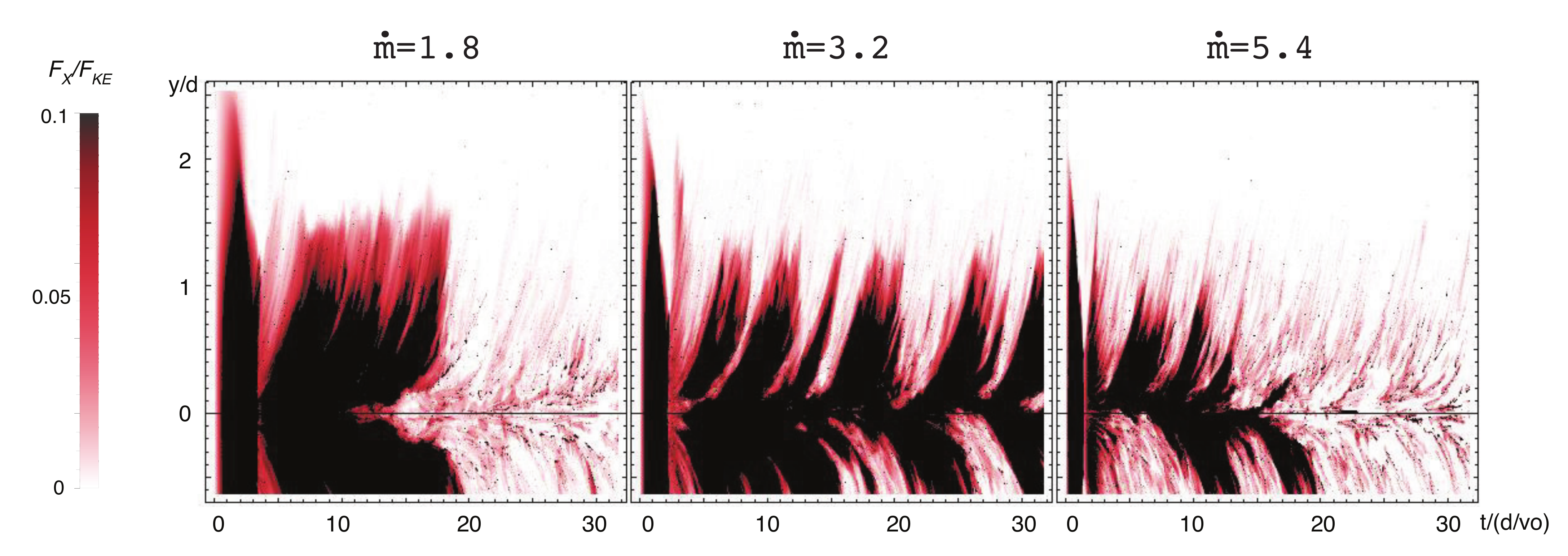}
\caption{Time-height variation of  laterally integrated X-rays for  transition cases ($\dot{m} = $ 1.8, 3.2 and 5.4), showing a bi-stability between relatively smooth states with high $L_X$ and highly structured states with reduced $L_X$.}
\label{fig:bistab}
\end{figure*}

Between these two limits, for the 3 intermediate cases with $ \dot{m} = $ 3.5, 6.3 and 5.6, the  ${\bar L}_X$ vary somewhat erratically, trending below the bridging curve, with large error bars indicating strong time variability.
Figure \ref{fig:bistab} illustrates that this apparently stems from the tendency for these intermediate cases to undergo switching between high X-ray states -- with the extended compression and emission of the low-$\dot{m}$, adiabatic limit --  and a low X-ray state -- with the small-scale shear and dense fingers typical of the unstable structure in the high-$\dot{m}$, radiative limit.

The overall result is thus that colliding-flow shock X-rays follow close the simple adiabatic-to-radiative bridging law for low to moderate density shocks with $\dot{m} \lesssim 1$, then show a somewhat random modest declining trend for the moderately radiative cases $2 < \dot{m} < 8$, and finally again increase linearly with $\dot{m}$ for the strongly radiative cases $\dot{m} > 10$, but at about 1/50 lower level than the expected linear scaling for the radiative limit without instabilities.

This factor 1/50 thus seems to represent a kind of  fixed ``shear/mixing penalty'' for X-ray production in the limit of strongly radiative shocks.

\section{Discussion \& Future Outlook}
\label{sec:summary}

All the above 2D planar expansion simulations  were computed with a fixed spatial mesh of grid size $\Delta x = 0.0026 \,d$.
For the highest-density model, $\dot{m} = 100 = 2d/\ell_c$, this is roughly half the 1D cooling scale $\ell_c$, indicating that this densest model could be resolution limited. Nonetheless,  note that it does still give roughly the same factor 1/50 reduction in X-ray emission seen in the 2D laminar-collision case, for which the grid is set to provide a quite high resolution of this cooling length, $\Delta x = 
 0.005\, \ell_c$.  
Additional planar expansion simulations done at a factor 2 lower resolution still give good general agreement to the ${\bar L}_X$ plotted in figure \ref{fig:masterPlot}, except that for the two highest-density models, with $\dot{m} =$ 56 and 100, there is a precipitous drop in X-ray emission. Since now the grid zones with  $\Delta x > \ell_c$ are no longer adequate to resolve the cooling length, even in the finger tips and troughs with compressive shocks the numerical hydrodynamics using the \citet{Tow09} exact integration scheme for radiative cooling now simply jumps to the fully cooled post-shock equilibrium, without showing any of the intermediate shock heating.

Such resolution issues are an inherent challenge for any numerical effort to examine the effects of thin-shell instability on X-ray emission. As emphasized by \citet{ParPit10}, there are many associated numerical effects, for example that of ``numerical conduction'', that can reduce and soften the associated X-ray emission. Our study here has not included any specialized attempts to mitigate such effects, but in the 2D planar advection model here, the fixed spatial grid has a resolution of the cooling length that is comparable to the adaptive mesh refinement models by \citet{ParPit10}. Moreover, the vertical advection makes the vertical structure a convenient proxy for the time-dependent evolution. This illustrates how the thin-shell instability transforms  a laminar flow compression into an extensive shear structure, with associated reduction in X-ray emission that seems {\em not}  just a numerical effect, but grounded in the fundamentally different shear vs.\ compressive flow structure.  By using a systematic parameter study in the cooling efficiency associated with the mass source rate $\dot{m}$, the planar divergence models allow us to examine how the reduction of X-rays depends on the relative strength of adiabatic vs.\ radiative cooling.

A key result is the indication here that this thin-shell instability transition to complex structure might simply lead to a fixed reduction in X-ray emission. But further work will be needed to determine the origin of this reduction, and how it may be affected by either numerical or physical parameters.  For example, the factor 1/50 found here is roughly comparable to the pre-shock vs.\ post-shock sound speed, or the inverse Mach number. Future work should thus explore how this  factor depends on the assumed input/floor temperature $T_o$, or the shock inflow speed $ v_o$. In principle, this can be done within the simple laminar flow model, using perhaps adaptive mesh refinement to further increase the effective resolution. Such adaptive mesh methods would also be particularly useful in testing the inferred linear trend of X-rays at high $\dot{m}$.

Apart from this study of how colliding wind X-rays scale with cooling efficiency, it will be of interest to apply the insights here toward understanding the scaling of X-rays in other contexts, such as from the embedded wind shocks arising from the intrinsic instability of radiative driving of hot-star winds. In particular, \citet{OwoSun13} have argued that the observed linear scaling between X-ray and stellar bolometric luminosity ($L_X \sim L_{bol}$) in single O-stars might be explained if thin-shell mixing of such embedded shocks reduces their X-ray emission by some power -- the ``mixing exponent'' $m \approx 0.4$ -- of the mass loss rate.
This is distinct from the nearly constant X-ray reduction factor found here in the context of direct flow collision, and so future work should explore how the scaling of any X-ray reduction might depend on the specific geometrical and physical context of the shock production, e.g. effect of a large-scale or turbulent magnetic field \citep{HeiSly07}.

Indeed, even within the context of CWB's, there are additional effects not considered here that could significantly alter how thin-shell instability affects X-ray production. For example, how will the factor 1/50 reduction found in these 2D models differ in a more realistic 3D wind collision? How might this be affected by the shear in an interaction betweens winds with different speed, on in the bow-shaped interaction of winds with differing momenta? Even in the planar interaction of winds with equal momenta, but unequal speed and density, the cooling parameter will be different on each side of the interaction front. How will this affect thin-shell structure and the reduction of X-ray emission? Clearly, there are many remaining issues for understanding the X-ray properties of such CWB's. But hopefully the current 2D study of thin-shell effects can form a good basis for addressing such more complex cases.
\chapter{Summary, Conclusions, and Future Directions}\label{chap:conc}

The central focus of this dissertation has been to investigate the interplay of radiation forces with circumstellar gaseous disks. To conclude, we here summarize the key results and then lay out some possible directions for future research.

\section{Overview of Dissertation Results}

After an introductory background overview in chapter \ref{chap:intro}, chapters \ref{chap:line} and \ref{chap:disks} review the general mathematical formalism for the two components of the dissertation: line-driven winds, and Keplerian, gaseous disks. In addition to an analytic scaling of wind mass loss rate with stellar parameters, chapter \ref{chap:line} gives particular emphasis to the 3D, vector line-acceleration formalism. Chapter \ref{chap:disks} and its discussion of circumstellar disks pointed out the scenarios in which disks occur, some key properties of circumstellar, gaseous disks including their density structure and total mass, and reviewed prior attempts to model their dynamics using viscosity.

By taking the results of the background chapters, chapter \ref{chap:ab_thin} presents the first results of line-driven ablation of a circumstellar disk. Not only does this confirm the prediction that the same line-accelerations that drive stellar winds can also significantly affect Keplerian disks, but it also investigates the roles played by non-radial velocity gradients and non-radial acceleration components in line-driven ablation. By further demonstrating that the effects of rapid rotation on ablation can be understood as scaling nearly linearly with equatorial brightness, it sets the stage for the parameter study completed in chapter \ref{chap:ab_spec_type}. 

Chapter \ref{chap:ab_spec_type} then discusses two of the central results of this dissertation:
\begin{enumerate}
\item The lack of observed Oe stars can be understood as a byproduct of their extreme luminosity.
\item The time scale for disk destruction can be well approximated by the ratio of disk mass to spherically symmetric mass loss rate, $M_{disk}/\dot{M}_{wind}$.
\end{enumerate}
These two results address long standing issues in understanding the Classical Be phenomena. Additionally, as both are derived from models that ignore viscosity, we now can understand the observationally inferred short disk decay times without appealing to anomalously strong viscous diffusion.
% Additionally, by comparison with observations, we show that such ablation can significantly reduce the necessity for anomalously strong viscous transport in Classical Be star disks and, as shown in chapter \ref{chap:ab_spec_type}, completely dominate the evolution of such a disk in earlier spectral types. By taking the ratio of the mass of the circumstellar disk to the spherically symmetric mass loss rate for each star, chapter \ref{chap:ab_spec_type} also showed that, within a factor of a few, such a simple scaling for the characteristic disk time scale matched within a factor of order unity the simulation times for disk removal.

In addition to questions regarding disk decay, there are also several open questions related to the growth of Classical Be disks. Chapter \ref{chap:pdome} presents a potential model for this growth by combining rapid rotation with non-radial pulsation modes. Here, the results show that it is possible to launch large-scale, nearly Keplerian disks using non-radial pulsations, consistent with those seen in observations \citep[see e.g.][]{RivBaa03}. Several variations on the non-radial pulsations are also investigated there in order to begin to probe the available parameter space.
%With these results in hand, chapter \ref{chap:pdome} took a brief diversion to investigate the growth portion of the Classical Be star phenomena of disk growth and destruction. By imposing simple mock-ups to g-mode pulsations, circumstellar disks could be built in the equatorial plane. Additionally, comparisons of different combinations of phase velocity and energy propagation direction show that such disks can be constructed using pulsations consistent with the retrograde phase velocities observed in many Classical Be stars. 

This dissertation returned to disk ablation in chapter \ref{chap:ab_thick}, but there for densities comparable to those of optically thick, star forming accretion disk. To handle their extreme optical depths, we introduce and derive a semi-analytic approximation for treating continuum optical depths under the assumption of a geometrically thin disk. By considering a model with pure absorption (i.e. no scattering or re-emission), and comparing it to one which omits continuum optical depths all together, we bracket the expected ablation of a disk for which the continuum opacity is dominated by electron scattering. These limits show that, even for a very optically thick disk, the simple scaling law derived in chapter \ref{chap:ab_spec_type} can still be applied.

Finally, chapter \ref{chap:thin_shell} presented some additional work also related to massive star winds, and particularly to their X-ray properties. Here the central purpose is the investigation of the scaling of X-ray emission from geometrically thin, shocked layers. Though there are many astrophysical scenarios in which such shocks occur, this work focuses on X-rays as mass loss diagnostics for colliding wind binaries. For the collision of high density winds, such as would be expected both for close binaries and binaries with two early O stars, the net effect of the so called ``thin-shell instability'' is to reduce X-ray emission by a factor $\sim 50$. This work has broad implications for the description of X-ray properties of colliding wind binaries as evidenced by the recent work of \cite{RauMos16}, and perhaps even for such seemingly far removed fields as habitability of planets as evidenced by the work of \cite{JohZhi15}

\section{Future Directions}

While this dissertation presented the first simulations of line-driven ablation of disks around massive stars, it is far from a complete investigation of the problem. Each of the following subsections briefly outlines a remaining open question which should be addressed by extensions of this work.

\subsection{Immediate extensions}

Directly from the work included in this dissertation, there are some obvious immediately available future projects. One particularly pressing and immediately available project is to introduce the competing mechanism for ablation. For Classical Be stars, we can use the pulsation models in chapter \ref{chap:pdome}, while for star forming disks we would need a prescription for how mass enters from the outer disk as well as potentially a viscosity prescription (discussed more in the next subsection). In both cases, this would provide us with the appropriate competing mechanism which will act against ablation to replenish the disk. The effects of this competition both on ablation rate and the resulting ability to more accurately discuss disk decay rates would be potential results of this study that would be of particular interest.

We also are already set-up to more thoroughly investigate the behavior of an optically thick disk. As discussed in chapter \ref{chap:ab_thick}, the model there can be seen as an initial proof of concept for our ability to calculate the ablation of optically thick disks. Now that we have proven the method, though, there are a variety of directions we can go next. For instance, we could repeat the spectral type parameter study from chapter \ref{chap:ab_spec_type} but at $\tau=400$ to test whether all disks have the factor $4-5\times\dot{M}_{wind}$ ablation rate shown by the O7 star. We could also fill in the parameter range in $\tau$ with intermediate models to test how the strong wind of an O7 star reacts to different densities. Finally, and probably most importantly, we should do the proposed test at the end of chapter \ref{chap:ab_thick} and change the density power-law index to 1.5 to more accurately model the ablation of an accretion disk.

As briefly mentioned in the introduction, one potential area of interest for application of this work is B[e] stars \citep{KraMir06}. With O star-like luminosities and B-like effective temperatures, these post-main-sequence supergiants provide an interesting test bed for investigating the relative effects of stellar luminosity and stellar temperature for disk ablation. Additionally, the optically thick, radially extended, dusty B[e] disks provide an opportunity to study ablation of pre-main-sequence density disks without the need to contend with extensive extinction from the star's natal environment.

Finally, we should also attempt to more accurately treat the thermal effects associated with radiation transport. For all of the work here we have assumed purely isothermal disks and winds. While this is a good first order approximation, it is almost certainly not a realistic assumption in any of the cases treated. As material is irradiated by the stellar flux it should be expected to heat up to some fraction of the stellar effective temperature, here approximated by holding all material at the expected stellar effective temperature. However, in regions that are shaded from the star and regions that are dense, near the disk mid-plane in both cases, material can cool to much lower temperatures. This could substantially affect the structure of the disk, potentially removing or reversing the flaring of the disk which exposes high-lying material to direct stellar flux. It also could begin to introduce variations in the types of opacity which need to be addressed, particularly for the case of the star in formation where the UV resonance lines we address here will disappear as elements re-combine and begin to precipitate out as dust with its own associated opacity. This study will likely be a major focus of postdoctoral work by this doctoral candidate.

\subsection{Comparison of viscous and radiative disk destruction}

The results of this dissertation are all derived in the absence of an explicit viscosity prescription. Inclusion of such a prescription in VH-1 would be possible and would allow a direct comparison of numerical results using only radiation, using only viscosity, and using both. Particularly of interest in this line of questioning would be whether a transition occurs in the Classical Be star domain where viscous effects would be expected to outperform radiative effects, for normal viscosity parameters $\alpha\sim0.01$. A key challenge in this, however, would be to reconcile the long time-scales associated with viscosity to the rapid outflows associated with radiative acceleration.

\subsection{Confronting theory with observations}

Owing to the nearly 150 year history of interest in Classical Be stars, there is an extensive database of legacy observations that are freely available. One particularly accessible characteristic in these observations is stellar magnitude. As was shown by \cite{CarBjo12}, the transition between the presence and absence of a circumstellar disk is accompanied by a marked decline in V-band magnitude. By mining the vast number of observations, it should be possible to identify a large number of such declines as well as to measure their durations. The results of chapter \ref{chap:ab_spec_type} predict a trend of disk duration with spectral type. Such a trend is an unavoidable consequence of line-driven ablation, so its appearance in observation would provide a strong indicator for the presence and strength of radiative ablation. One possible issue here, however, is the dependence of the ejection mechanism on time and spectral type. If the ejection mechanism simply shuts off periodically, the trend should behave as predicted here. If it simply gets weaker, the expected trend is as of yet unknown.

As briefly addressed in chapter \ref{chap:disks} there are also direct observational signatures of ablation \citep{GraBjo87,GraBjo89}. By modeling individual systems, such signatures provide direct observational constraints for the radiative ablation work conducted here. Such constraints would allow us to further investigate the strength and efficiency expected for radiative ablation, as well as guiding us toward the potential need for inclusion of physics beyond the scope of what was done in this dissertation.

\subsection{An upper mass limit for star formation}

While the question of whether the upper mass limit of stars is controlled in part by their formation mechanics is briefly touched on in the introduction, the results of chapter \ref{chap:ab_thick} only begin to address this. To investigate such a question more thoroughly would require including mass feeding from the outer simulation boundary according to predictions of massive star formation theory, as well as a handling of the expected stellar structure which exists below the lower simulation boundary. The former issue of mass feeding rate and spatial distribution is already being investigated in the presence of continuum radiation forces by, for instance, \cite{KruKle07}, \cite{ComHen11}, and \cite{KuiYor15}. The later issue of underlying stellar properties and structure is under investigation by, for instance, the MESA \citep{PaxCan13} and Geneva \citep{GeoGra14} collaborations, as well as \cite{HosOmu09} who have put particular emphasis on the structure of a star still undergoing accretion. Therefore, the challenge is in combining accretion, stellar structure, and line-driven radiative ablation into a single simulation.

\subsection{High binary fraction in massive stars}

Line-driven ablation may even play a role in describing the high binary fraction observed for luminous, massive stars. As first introduced in \cite{KraMat06}, and then later confirmed by numerical models carried out by \cite{KraMat10}, circumstellar disks become subject to gravitational instabilities at sufficiently high accretion rates. As the accretion rate onto a luminous massive star must be high in order to compete with the very efficient radiative ablation, forming a massive star may often require accretion rates which come up against this gravitational instability upper limit, at which point the disk fragments and forms a binary companion.

\bibliography{dissertation}
\bibliographystyle{apj}

\appendix{Properties of rapidly rotating stars}\label{app:ob_gravdark}
%\begin{itemize}
%\item Because rapidly rotating stars are to be used, the shape and intensity profile of the star are needed.
%\item Jack this largely from old version
%\end{itemize}

Given that the population of stars to be studied is known to be rapidly rotating, understanding how this affects the underlying star is necessary. For the surface properties of the star, these effects predominantly reduce to a change of stellar shape and a redistribution of surface flux. Both effects are strongly dependent on knowing the rotation rate of the star.

\section{Stellar oblateness and the parameterization of rotation rate}

As a star is spun up, centrifugal forces reduce its effective gravity away from the poles. The net effect of this is to make the equatorial regions of the star less tightly bound than the polar regions, which allows the equator to swell to a larger size than the poles. In the most extreme case where the stellar equator is traveling rapidly enough to feel no net gravity (i.e. the equator is effectively in Keperian orbit) the equatorial radius can grow to 1.5 times the stellar radius at the pole.

To describe the shape that the star takes, one has to know how rapidly the star is actually rotating. In addition to the observational difficulty of disentangling rotation rate from viewing angle\footnote{Observations recover rotation velocity times sine of $i$, the inclination angle between the viewer and the star's rotation axis. This product is normally reported as $v\sin i$. To get back the actual equatorial rotation velocity requires auxiliary knowledge of $i$ and to get to rotation rate requires knowledge of stellar parameters.}, there are several different ways to parameterize the quantity that one gets out in terms of either a rotational or linear velocity. These velocities are always defined at the stellar equator but can either be defined for a ``critically-rotating'' object, for which the equatorial rotation velocity is the same as the circular Keplerian orbital velocity at the equator, or for the actual star in question. In the former case, the relevant quantities are

\begin{align}
v_{crit}& =\sqrt{\frac{2}{3}\frac{G M_\ast}{R_{pole}}}\label{eq:vcrit} \\
\Omega_{crit}&=\sqrt{\frac{8}{27}\frac{G M_\ast}{R_{pole}^3}}\label{eq:omcrit}\, ,
\end{align}
while, in the later case, the relevant quantities are
\begin{align}
v_{orb}& =\sqrt{\frac{G M_\ast}{R_{eq}}}\label{eq:vorb} \\
\Omega_{orb}&=\sqrt{\frac{G M_\ast}{R_{eq}^3}}\label{eq:omorb}\, .
\end{align}
Note that the factor of 2/3 in the first equations arises from the oblateness of a rotating star causing $R_{eq}=2/3R_{pole}$ for critical rigid body rotation under the Roche Approximation \citep[see, e.g.][chapter 2]{Mae09}.

Given these definitions, three common parameters exist in the literature to define the rotation rate of a star. The first two, based on the critical angular and linear velocity respectively are
\begin{align}
\omega&\equiv\frac{\Omega_{rot}}{\Omega_{crit}}\label{eq:omega} \\
\Upsilon&\equiv\frac{v_{rot}}{v_{crit}}\label{eq:upsilon}\, ,
\end{align}
which are more commonly used for stating statistical properties of the rotation rates of stars. The third, based on the equatorial Keplerian velocity,
\beq
W\equiv\frac{v_{rot}}{v_{orb}}\, ,
\eeq
will be the parameter predominantly used in this work. The choice is based on the additional information encoded in this parameter in the form of the necessary velocity boost to get from the stellar surface into orbit which has particular physical relevance in the case of Classical Be stars.

Since all three of these parameters are used in the literature, transformations between the three and the physical parameters of the star are essential. Using the relations provided by \cite{ColHar66}, gives \citep{RivCar13}
\beq
\Upsilon(\omega) = 2\cos\left(\frac{\pi +\arccos(\omega)}{3}\right)\, ,
\eeq
and
\beq
\omega(\Upsilon) = \cos\left(3\left[\arccos\left(\frac{\Upsilon}{2}\right) -\pi \right]\right)\, .
\eeq
Combining these two relations allows for calculation of the ratio of equatorial to polar radius as
\beq
\frac{R_{eq}}{R_{pole}}=\frac{3}{2} \frac{\Upsilon}{\omega}\, .
\eeq
Furthermore, by combining equations \ref{eq:vcrit} and \ref{eq:vorb} with this relation
\beq
\frac{v_{crit}}{v_{orb}}=\sqrt{\frac{2}{3}\frac{R_{eq}}{R_{pole}}}=\sqrt{\frac{\Upsilon}{\omega}}\, ,
\eeq
and finally
\beq
W=\frac{v_{rot}}{v_{orb}}=\frac{v_{rot}}{v_{crit}}\frac{v_{crit}}{v_{orb}}=\sqrt{\frac{\Upsilon^3}{\omega}}\, .
\eeq

Under the assumption that the star's gravity comes from a centrally concentrated point source, the potential felt by a parcel of gas is the sum of this gravity and a centripetal contribution
\beq
\Phi(r,\theta) = - \frac{G M_\ast}{r} - \frac{1}{2} \Omega^2 r^2 \sin(\theta)\, .
\eeq
The surface over which $\Phi$ is constant is referred to as the Roche equipotential. Since more detailed modeling of the stellar shape for a non-point mass shows that polar radius stays constant with increasing rotation rate to the level of around 1\% \citep{Orl61}, an analytic solution can be undertaken by using
\beq
\frac{GM_\ast}{R_{pole}}= \frac{G M_\ast}{R(\theta)} + \frac{1}{2} \Omega^2 R(\theta)^2 \sin(\theta)\, ,
\eeq
such that \citep[see, e.g.][]{Col63,ColHar66}
\beq
\frac{R(\theta)}{R_{pole}}=\frac{3}{\omega \sin\theta}\cos\left[\frac{\pi+\cos^{-1}(\omega \sin\theta)}{3}\right]\, .
\eeq

\section{Gravity darkening}

As shown by \cite{Zei24} and \cite{Cha33}, the radiative flux of a distorted star, such as the oblate, rapidly-rotating stars discussed here, is proportional to its local effective surface gravity. This effect is commonly referred to as ``gravity darkening''. Since the components of gravity are given by
\begin{align}
g_{r}&=-\frac{\partial \Phi}{\partial r}=-\frac{GM_\ast}{r^2}+\Omega^2r\sin^2\theta \\
g_{\theta}&=-\frac{1}{r}\frac{\partial \Phi}{\partial \theta}=\Omega^2r\sin\theta\cos\theta\, ,
\end{align}
and $g_{\phi}=0$ by azimuthal symmetry, the flux as a function of latitude can be established to be proportional to
\beq
|g(\theta)|=\sqrt{\frac{GM_\ast}{r^4}+\Omega^4r^2\sin^4\theta}\, .
\eeq
The factor that relates flux and surface gravity, referred to as von Zeipel's constant and commonly notated $K_{vz}$, then arises from the necessity that the total integral of flux be equal to bolometric luminosity such that
\beq
F = K_{vz} |g| = \frac{L_\ast}{\oint g dS} |g|\, .
\eeq
While not difficult to express, this integral can prove difficult to evaluate and must be handled numerically on a case by case basis.

\appendix{Translating Between the $\{\lowercase{q},\bar{Q},Q_0\}$, $\{\kappa,\kappa_0,\kappa_{\lowercase{max}}\}$, and $\{\lowercase{k}_L,\lowercase{k}_{\lowercase{max}}\}$ Notations}\label{app:kappa_to_q}

While much of this dissertation has bypassed this issue, the development of CAK theory and the addenda to it have caused the introduction of a variety of notations as well as the inclusion of several additional effects within the formalism. Of particular importance for the discussions in this dissertation is the treatment of the line distribution cutoff as introduced by OCR. While the formalism for the inclusion of this effect in the $\kappa$ notation of OCR is clear, the translation of the effect into the $q$ notation of \cite{Gay95} is not obvious. Additionally, as the line force parameters used in this work are drawn from \cite{PulSpr00}, it is essential that the line force parameters as defined therein are consistent with the definitions in OCR and \cite{Gay95}.

To begin, I first want to determine the inter-relation between the notations of Gayley and OCR. By comparing their definitions of the Sobolev optical depth, it can be seen that $q=\kappa v_{th}/(\kappa_e c)$. Using this, it is possible to compare the number distributions in the two notations. In OCR the number distribution is given by

\beq
\left| \frac{dN}{d\kappa}\right| = \frac{1}{\kappa_0}\left(\frac{\kappa}{\kappa_0}\right)^{\alpha-2} e^{-\kappa/\kappa_{max}} \, ,
\eeq
where Gayley provides

\beq
\left| \frac{dN}{dq}\right| = \frac{\bar{Q}}{\Gamma(\alpha)Q_0^2}\left(\frac{q}{Q_0}\right)^{\alpha-2}e^{-q/Q_0} \, ,
\eeq
to be his number distribution. For these to be compatible, their exponential cutoffs must occur in the same place, imposed by $Q_0=\kappa_{max}v_{th}/(\kappa_e c)$. By using these inter-relations of $q$ and $\kappa$ and $Q_0$ and $\kappa_{max}$, the number distributions can be shown to be compatible with one another if

\beq
\bar{Q}Q_0^{-\alpha} = \left(\frac{\kappa_0 v_{th}}{\kappa_e c}\right)^{1-\alpha} \Gamma(\alpha)\, ,
\eeq
which implies

\beq
\frac{\kappa_{max}}{\kappa_0}=\left(\frac{Q_0}{\bar{Q}}\Gamma(\alpha)\right)^{1/(1-\alpha)}\, .
\eeq

Given these relations, the remaining task is that of determining whether the definitions of Puls et al. are consistent with those derived above. Using their line distribution function

\beq
\left|\frac{dN}{dk_L}\right|=N_0 k_L^{\alpha-2} e^{-k_L/k_{max}}\, ,
\eeq
and their given relations of $q$, $Q_0$, and $Q_{max}$ to $k_L$ and $k_{max}$

\begin{align}
q &= k_L \frac{v_{th}}{c} \\
Q_0 &= k_{max} \frac{v_{th}}{c} \\
\bar{Q} &= N_0 \frac{v_{th}}{c} \Gamma(\alpha)k_{max}^\alpha \, ,
\end{align}
exactly reproduces the line distribution function of \cite{Gay95}, closing the loop and allowing the $\bar{Q}$ and $Q_0$ values of \cite{PulSpr00} to be used in the formalism of OCR with the translations given above.
\appendix{Inclusion of the effects of ionization and recombination as a function of position in the wind}\label{app:ionization}

An additional correction that can be made to the radiation force and mass loss rate scaling, as has been done in this work, is accounting for the competition of ionization and recombination in the stellar wind. Since the line force depends on the number and strength of lines, a full treatment would require calculating the Non-Local Thermodynamic Equilibrium (NLTE) ionization structure. However, the effects of photoionization can be assumed proportional the dilution factor
\beq
W\equiv\frac{1-\mu_\ast}{2}
\eeq
and the effects of recombination proportional to the number density of electrons $n_e$. The usual treatment is the use the ratio $n_{e,11}/W$ as the parameter of interest \citep{Abb82} where $n_{e,11}\equiv n_e/(10^{11} \mathrm{cm}^{-3})$. As neither the force nor mass loss rate are found to vary linearly with this term, a power-law fit is carried out using the free parameter $\delta$ as the power-law index such that
\beq
g_{line,ion}=g_{line}\left(\frac{n_{e,11}}{W}\right)^\delta=g_{line}\left(\frac{2\dot{M}(1+\mu_\ast)}{4\pi R_\ast^2 v_r 10^{11} \mu_e}\right)^\delta\, ,
\eeq
where the second equality is achieved by making the substitutions $n_e=\rho/\mu_e$ and $1-\mu_\ast^2=(R_\ast/r)^2$. Given that $\dot{M}$ is a constant this leaves us needing to calculate both $v_r$ and $\mu_\ast$. In the course of handling the finite sound speed correction to the mass loss rate, \cite{OwoudD04} did just this. At what they refer to as the ``critical point'' where radiative driving will be the most difficult,
\beq
r_c=\frac{R_\ast}{1-\left(a\sqrt{1-\alpha}\right)/\left(2v_{esc}\right)}\,,
\eeq
which means that
\beq
\mu_{\ast,c}=\frac{a}{v_{esc}}\sqrt{1-\alpha}-\left(\frac{a}{v_{esc}}\right)^2(1-\alpha)\,.
\eeq
At this same point, the wind velocity is
\beq
v_c = \sqrt{\frac{\alpha}{2}}(1-\alpha)^{-1/4}\sqrt{av_{esc}}\,.
\eeq
With the necessary values for $\mu_\ast$ and $v_c$ in hand, the effects of ionization and recombination will enter $\dot{M}$ such that
\beq
\dot{M} = \frac{\bar{Q}^{\frac{1}{\alpha-\delta}}}{(1 + \alpha)^{\frac{1}{\alpha}}}\left(\frac{\alpha}{1-\alpha}\frac{L_*}{Q_\mathrm{o} c^2}\right)^{\frac{\alpha}{\alpha-\delta}} \left(\frac{\Gamma_e}{1-\Gamma_e}\right)^{\frac{1-\alpha}{\alpha-\delta}}\left(\frac{2(1+\mu_{\ast,c})}{4\pi R_\ast^2 v_c 10^{11} \mu_e}\right)^{\frac{\delta}{\alpha-\delta}}\, .
\eeq

\appendix{Permission for Use of Figures}

\begin{figure}
\centering
\includegraphics[width=\textwidth]{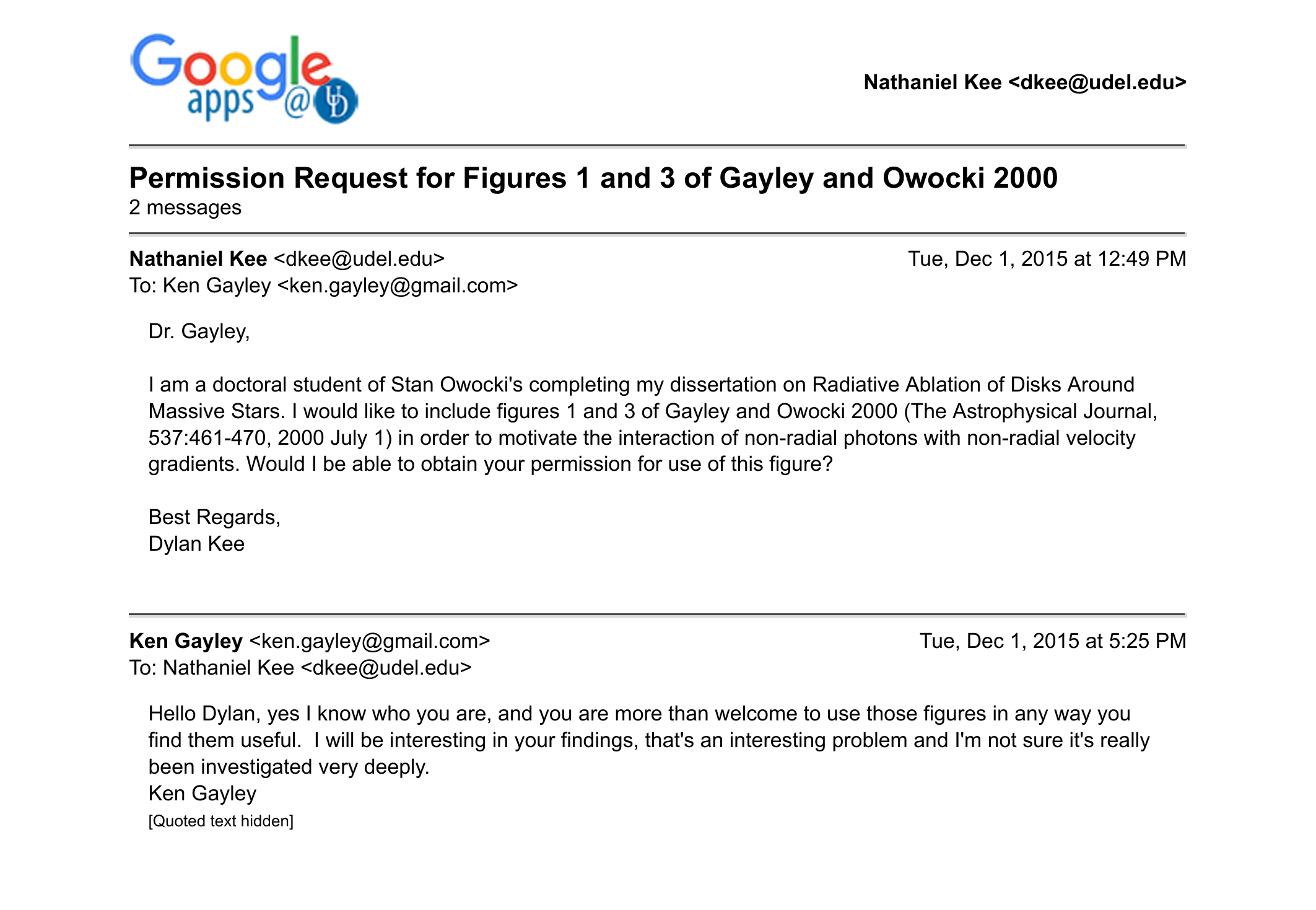}
\caption{Permission Letter \#1}
\end{figure}

\begin{figure}
\centering
\includegraphics[width=\textwidth]{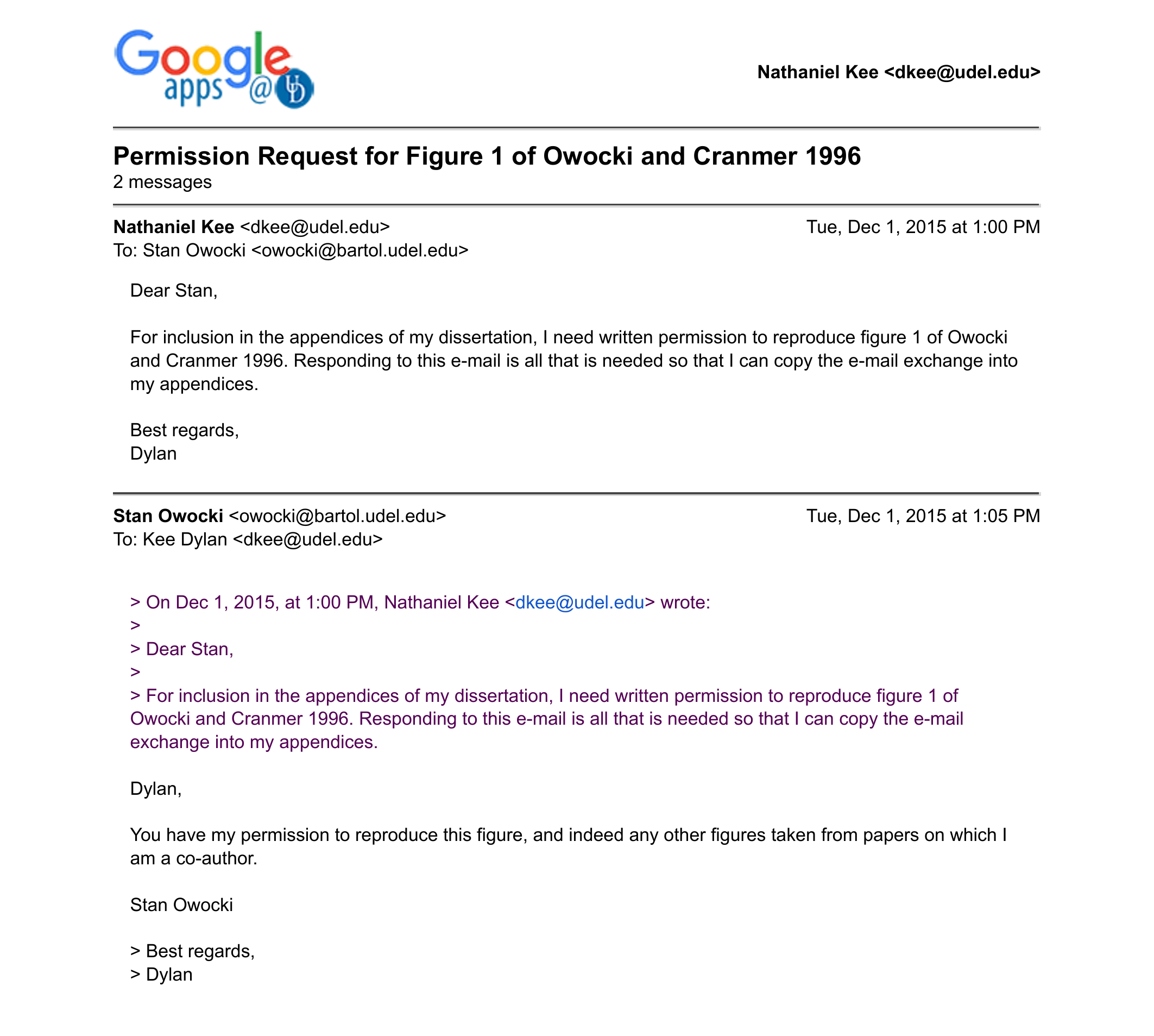}
\caption{Permission Letter \#2}
\end{figure}

\begin{figure}
\centering
\includegraphics[width=\textwidth]{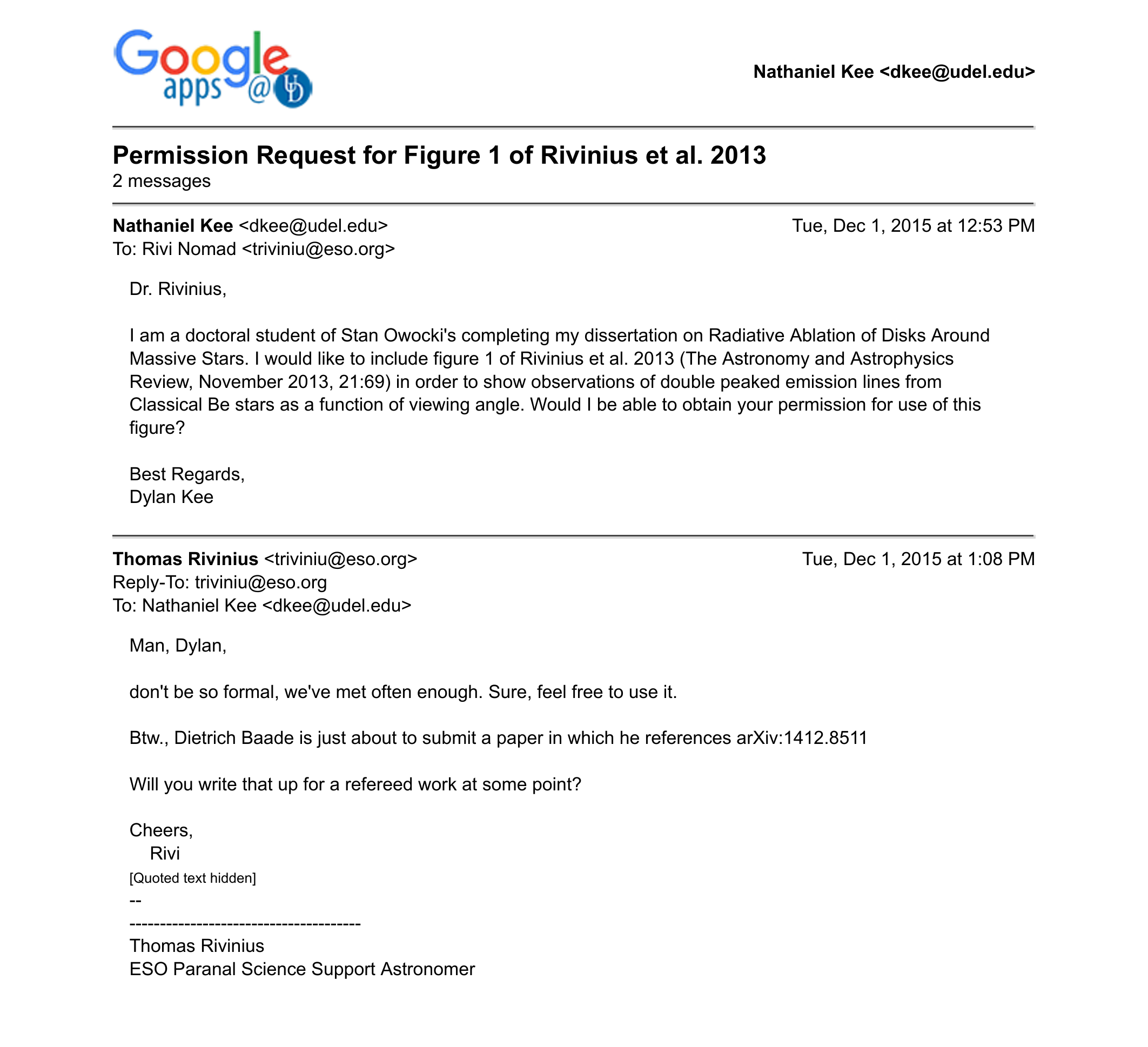}
\caption{Permission Letter \#3}
\end{figure}

\begin{figure}
\centering
\includegraphics[width=\textwidth]{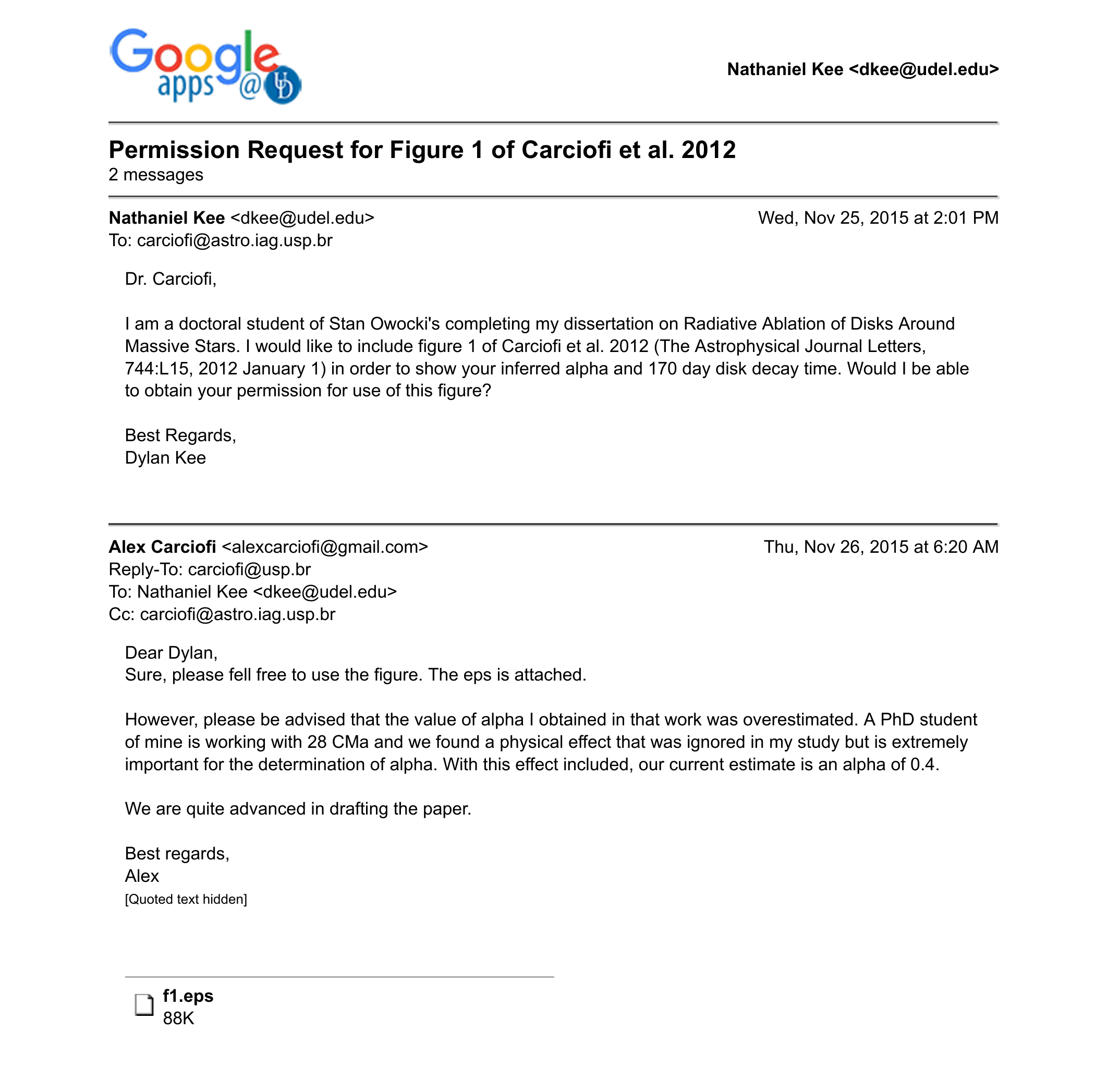}
\caption{Permission Letter \#4}
\end{figure}

\end{document}